\documentclass[journal]{IEEEtran}

\makeatletter
\def\ps@headings{%
\def\@oddhead{\mbox{}\scriptsize\rightmark \hfil \thepage}%
\def\@evenhead{\scriptsize\thepage \hfil \leftmark\mbox{}}%
\def\@oddfoot{}%
\def\@evenfoot{}}
\makeatother
\usepackage{subfigure}
\usepackage{graphicx, amssymb, amsbsy, amsmath}
\usepackage{subfigure}
\usepackage{multirow}
\usepackage{cite,url}
\usepackage{color,soul}
\usepackage[font=footnotesize,labelfont=sf,textfont=sf]{caption}
\usepackage{algorithm}
\usepackage{algorithmic}
\DeclareGraphicsExtensions{.eps}

\def\squareforqed{\hbox{\rlap{$\sqcap$}$\sqcup$}}
\def\qed{\ifmmode\squareforqed\else{\unskip\nobreak\hfil
\penalty50\hskip1em\null\nobreak\hfil\squareforqed
\parfillskip=0pt\finalhyphendemerits=0\endgraf}\fi}

\ifodd 1
\newcommand{\com}[1]{\textbf{\color{red} (COMMENT: #1)}} %comment of the text
\newcommand{\comg}[1]{\textbf{\color{green} (COMMENT: #1)}}
\newcommand{\response}[1]{\textbf{\color{magenta} (RESPONSE: #1)}} %response to comment
\else

\newcommand{\com}[1]{}
\newcommand{\comg}[1]{}
\newcommand{\response}[1]{}
\fi

\begin{document}

\title{Edge Intelligence: Paving the Last Mile of Artificial Intelligence with Edge Computing}

\author{Zhi Zhou, Xu Chen, En Li, Liekang Zeng, Ke Luo, Junshan Zhang\vspace{-0.8cm}
%\thanks{This work was supported in part by the National Key Research and Development Program of China under Grant 2017YFB1001703, National Science Foundation of China under Grants U1711265 and 61802449, the Program for Guangdong Introducing Innovative and Enterpreneurial Teams under Grant 2017ZT07X355, the Guangdong Natural Science Funds under Grant 2018A030313032, the Fundamental Research Funds for the Central Universities under Grant 17lgjc40, the U.S. Army Research Office under Grant W911NF-16-1-0448 and DTRA under Grant HDTRA1-13-1-0029. }
%(Corresponding author: Xu Chen.)}
\thanks{{Z. Zhou}, {X. Chen}, {E. Li}, {L. Zeng} and {K. Luo} are with the School of Data and Computer Science, Sun Yat-sen University (SYSU), Guangzhou 510006, China. E-mail: {\tt \{zhouzhi9, chenxu35\}@mail.sysu.edu.cn}, {\tt \{lien5, luok7, zenglk3\}@mail2.sysu.edu}.}

\thanks{{J. Zhang} is with the School of Electrical, Computer and Energy Engineering, Arizona State University, Tempe, AZ 85287-7206, USA. E-mail: {\tt Junshan.Zhang@asu.edu}.}}

\maketitle

\begin{abstract}
%\boldmath

With the breakthroughs in deep learning, the recent years have witnessed a booming of artificial intelligence (AI) applications and services, spanning from personal assistant to recommendation systems to video/audio surveillance. More recently, with the proliferation of mobile computing and Internet-of-Things (IoT), billions of mobile and IoT devices are connected to the Internet, generating zillions Bytes of data at the network edge. Driving by this trend, there is an urgent need to push the AI frontiers to the network edge so as to fully unleash the potential of the edge big data. To meet this demand, edge computing, an emerging paradigm that pushes computing tasks and services from the network core to the network edge, has been widely recognized as a promising solution. The resulted new inter-discipline, edge AI or edge intelligence, is beginning to receive a tremendous amount of interest. However, research on edge intelligence is still in its infancy stage, and a dedicated venue for exchanging the recent advances of edge intelligence is highly desired by both the computer system and artificial intelligence communities. To this end, we conduct a comprehensive survey of the recent research efforts on edge intelligence. Specifically, we first review the background and motivation for artificial intelligence running at the network edge. We then provide an overview of the overarching architectures, frameworks and emerging key technologies for deep learning model towards training/inference at the network edge. Finally, we discuss future research opportunities on edge intelligence. We believe that this survey will elicit escalating attentions, stimulate fruitful discussions and inspire further research ideas on edge intelligence.
\end{abstract}
\IEEEpeerreviewmaketitle

\section{Introduction}
\label{sec:intro}

\IEEEPARstart{W}{e} are living in an unprecedented booming era of artificial intelligence (AI). Driving by the recent advancements of algorithm, computing power and big data, deep learning \cite{Lecun2015Deep} --- the most dazzling sector of AI --- has made substantial breakthroughs in a wide spectrum of fields, ranging from computer vision, speech recognition, natural language processing to chess playing (e.g., AlphaGo) and robotics \cite{deng2014deep}. Benefited from these breakthroughs, a set of intelligent applications, as exemplified by intelligent personal assistants, personalized shopping recommendation, video surveillance and smart home appliances have quickly ascended to the spotlight and gained enormous popularity. It is widely recognized that these intelligent applications are significantly enriching people's lifestyle, improving human productivity, and enhancing social efficiency.

As a key driver that boosts AI development, big data has recently gone through a radical shift of data source from the mega-scale cloud datacenters to the increasingly widespread end devices, e.g., mobile devices and IoT devices. Traditionally, big data, such as online shopping records, social media contents and business informatics, were mainly born and stored at mega-scale datacenters. However, with the proliferation of mobile computing and IoT, the trend is reversing now. Specifically, Cisco estimates that nearly 850 ZB will be generated by all people, machines, and things at the network edge by 2021 \cite{ciscogci}. In sharp contrast, the global datacenter traffic will only reach 20.6 Zettabytes by 2021. Clearly, via bringing the huge volumes of data to AI, the edge ecosystem will present many novel application scenarios for AI and fuel the continuous booming of AI.

Pushing the AI frontier to the edge ecosystem that resides at the last mile of the Internet, however, is highly non-trivial, due to the concerns on performance, cost and privacy. Towards this goal, the conventional wisdom is to transport the data bulks from the IoT devices to the cloud datacenters for analytics \cite{ben2015group}. However, when moving a tremendous amount of data across the wide-area-network (WAN), both monetary cost and transmission delay can be prohibitively high, and the privacy leakage can also be a major concern \cite{pu2015low}. An alternative is on-device analytics that run AI applications on the device to process the IoT data locally, which, however, may suffer from poor performance and energy efficiency. This is because many AI applications require high computational power that greatly outweighs the capacity of resource- and energy-constrained IoT devices.

%\begin{figure*}[!t]
%	\centering %\vspace{-100pt}
%	\includegraphics[width=5.6in]{fig/overview.eps}
%	 \caption{An overview of the organization of this paper}
%	\label{overview} 
%\end{figure*}

\begin{figure}[!t]  
	\centering %\vspace{-100pt}
	\includegraphics[width=3.0in]{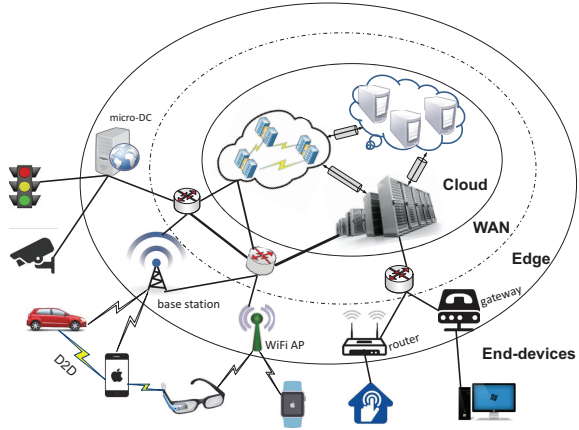}
	 \caption{An illustration of edge computing}
	\label{fig-edge} \vspace{-15pt}
\end{figure}

To address the above challenges, edge computing \cite{shi2016edge} has recently been proposed, which pushes cloud services from the network core to the network edges that are in closer proximity to IoT devices and data sources. As shown in Fig. \ref{fig-edge}, here an edge node can be nearby end-device connectable by device-to-device (D2D) communications \cite{chen2017exploiting}, a server attached to an access point (e.g., WiFi, router, base station), a network gateway, or even a micro-datacenter available for use by nearby devices. While edge nodes can be varied in size: ranging from a credit-card-sized computer to a micro-datacenter with several server racks, physical proximity to the information-generation sources is the most crucial characteristic emphasized by edge computing. Essentially, the physical proximity between the computing and information-generation sources promises several benefits compared to the traditional cloud-based computing paradigm, including low-latency, energy-efficiency, privacy protection, reduced bandwidth consumption, on-premises and context-awareness \cite{shi2016edge,mao2017survey}.

%promising significant improvements on performance, cost-efficiency and privacy protection. 

Indeed, the marriage of edge computing and AI has given rise to a new research area, namely "edge intelligence" or "edge AI" \cite{wang2018edge, Li2018edgeintelligence}. Instead of entirely relying on the cloud, edge intelligence makes the most of the widespread edge resources to gain AI insight. Notably, edge intelligence has garnered much attention from both the industry and academia. For example, the celebrated Gartner hype cycle has incorporated edge intelligence as an emerging technology that will reach a plateau of productivity in the following 5 to 10 years \cite{gartneredgeai}. Major enterprises, including Google, Microsoft, Intel and IBM, have put forth pilot projects to demonstrate the advantages of edge computing in paving the last mile of AI. These effort have boosted a wide spectrum of AI applications, spanning from live video analytics \cite{ananthanarayanan2017real}, cognitive assistance \cite{ha2014towards} to precision agriculture, smart home \cite{Jie2017EdgeOS} and industrial internet-of-things (IIoT) \cite{li2018deep}.

Notably, research and practice on this emerging inter-discipline --- edge intelligence, is still in a very early stage. There is,, in general a lack of venue dedicated for summarizing, discussing, and disseminating the recent advances of edge intelligence, in both industrial and academia. To bridge this gap, in this paper we conduct a comprehensive and concrete survey of the recent research efforts %as well as industrial progresses 
on edge intelligence. Specifically, we will first review the background of artificial intelligence. We will then discuss the motivation, definition and rating of edge intelligence. Next, we will further review and taxonomically summarize the emerging computing architectures and enabling technologies for edge intelligence model training and inference. Finally, we will discuss some open research challenges and opportunities on edge intelligence. The paper is organized as follows:
\begin{itemize}
\item Sec. II gives an overview of the basic concepts of artificial intelligence, with a focus on deep learning --- the most popular sector of AI.
\item Sec. III discusses the motivation, definition, and rating of edge intelligence.
\item Sec. IV reviews the architectures, enabling techniques, systems and frameworks for training edge intelligence models.
\item Sec. V reviews the architectures, enabling techniques, systems and frameworks for edge intelligence model inference.
\item Sec. VI discusses future directions and challenges of edge intelligence.
\end{itemize}

For this survey, we hope it can elicit escalating attentions, stimulate fruitful discussions and inspire further research ideas on edge intelligence.

\section{A Primer on Artificial Intelligence}
\label{dnnbackground}
In this section, we review the concepts, models and methods for artificial intelligence, with a particular focus on deep learning --- the most popular sector of artificial intelligence.

\subsection{Artificial Intelligence}
While AI has recently ascended to the spotlight and gained tremendous attention, it is not a new term and it was first coined in 1956. Simply put, AI is an approach to building intelligent machines capable of carrying out tasks as humans do. This is obviously a very broad definition, and it can refer from Apple Siri to Google AlphaGo and too powerful technologies yet to be invented. In simulating human intelligence, AI systems typically demonstrate at least some of the following behaviors associated with human intelligence: planning, learning, reasoning, problem-solving, knowledge representation, perception, motion, and manipulation and, to a lesser extent, social intelligence and creativity. During the past 60 year's development, AI has experienced rise, fall and again rise and fall. The latest rise of AI after 2010's was partially due to the breakthroughs made by deep learning, a method that has achieved human-level accuracy in some interesting areas.  

\subsection{Deep Learning and Deep Neural Networks}
\begin{figure}[htbp]
	\centering
	\subfigure[The layers in a DL model]{
		\includegraphics[scale=0.33]{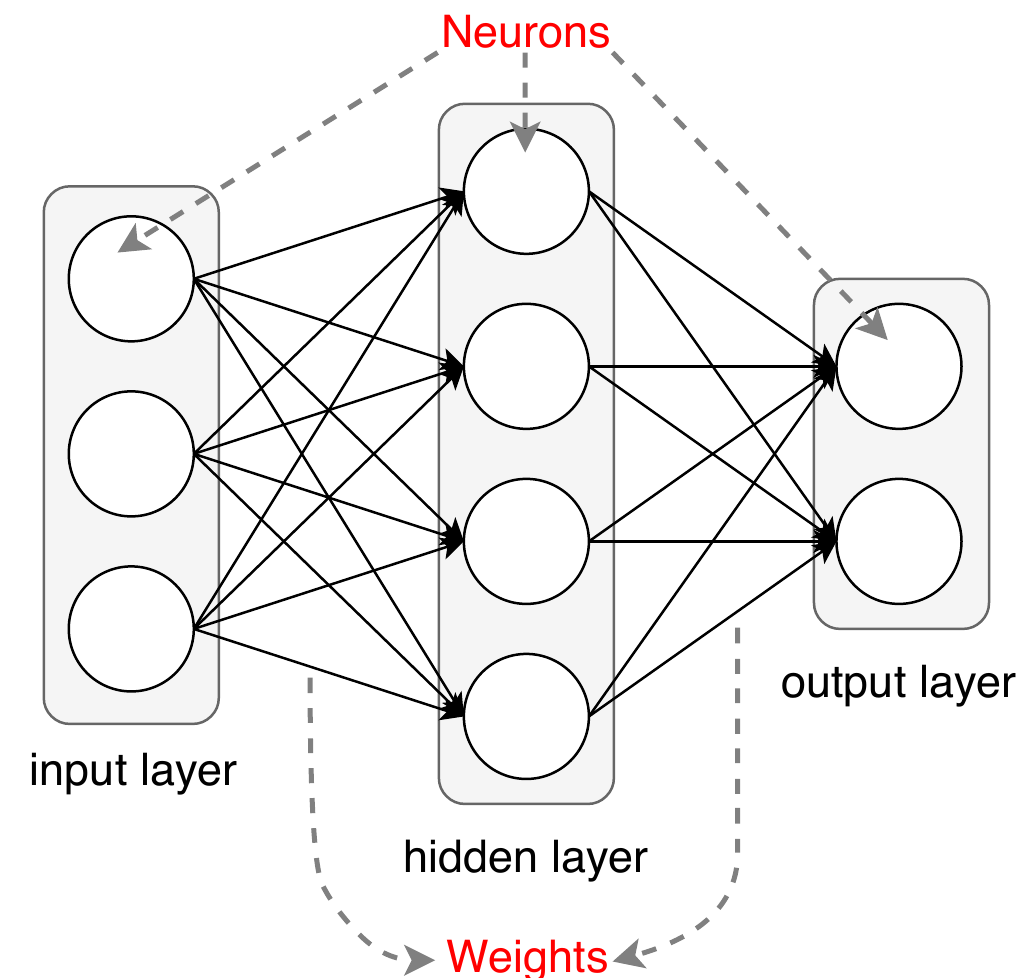}\label{dldnn}
	}
	\subfigure[The architecture of a neuron]{
		\includegraphics[scale=0.43]{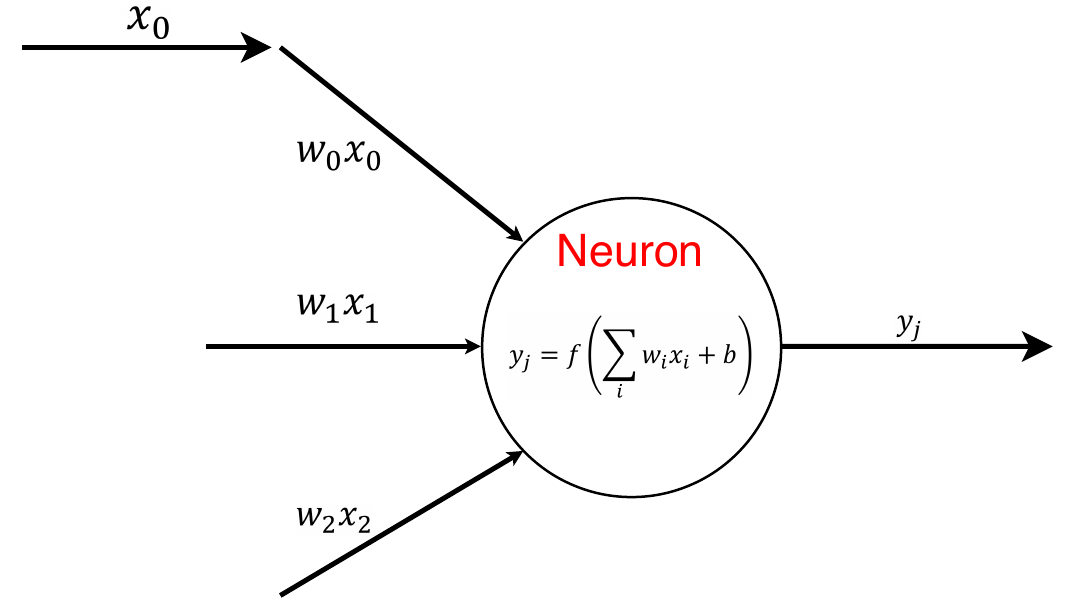}\label{dlneuron}
	}
	\caption{A standard composition of DL model}\label{dl composition}
\end{figure}

Machine learning (ML) is an effective method to achieve the goal of AI. %The relationship between deep learning and AI is illustrated in Fig. \ref{dlai}. 
Many machine learning methodologies as exemplified by decision tree, K-means clustering and Bayesian network, etc. have been developed to train the machine to make classifications and predictions, based on the data obtained from the real world. Among the existing machine learning methods, deep learning, by leveraging artificial neural network (ANN) \cite{svozil1997introduction} to learn the deep representation of the data, have resulted in an amazing performance in multiple tasks, including image classification, face recognition, etc. Since the ANN adopted by deep learning model typically consists of a series of layers, the model is called deep neural network (DNN). As shown in Fig. \ref{dl composition}, each layer of a DNN is composed of neurons that are able to generate the non-linear outputs based on the data from the input of the neuron. 

\begin{figure}[htbp]
	\centering
	\subfigure[Multilayer Perceptrons]{
		\includegraphics[scale=0.3]{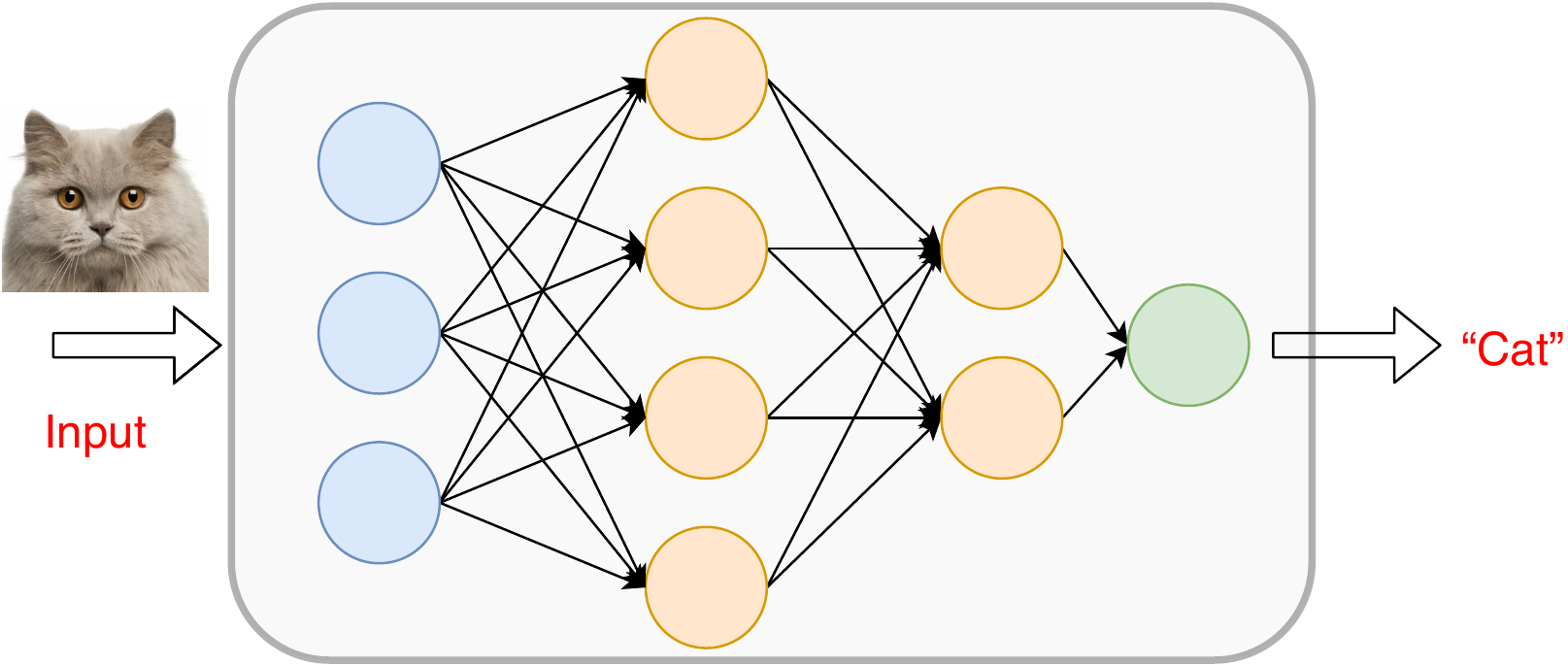}\label{mlp}
	}
	\subfigure[Convolution Neural Network]{
		\includegraphics[scale=0.43]{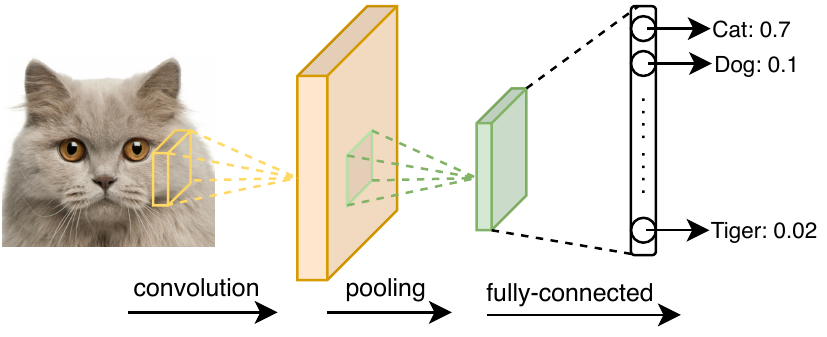}\label{cnn}
	}
	\subfigure[Recurrent Neural Network]{
		\includegraphics[scale=0.37]{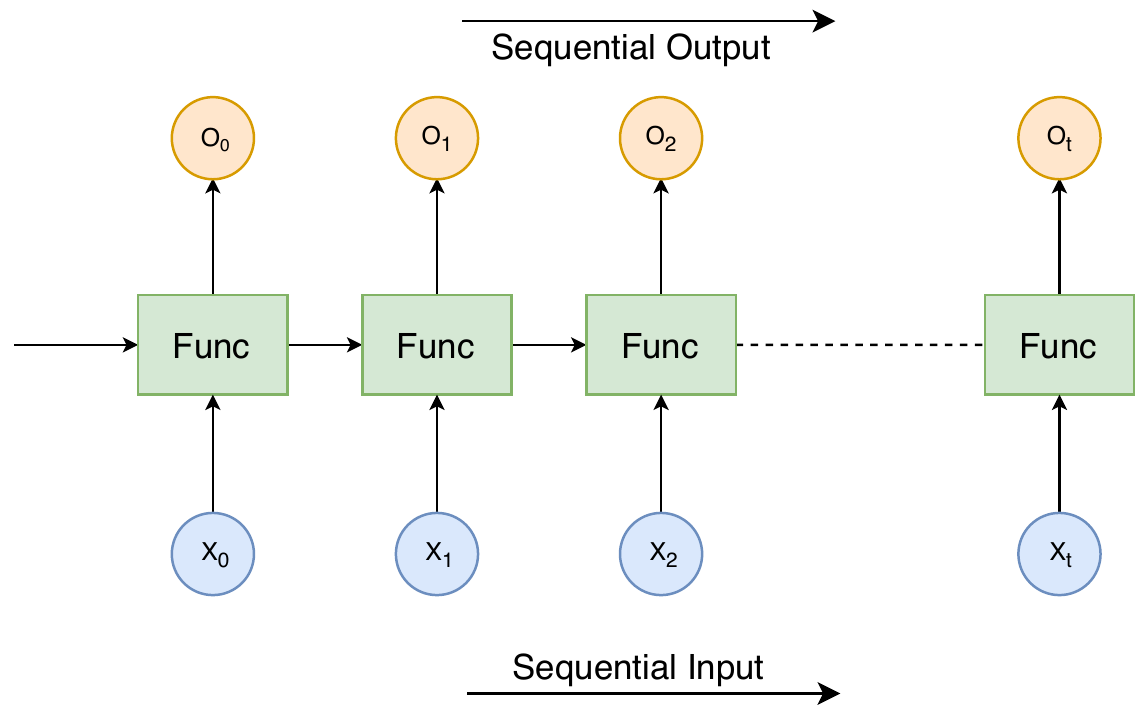}\label{rnn}
	}
	\caption{Three typical structures of DL models}\label{dl structure}
\end{figure}

The neurons in the input layer receive the data and propagate them to the middle layer (a.k.a the hidden layer). Then the neurons in the middle layer generate the weighted sums of the input data and output the weighted sums using the specific activation functions (e.g., tanh), and the outputs are then propagated to the output layer. The final results are presented at the output layer. With more complex and abstract layers than a typical model, DNNs are able to learn the high-level features, enabling high precision inference in tasks. Fig. \ref{dl structure}  presents three popular structures of DNNs: Multilayer Perceptrons (MLPs), Convolution Neural Network (CNN) and Recurrent Neural Network (RNN).

MLPs models are the most basic deep neural network, which is composed of a series of fully-connected layers \cite{collobert2011natural}. Different from fully-connected layers in MLPs, in CNN models, the convolution layers extract the simple features from input by executing convolution operations. Applying various convolutional filters, CNN models can capture the high-level representation of the input data, making it most popular  for computer vision tasks, e.g., image classification (e.g., AlexNet \cite{krizhevsky2012imagenet}, VGG network \cite{simonyan2014very}, ResNet \cite{he2016deep}, MobileNet \cite{howard2017mobilenets}) and object detection (e.g., Fast R-CNN \cite{mao2018towards}, YOLO \cite{redmon2016you}, SSD \cite{liu2016ssd}). RNN models are another type of DNNs, which use sequential data feeding. Shown in Fig. \ref{rnn}, the basic unit of RNN is called cell; and further, each cell consists of layers and a series of cells enables the sequential processing of RNN models. RNN models are widely used in the task of natural language processing, e.g., language modeling, machine translation, question answering and document classification.

Deep learning represents the state-of-the-art AI technology as well as a highly resource-demanding workload that naturally suits for edge computing. Therefore, due to space limitation, in the remaining of this paper, we will focus on the interaction between deep learning and edge computing. We believe that the techniques discussed can also have meaningful implications for other AI models and methods, i.e., stochastic gradient descent is a popular training method for many AI/ML algorithms (e.g., k-means, support vector machine, lasso regression) \cite{bottou2010large}, the optimization techniques of stochastic gradient descent training introduced in this paper can be also deployed on other AI models training process.

\subsection{From Deep Learning to Model Training and Inference}

For each neuron in a DNN layer, it has a vector of weights associated with the input data size of the layer. Needless to say, the weights in a deep learning model need to be optimized through a training process. 

In a training process for a deep learning model, the values of weights in the model are often randomly assigned initially. Then the output of the last layer represents the task result, and a loss function is set to evaluate the correctness of the results by calculating the error rate (e.g., root mean squared error) between the results and the true label. To adjust the weights of each neuron in the model, an optimization algorithm, such as Stochastic Gradient Descent (SGD) \cite{bottou2010large}, is used and the gradient of the loss function is calculated. Leveraging the Back Propagation mechanism \cite{rumelhart1986learning,chauvin2013backpropagation}, the error rate is propagated back across the whole neural network and the weights are updated based on the gradient and the learning rate. By feeding a large number of training samples and repeating this process until the error rate is below a predefined threshold, a deep learning model with high precision is obtained.

%

%\subsection{From Deep Learning to Model Inference}
DNN model inference happens after training. For instance, for an image classification task, with the feeding of a large amount of training samples,  the deep neural network is trained to learn how to recognize an image, and then inference takes real-world images as inputs and quickly draws the predictions/classifications of them. The training procedure consists of the feed-forward process and the backpropagation process. Note that the inference involves the feed-forward process only, i.e., the input from real-world is passed through the whole neural network and the model outputs the prediction. %Typically, the training process requires significant computational resources for iterative parameter updates and thus is often performed in the cloud data center. The inference process, on the other hand, is executed either at the local devices (e.g., mobile phones) or in the cloud data center. Due to the resource constraint on mobile devices, many current applications (e.g., Apple Siri, Microsoft Cortana) run in the cloud data center. 

%\subsection{Deep Learning Applications}
%Deep learning has been demonstrated to be effective in a wide spectrum of applications, including:

%\begin{itemize}
%	\item \textbf{Computer Vision}: Given a series of images or video from real-world, the AI system learns to automatically extract the features of these inputs to complete a specific task, e.g, image classification, face authentication, image semantic segmentation.
%	\item \textbf{Natural Language Processing}: In this field, the task of AI is to build the system that can understand and comprehend natural language spoken by humans, e.g., natural language modeling, word embedding, machine translation.
%	\item \textbf{Speech Recognition}: AI system accepts the human voice as input, hearing and comprehending the language in terms of sentences and meanings. It also can handle the accents from different regions, noise in background, and trendy words.
%	\item \textbf{Intelligent Robots}: There are multiple sensors on a robot, e.g., light, heat, temperature, sound. AI system aggregates the sensing inputs and learns to guide the action  of the robots in real-world environment.
%\end{itemize}

\subsection{Popular Deep Learning Models}
For a better understanding of the deep learning and their applications, in this subsection we give an overview of various popular deep learning models.

\noindent \textbf{Convolution Neural Network (CNN):} For image classification, as the first CNN to win the ImageNet Challenge in 2012, AlexNet \cite{krizhevsky2012imagenet} consists of 5 convolution layers and 3 fully-connected layers. AlexNet requires 61 million weights and 724 million MACs (Multiply-Add Computation) to classify the image with a size of 227*227. To achieve higher accuracy, VGG-16 \cite{simonyan2014very} is trained to a deeper structure of 16 layers consisting of 13 convolution layers and 3 fully-connected layers, requiring 138 million weights and 15.5G MACs to classify the image with a size of 224*224. To improve accuracy while reducing the computation of DNN inference, GoogleNet \cite{szegedy2015going} introduces an inception module composed of different sized filters. GoogleNet achieves a better accuracy performance than VGG-16, while only requiring 7 million weights and 1.43G MACs to process the image with the same size. ResNet \cite{he2016deep} , the state-of-the-art effort, uses the "shortcut" structure to reach a human-level accuracy with a top-5 error rate below 5\%. The ``shortcut" module is used to solve the gradient vanishing problem during the training process, making it possible to train a DNN model with deeper structure. Convolution neural network typically employed in computer vision. Given a series of images or video from real-world, with the utilization of CNN, the AI system learns to automatically extract the features of these inputs to complete a specific task, e.g.,  image classification, face authentication, image semantic segmentation.

%\noindent \textbf{VGG:} To achieve a higher accuracy, VGG-16 \cite{simonyan2014very} is trained to a deeper structure of 16 layers consisting of 13 convolution layers and 3 fully-connected layers, requiring 138 million weights and 15.5G MACs to classify the image with a size of 224*224. 

%\noindent \textbf{GoogleNet:} To improve the accuracy while reducing the computation of DNN inference, GoogleNet \cite{szegedy2015going} introduces an inception module composed of different sized filters. GoogleNet achieves a better accuracy performance than VGG-16, while only requiring 7 million weights and 1.43G MACs to process the image with the same size.
%\begin{figure}[H]
%	\centering
%	% Requires \usepackage{graphicx}
%	\includegraphics[scale=0.4]{fig/dnn_inference/googlenet.pdf}\%\
%	\caption{Inception module of GoogleNet}\label{googlenet}
%\end{figure}

%\noindent \textbf{ResNet:}  ResNet \cite{he2016deep} , the state-of-the-art effort, uses the "shortcut" structure to reach a human-level accuracy with a top-5 error rate below 5\%. The ``shortcut" module %showed in Fig. \ref{resnet} 
%is used to solve the gradient vanishing problem during the training process, making it possible to train a DNN model with deeper structure.
%\begin{figure}[H]
%	\centering
%	% Requires \usepackage{graphicx}
%	\includegraphics[scale=0.4]{fig/dnn_inference/resnet.pdf}\\
%	\caption{"Shortcut" structure of ResNet}\label{resnet}
%\end{figure}

\begin{figure}[htbp]
	\centering
	\includegraphics[scale=0.65]{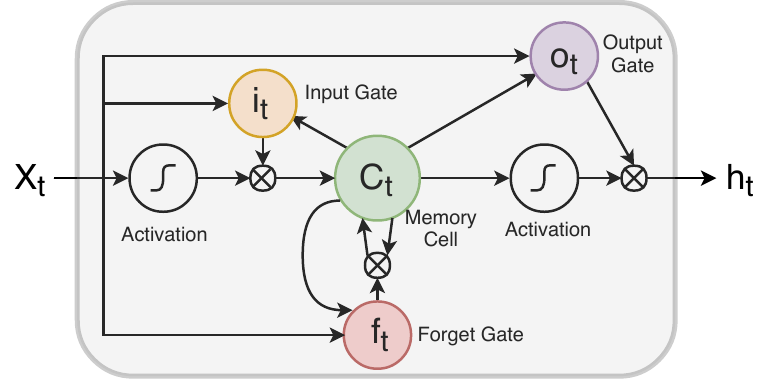}
	\caption{Structure of a LSTM memory cell}\label{lstm}
\end{figure}

\noindent \textbf{Recurrent Neural Network (RNN):} For sequential input data, recurrent neural networks (RNNs) have been developed to address the time-series problem. The input of RNN consists of the current input and the previous samples. Each neuron in a RNN owns an internal memory that keeps the information of the computation from the previous samples. The training of RNN is based on Backpropagation Through Time (BPTT) \cite{werbos1990backpropagation}. Long Short Term Memory (LSTM) \cite{hochreiter1997long} is an extended version of RNNs. In LSTM, the gate is used to represents the basic unit of a neuron. As shown in Fig. \ref{lstm}, each neuron in LSTM is called memory cell and includes a multiplicative forget gate, input gate, and output gate. These gates are used to control the access to memory cells and to prevent them from perturbation by irrelevant inputs. Information is added or removed through the gate to the memory cell. Gates are different neural networks that determine what information is allowed on the memory cell. The forget gate can learn what information is kept or forgotten during training. Recurrent neural network has been widely used in natural language processing due to the superior of processing the data with an input length that is not fixed. The task of the AI here is to build a system that can comprehend natural language spoken by humans, e.g., natural language modeling, word embedding, machine translation.

%\noindent \textbf{Long Short Term Memory:} Long Short Term Memory (LSTM) \cite{hochreiter1997long} is an extended version of RNNs. In LSTM, the gate is used to represents the basic unit of a neuron. As shown in Fig. \ref{lstm}, each neuron in LSTM is called memory cell and includes a multiplicative forget gate, input gate, and output gate. These gates are used to control the access to memory cells and to prevent them from perturbation by irrelevant inputs. Information is added or removed through the gate to the memory cell. Gates are different neural networks that determine what information is allowed on the memory cell. The forget gate can learn what information is kept or forgotten during training.

\begin{figure}[htbp]
	\centering
	\includegraphics[scale=0.3]{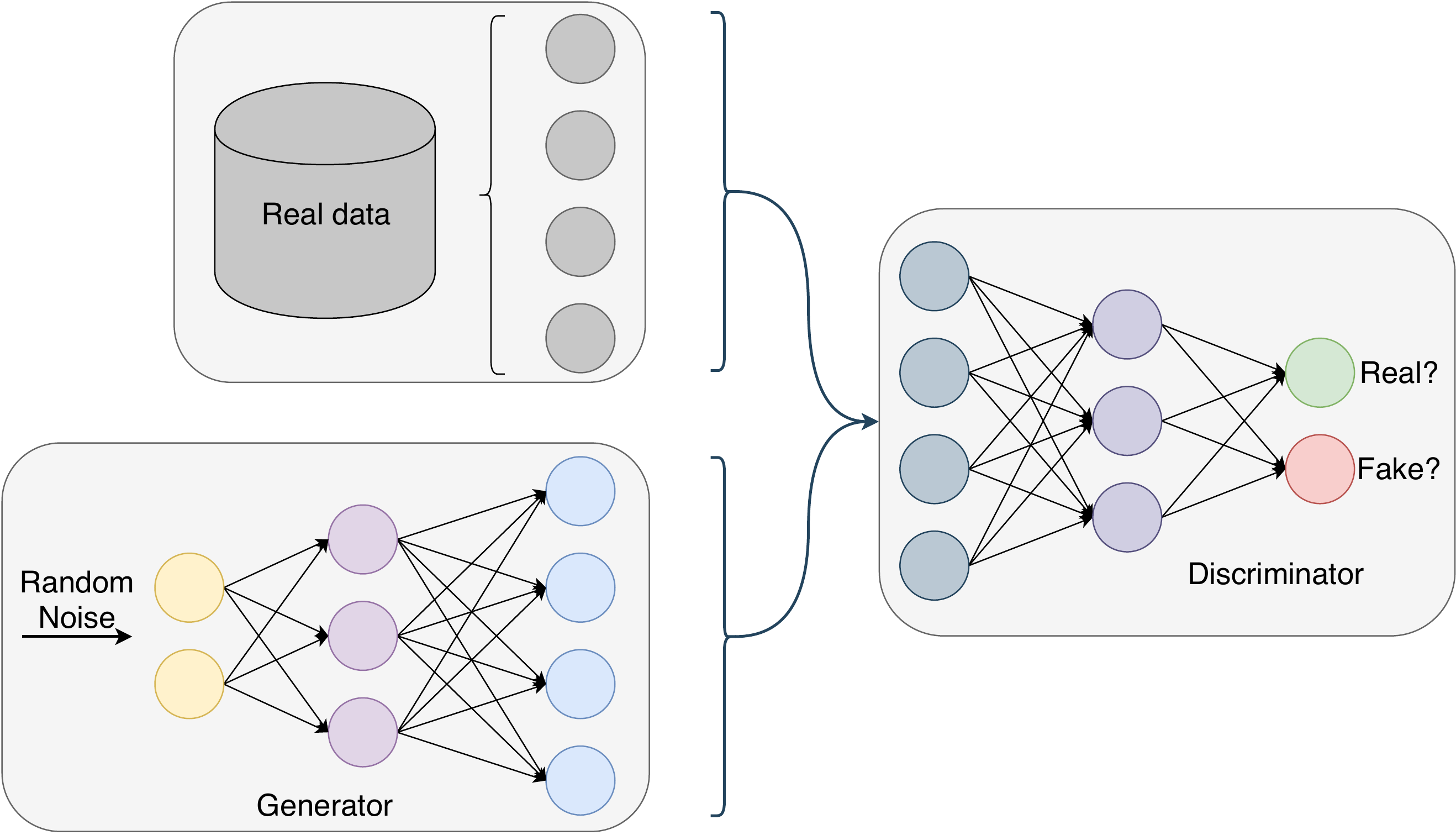}
	\caption{Composition of a generative adversarial network}\label{gan}
\end{figure}

\noindent \textbf{Generative Adversarial Network (GAN):} Shown in Fig. \ref{gan}, generative adversarial networks (GANs)\cite{goodfellow2014generative} consists of two main components, namely the generative and discriminator network (i.e., generator and discriminator). The generator is responsible for generating new data after it learns the data distribution from a training dataset of real data. The discriminator is in charge of classifying the real data from the fake data generated by the generator. GAN is often deployed in image generation, image transformation, image synthesis, image super-resolution and other applications.
%The target of GANs is that the generator is optimized to generate input data that is deceiving the discriminator, i.e., the discriminator cannot recognize the data whether it is fake or real.

\noindent \textbf{Deep Reinforcement Learning (DRL):} Deep Reinforcement Learning (DRL) is composed of DNNs and reinforcement learning (RL). The goal of DRL is to create an intelligent agent that can perform efficient policies to maximize the rewards of long-term tasks with controllable actions. The typical application of DRL is to solve various scheduling problems, such as decision problems in games, rate selection of video transmission, etc.

\begin{figure}[htbp]
	\centering
	\includegraphics[scale=0.2]{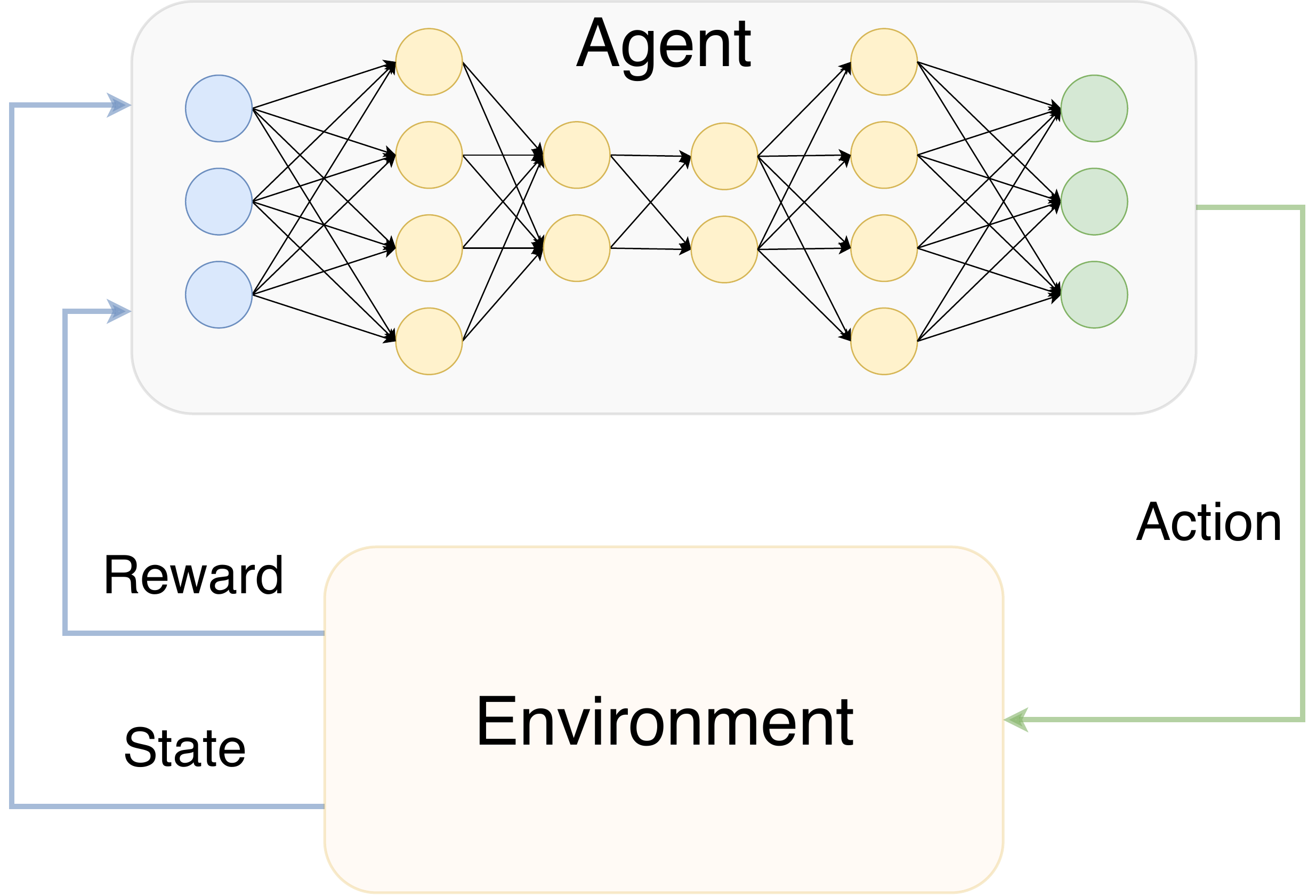}
	\caption{Concept of a deep reinforcement learning model}\label{drl}
\end{figure}

In the DRL approach, the reinforcement learning searches for the optimal policy of actions over states from the environment, and the DNN is in charge of representing a large number of states and approximating the action values to estimate the quality of the action in the given states. The reward is a function to represent the distance between the predefined requirement and the performance of an action. Through continuous learning, the agent of DRL model can be used for various tasks, e.g., gaming \cite{mnih2015human}.
\section{Edge Intelligence}

The marriage of edge computing and artificial intelligence gives the birth of edge intelligence. In this section, we discuss the motivation, benefits and definition of edge intelligence.

\subsection{Motivation and Benefits of Edge Intelligence}
The fusion of AI and edge computing is natural, since there is a clear intersection between them. Specifically, edge computing aims at coordinating a multitude of collaborative edge devices and servers to process the generated data in proximity; and AI strives for simulating intelligent human behavior in devices/machines by learning from data. Besides enjoying the general benefits of edge computing (e.g., low-latency, reduced bandwidth consumption), pushing AI to the edge further benefits each other in the following aspects. 

\textbf{On one hand, data generated at the network edge need AI to fully unlock their potential.} %In recent years, we have witnessed a skyrocketing number and types of end devices ranging from surveillance cameras, smart sensors and wearables to Internet-of-Vehicles being connected to the Internet. 
As a result of the proliferation of the skyrocketing number and types of mobile and IoT devices, large volumes of multi-modal data (e.g., audio, picture and video) of physical surroundings are continuously sensed at the device side. In this context, AI will be functionally necessary due to its ability to quickly analyze those huge data volumes and extract insights from them for high-quality decision making. As one of the most popular AI techniques, deep learning brings the ability to automatically identify patterns and detect anomalies in the data sensed by the edge device, as exemplified by population distribution, traffic flow, humidity, temperature, pressure and air quality. The insights extracted from the sensed data are then fed to the real-time predictive decision-making (e.g., public transportation planning, traffic control and driving alert) in response to the fast-changing environments, increasing the operational efficiency. %Compared to traditional intelligence approaches based on the monitoring of numeric thresholds to be crossed, deep learning approaches improve the optimal decision making with greater accuracy. Inspired by the efficacy, predictive AI capability has been integrated with major general-purpose and industrial IoT platforms, such as Microsoft Azure IoT, IBM Watson IoT, Amazon AWS IoT, GE Predix, and PTC ThingWorx. 
As forecasted by Gartner \cite{gartner3ai}, more than 80 percent of enterprise IoT projects will include an AI component by 2022, up from only 10 percent today. 

\textbf{On the other hand, edge computing is able to prosper AI with richer data and application scenarios.} %Over the past decade, deep learning has profoundly revolutionized the ICT horizon and quickly ascended as a backbone of modern digital, by substantial breakthroughs in a wide range of fields, especially computer vision, speech recognition and natural language processing. 
It is widely recognized that the driving force behind the recent booming of deep learning is four-folds: algorithm, hardware, data and application scenarios. While the effect of algorithm and hardware on the development of deep learning is intuitive, the role of data and application scenarios have been mostly overlooked. Specifically, to improve the performance of a deep learning algorithm, the most commonly adopted approach is to refine the DNN with more layers of neurons. By doing this, we need to learn more parameters in the DNN, and so does the data required for training increase. %By reviewing the timing of the most publicized AI breakthroughs over the past 30 years, it is observed that the average elapsed time between key algorithms and corresponding advances was about 18 years, whereas the average elapsed time between key datasets and corresponding advances was less than 3 years \cite{data4ai}. 
This definitely demonstrates the importance of data on the development of AI. Having recognized the importance of data, the next problem is, where is the data from. Traditionally, data is mostly born and stored in the mega-scale datacenters. Nevertheless, with the rapid development of IoT, the trend is reversing now. According to Cisco's report\cite{ciscogci}, in the near future, massive IoT data will be generated at the edge side. If these data are processed by AI algorithms at the cloud data center, it will consume a lot of bandwidth resources and bring great pressure to the cloud data center. To address these challenges, edge computing is proposed to achieve low latency data processing by sinking the computing capability from the cloud data center to the edge side, i.e., data generation source, which may enable AI processing with high performance.

%Cisco GCI estimates that nearly 850 ZB will be generated by all people, machines, and things by 2021, up from 220 ZB generated in 2016 \cite{ciscogci}. In contrast, the global datacenter traffic will only reach 20.6 Zettabytes by 2021, up from 6.8 ZB in 2016. Clearly, in the following years, IoT would continue to fuel the booming of AI.

While edge computing and AI complement each other from a technical perspective, their application and popularization are also mutually beneficial.

\textbf{On one hand, AI democratization requires edge computing as a key infrastructure.} AI technologies have witnessed great success in many digital products or services in our daily life, e.g., online shopping, service recommendation, video surveillance, smart home devices, etc. AI is also a key driving force behind emerging innovative frontiers, such as self-driving cars, intelligent finance, cancer diagnosis and medicine discovery. Beyond the above examples, to enable a richer set of applications and push the boundaries of what's possible, AI democratization or ubiquitous AI \cite{aidemo} has been declared by major IT companies, with the vision of ``making AI for every person and every organization at everywhere''. To this end, AI should go `closer' to the people, data and end devices. Clearly, edge computing is more competent than cloud computing in achieving this goal. Firstly, compared to the cloud datacenter, edge servers are in closer proximity to people, data source and devices. Secondly, compared to cloud computing, edge computing is also more affordable and accessible. Finally, edge computing has the potential to provide more diverse application scenarios of AI than cloud computing. Due to these advantages, edge computing is naturally a key enabler for ubiquitous AI.

\textbf{On the other hand, edge computing can be popularized with AI applications.} During the early development of edge computing, there has always been the concern in the cloud computing community with which high-demand applications edge computing could take to the next level that cloud computing could not, and what are the killer applications of edge computing. To clear up the doubt, %the research group in Microsoft, who co-introduced the concept of cloudlet \cite{Satyanarayanan2009The}, 
Microsoft has conducted continuous exploration on what kinds should be moved from the cloud to the edge since 2009 \cite{satyanarayanan2009case}, ranging from voice command recognition, AR/VR and interactive cloud gaming \cite{mscloudgame} to real-time video analytics. By comparison, real-time video analytics is envisioned to be a killer application for edge computing \cite{ananthanarayanan2017real,zhang2017live,hung2018videoedge}. As an emerging application built on top of computer vision, real-time video analytics continuously pulls high-definition videos from surveillance cameras and requires high computation, high bandwidth, high privacy and low-latency to analyze the videos. The one viable approach that can meet these strict requirements is edge computing. %Interdependently, in CMU, to popularize the concept of cloudlet, researchers have mainly focused on a dozen of cognitive assistance applications that bring AI technologies such as computer vision, speech recognition, natural language processing to the inner loop of human cognition and interaction. 
Looking back to the above evolution of edge computing, it can be foreseen that novel AI applications emerged from the sectors such as industrial IoT, intelligent robots, smart cities and smart home will play a crucial role in the popularization of edge computing. This is mainly due to the fact that many mobile and IoT related AI applications represent a family of practical applications that are computation- and energy- intensive, privacy- and delay- sensitive, and thus naturally align well with edge computing.

Due to the superiority and necessity of running AI application on the edge, edge AI has recently received great attention. In December 2017, in a white paper, ``A Berkeley View of Systems Challenges for AI'' \cite{stoica2017berkeley} published by UC Berkeley, the cloud-edge AI system is envisioned as an important research direction to achieve the goal of mission-critical and personalized AI. In August 2018, edge AI emerges in the Gartner Hype Cycle for the first time \cite{gartner5ei}. According to Gartner’s prediction, edge AI is still in the innovation trigger phase, and it will reach a plateau of productivity in the following 5 to 10 years. In the industry, many pilot projects have also been carried out towards edge AI. Specifically, on the edge AI service platform, the traditional cloud providers, such as Google, Amazon and Microsoft, have launched service platforms to bring the intelligence to the edge, through enabling end devices to run ML inferences with pre-trained models locally. On edge AI chips, various high-end chips designated for running ML models have been made commercially available on the market, as exemplified by Google Edge TPU, Intel Nervana NNP, Huawei Ascend 910 \& Ascend 310. 

\subsection{Scope and Rating of Edge Intelligence}
While the term edge AI or edge intelligence is brand new, explorations and practices in this direction have begun early. As aforementioned, in 2009, to demonstrate the benefits of edge computing, Microsoft has built an edge-based prototype to support mobile voice command recognition, an AI application. Albeit the early begin of exploration, there is still not a formal definition for edge intelligence. 

Currently, most organizations \cite{eiwhite,eitookit} and presses \cite{eipaper} refer to edge intelligence as the paradigm of running AI algorithms locally on an end device, with data (sensor data or signals) that are created on the device. While this represents the current most common approach (e.g., with high-end AI chips) towards edge intelligence in the real world, it is crucial to note that this definition greatly narrows down the solution scope of edge intelligence. Running computation intensive algorithms as exemplified by DNN models locally is very resource-intensive, requiring high-end processors to be equipped in the device. Such stringent requirement not only increases the cost of edge intelligence but is also incompatible and unfriendly to existing legacy end devices that have limited computing capacities. 

In this paper, we submit that the scope of edge intelligence should not be restricted to running AI models solely on the edge server or device. In fact, as demonstrated by a dozen recent studies, for DNN models, running them with edge-cloud synergy can reduce both the end-to-end latency and energy consumption, when compared to the local execution approach. Due to these practical advantages, we believe that such collaborative hierarchy should be integrated into the design of efficient edge intelligence solutions. Further, existing thoughts on edge intelligence mainly focus on the inference phase (i.e., running the AI model), assuming that the training of the AI model is performed in the power cloud datacenters, since the resource consumption of the training phase significantly overweights the inference phase. However, this means that the enormous amount of training data should be shipped from devices or edges to the cloud, incurring prohibitive communication overhead as well as the concern on data privacy. 

\begin{figure}[!t]  
	\centering %\vspace{-100pt}
	\includegraphics[width=3.4in]{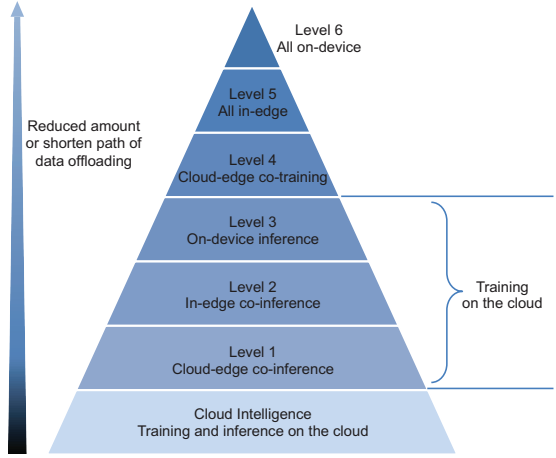}
	 \caption{A 6-level rating for edge intelligence}
	\label{fig-rating} \vspace{-15pt}
\end{figure}

Instead, we believe that edge intelligence should be the paradigm that fully exploits the available data and resources across the hierarchy of end devices, edge nodes and cloud datacenters to optimize the overall performance of training and inferencing a DNN model. This indicates that edge intelligence does not necessarily mean that the DNN model is fully trained or inferenced at the edge, but can work in a cloud-edge-device coordination manner via data offloading. Specifically, according to the amount and path length of data offloading, we rate edge intelligence into 6 levels, as shown in Fig. \ref{fig-rating}. Specifically, the definition of various levels of edge intelligence is given as follows:

\begin{itemize}
\item Cloud Intelligence: training and inferencing the DNN model fully in the cloud.
\item Level-1 -- Cloud-Edge Co-Inference and Cloud Training: training the DNN model in the cloud, but inferencing the DNN model in an edge-cloud cooperation manner. Here edge-cloud cooperation means that data is partially offloaded to the cloud. 
\item Level-2 -- In-Edge Co-Inference and Cloud Training: training the DNN model in the cloud, but inferencing the DNN model in an in-edge manner. Here in-edge means that the model inference is carried out within the network edge, which can be realized by fully or partially offloading the data to the edge nodes or nearby devices (via D2D communication).
\item Level-3 -- On-Device Inference and Cloud Training: training the DNN model in the cloud, but inferencing the DNN model in a fully local on-device manner. Here on-device means that no data would be offloaded.
\item Level-4 -- Cloud-Edge Co-Training \& Inference: training and inferencing the DNN model both in the edge-cloud cooperation manner. 
\item Level-5 -- All In-Edge: training and inferencing the DNN model both in the in-edge manner.
\item Level-6 -- All On-Device: training and inferencing the DNN model both in the on-device manner.
\end{itemize}

As the level of edge intelligence goes higher, the amount and path length of data offloading reduce. As a result, the transmission latency of data offloading decreases, the data privacy increases and the WAN bandwidth cost reduces. However, this is achieved at the cost of increased computational latency and energy consumption. This conflict indicates that there is no ``best-level'' in general; instead, the ``best-level'' edge intelligence is application-dependent and it should be determined by jointly considering multi-criteria such as latency, energy efficiency, privacy and WAN bandwidth cost. In the later sections, we will review enabling techniques as well as existing solutions for different levels of edge intelligence.

\section{Edge Intelligence Model Training}
\label{sec:training}

%Due to the superior recognition ability of DNN models, deep learning methods have been deployed in a wide range of scenarios. In general, the data for training DNN may come from everywhere data is generated, and the result inferred by DNN will send back everywhere it is needed. This trend consequently results in a distributed implementation of DNN, spanning from distributed placement to distributed training to distributed inference. In particular, there are a variety of motivations for distributed DNN training  in the edge computing paradigm,  such as maximizing the utilization of computing resources, minimizing the processing latency of computation tasks and protecting personal privacy. 

With the proliferation of mobile and IoT devices, data which is essential for AI model training is increasingly generated at the network edge. In this section, we focus on distributed training of DNN at the edge, including the architectures, key performance indicators, enabling techniques and existing systems \& frameworks.

\subsection{Architectures}

\begin{figure}[htbp]
	\centering
	\includegraphics[scale=0.6]{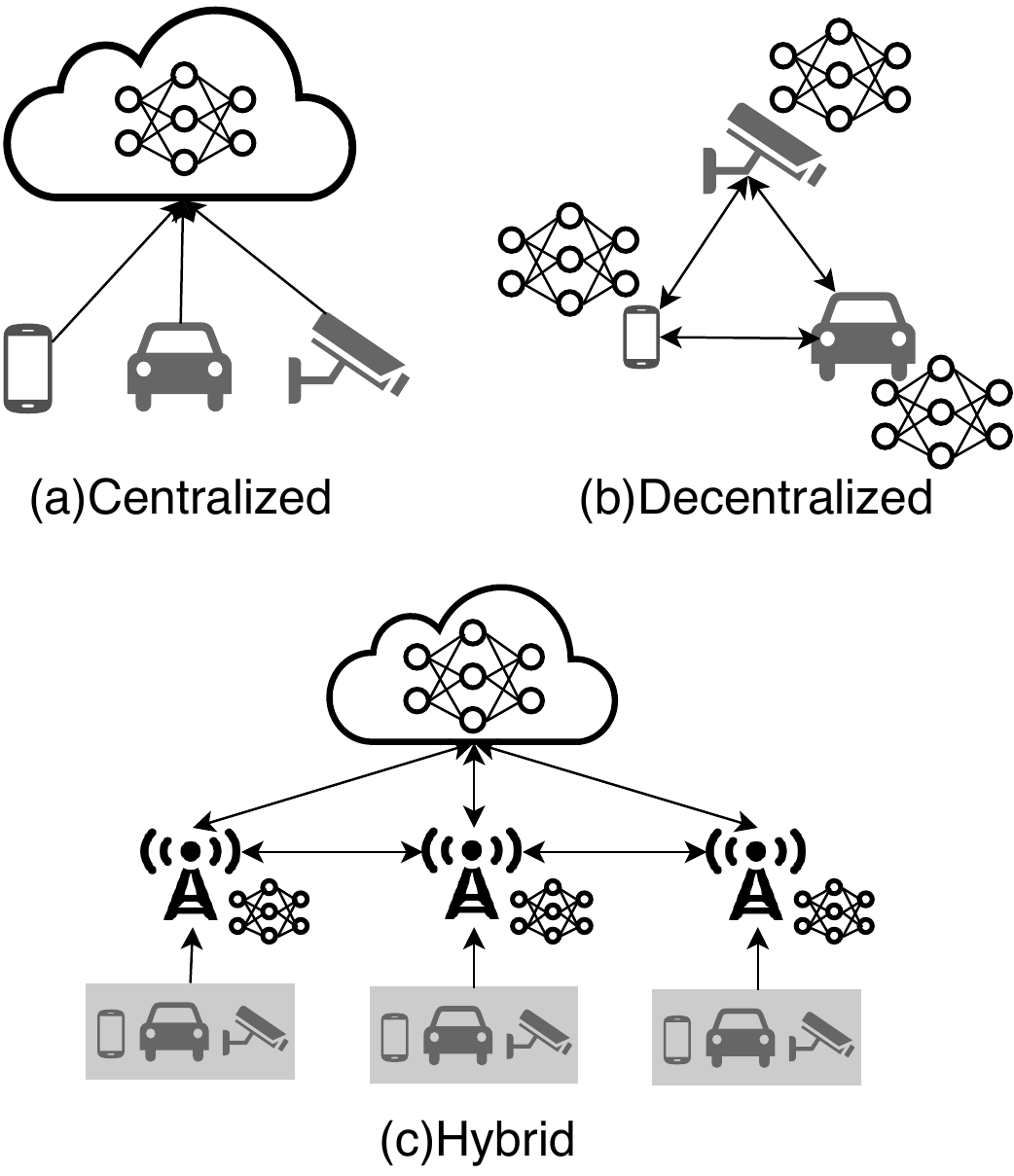}
	%\vspace{-10pt}
	\caption{The architecture modes of distributed training}\label{fig:Architecture}
	%\vspace{-5pt}
\end{figure}

The architectures of distributed DNN training at the edge can be divided into three modes, Centralized, Decentralized, Hybrid (Cloud-Edge-Device). Fig. \ref{fig:Architecture} shows the three architectures, illustrated by subfigures (a), (b) and (c), respectively. The cloud refers to the central datacenter whereas the end devices are represented by mobile phones, cars and surveillance cameras, which are also data sources. For the edge server, we use base stations as the legend.% Seen in Fig. \ref{fig:Architecture}, all three architectures involve cooperation between different kinds of devices. However, only the devices with a neural network mark process DNN training, either partially or wholly. 

\subsubsection{Centralized}

Fig \ref{fig:Architecture}(a) describes a centralized DNN training, where the DNN model is trained in the cloud datacenter. The data for training is generated and gathered from distributed end devices such as mobile phones, cars and surveillance cameras. Once the data arrived, the cloud datacenter will perform DNN training using these data. Therefore, the system based on the centralized architecture can be identified in Cloud Intelligence, Level-1, Level-2 or Level-3 in Fig. \ref{fig-rating} according to the specific inference mode that the system employs.

%Under this centralized mode, it is easy to deploy the DNN model since the DNN model is only placed on the cloud datacenter. However, the centralization of training data may cost a considerable communication overhead due to the large data size and the unpredictable network connection. Besides, the data with sensitive personal information is gathered in the cloud datacenter, which is owned by curious companies. This inevitably involves a privacy issue, especially when the data geo-distribute globally in various regions.

\subsubsection{Decentralized}

%Users expect to benefit from AI-based service while keeping their own personal privacy. Unfortunately, the centralized mode fails to provide these benefits at the same time. To avoid this dilemma, the decentralized mode is introduced. As Fig. \ref{fig:Architecture}(b) shows, there is no centralized node in the decentralized network, i.e., all the computing nodes perform equal roles. 

%It is remarkable that all the computing nodes are with a neural network mark, implying that each of them can perform DNN training. 

Under the decentralized mode as shown in \ref{fig:Architecture}(b), each computing node trains its own DNN model locally with local data, which preserves private information locally. To obtain the global DNN model by sharing local training improvement, nodes in the network will communicate with each other to exchange the local model updates. In this mode, the global DNN model can be trained without the intervention of the cloud datacenter, corresponding to the Level-5 edge intelligence defined in Fig. \ref{fig-rating}. 

\subsubsection{Hybrid}

The hybrid mode combines the centralized mode and the decentralized mode. As shown in Fig. \ref{fig:Architecture}(c), %each end device connects an edge server in the proximity, where the data is gathered for DNN training. 
as the hub of the architecture, the edge servers may train the DNN model by either decentralized updates with each other or centralized training with the cloud datacenter, thus the hybrid architecture covers Level-4 and Level-5 in Fig. \ref{fig-rating}. The hybrid architecture is also called as Cloud-Edge-Device training due to the involved roles. 

%The private data is only gathered in the edge servers, resulting in the privacy preservation, weaker than the decentralized architecture but stronger than the centralized architecture. Besides, the distributed training at the edge consumes less communication overhead comparing to the other two modes. Note that the hybrid architecture is not limited in the Fig. \ref{fig:Architecture}(c) in practical deployment. It is flexible to adapt the application scenario. For this reason, the hybrid architecture is more difficult to be deployed and applied in practice.

\subsection{Key Performance Indicators}
To better assess a distributed training method, there are six key performance indicators.

\subsubsection{Training Loss}
Essentially, the DNN training process solves an optimization problem that seeks to minimize the training loss. Since the training loss captures the gap between the learned (e.g., predicted) value and the labeled data, it indicates how well the trained DNN model fits the training data. Therefore, it is expected that the training loss can be minimized. Training loss is mainly affected by training samples and training methods.

\subsubsection{Convergence}
The convergence indicator is specialized for the decentralized methods. Intuitively, a decentralized method works only if the distributed training processes converge to a consensus, which is the training result of the method. The term convergence measures whether and how fast a decentralized method converges to such a consensus. Under the decentralized training mode, the convergence value depends on the way the gradient is synchronized and updated.

\subsubsection{Privacy}
When training the DNN model by using the data originated at a massive of end devices, the raw data or intermediate data should be transferred out of the end devices. Obviously, it is inevitable to deal with privacy issues in this scenario. To preserve privacy, it is expected that less privacy-sensitive data is transferred out of the end-devices. Whether privacy protection is implemented depends on whether the raw data is offloaded to the edge.
%Whether privacy protection is done depends on how to process the users' original data.

\subsubsection{Communication Cost}
Training the DNN model is data-intensive, since the raw data or intermediate data should be transferred across the nodes. Intuitively, this communication overhead increases the training latency, energy and bandwidth consumption. Communication overhead is affected by the size of the original input data, the way of transmission and the available bandwidth.

\subsubsection{Latency}
Arguably, latency is one of the most fundamental performance indicators of distributed DNN model training, since it directly influences when the trained model is available for use. The latency of the distributed training process typically consists of the computation latency and the communication latency. The computation latency is tightly dependent on the capability of the edge nodes. The communication latency may vary from the size of transmitted raw or intermediate data, and the bandwidth of network connection. 

\subsubsection{Energy Efficiency}
When training the DNN model in a decentralized manner, both the computation and communication process consume enormous energy. However, for most end-devices, they are energy-constrained. As a result, it is highly desirable that the DNN model training is energy-efficient. Energy efficiency is mainly affected by the size of the target training model and resources of the used devices.

It is worth noting that the performance indicators training loss and convergence are common objectives, thus they may not be explicitly claimed by some literature on DNN training.

%Note that the convergence of a method is influenced by many other factors such as the training dataset, the selected optimizer and the provided computing resources. Generally, for the decentralized methods, the faster the method converges, the better the method is.

\subsection{Enabling Technologies}
In this subsection, we review the enabling technologies for improving one or more of the aforementioned key performance indicators when training edge intelligence model. Table \ref{tech highlights1} summarizes the highlights of each enabling technology.

%introduce seven enabling technologies about distributed DNN training at the edge, 

\begin{table*}[htbp]
	\caption{Technologies for distributed DNN training at the edge}\label{tech highlights1}
	\centering
	\begin{tabular}{c|c|c}
		\hline
		\textbf{Technology} & \textbf{Highlights} & \textbf{Related Work} \\ \hline
		\hline
		Federated Learning &           \multicolumn{1}{l|}{\begin{minipage}{3.5in}
			\vskip 4pt
			\begin{itemize}
				\item Leave the training data distributed on the end devices
				\item Train the shared model on the server by aggregating locally-computed updates
				\item Preserve privacy
			\end{itemize}
		\vskip 4pt
		\end{minipage}  }       &         
		\cite{chen2019data,mcmahan2016communication, shokri2015privacy, konevcny2016federated, lalithafully, kim2018device}           \\ \hline %bonawitz2017practical, zhao2018privacy, lynskey2018intelligent,
		
		Aggregation Frequency Control &
		\multicolumn{1}{l|}{ \begin{minipage}{3.5in}
			\vskip 4pt
			\begin{itemize}
				\item Determine the best trade-off between local update and global parameter aggregation under a given resource budget
				\item Intelligent communication control
			\end{itemize}
			\vskip 4pt
		\end{minipage} }          &           
		\cite{hsieh2017gaia, wang2018adaptive, nishio2018client}         \\ \hline
		
		Gradient Compression      &           
		\multicolumn{1}{l|}{ \begin{minipage}{3.5in}
				\vskip 4pt
				\begin{itemize}
					\item Gradient quantization by quantizing each element of gradient vectors to a finite-bit low precision value 
					\item Gradient sparsification by transmitting only some values of the gradient vectors
				\end{itemize}
				\vskip 4pt
		\end{minipage} }           &           
		\cite{lin2017deep, tao2018esgd, stich2018sparsified, tang2018communication, amiri2019machine}         \\ \hline
		
		DNN Splitting                 &           
		\multicolumn{1}{l|}{ \begin{minipage}{3.5in}
				\vskip 4pt
				\begin{itemize}
					\item Select a splitting point to reduce latency as much as possible
					\item Preserve privacy
				\end{itemize}
				\vskip 4pt
		\end{minipage} }           &          \cite{mao2018privacy, wang2018not, osia2017hybrid, harlappipedream}          \\ \hline
		
		Knowledge Transfer Learning                &           
		\multicolumn{1}{l|}{ \begin{minipage}{3.5in}
				\vskip 4pt
				\begin{itemize}
					\item First train a base network (teacher network) on a base dataset and task and then transfer the learned features to a second target network (student network) to be trained on a target dataset and task
					\item The transition from generality to specificity
				\end{itemize}
				\vskip 4pt
		\end{minipage} }           &          
		\cite{wang2018not, osia2017hybrid, %yosinski2014transferable, 
		sharma2018existing}          \\ \hline
		
		Gossip Training             &           
		\multicolumn{1}{l|}{ \begin{minipage}{3.5in}
				\vskip 4pt
				\begin{itemize}
					\item Random gossip communication among devices
					\item Full asynchronization and total decentralization
					\item Preserve privacy
				\end{itemize}
				\vskip 4pt
		\end{minipage} }           &         \cite{boyd2006randomized, blot2016gossip, jin2016scale, daily2018gossipgrad}           \\ \hline

	\end{tabular}
\end{table*}

\subsubsection{Federated Learning}
%Due to the unprecedented accuracy, deep learning methods have turned themselves into an essential role in AI-based services on the Internet. While enjoying their advantage, it is inevitable to meet privacy issues. To obtain a superior performance in specific tasks, deep learning methods require massive data to train its models and learn features’ representation. 

%The data for training, however, may be users’ personal, highly sensitive data such as photos and voice recordings. Traditional centralized methods require users’ data to keep in companies that collect it, or users will not enjoy smart services powered by deep learning methods since they only run in the core of the network.
Federated learning is dedicated to optimizing privacy issue in the above key performance indicators. Federated learning is an emerging yet promising approach to preserve privacy when training the DNN model based on data originated by multiple clients. Rather than aggregating the raw data to a centralized datacenter for training, federated learning \cite{mcmahan2016communication} leaves the raw data distributed on the clients (e.g., mobile devices), and trains a shared model on the server by aggregating locally-computed updates. The main challenges of federated learning are optimization and communication. 

%\begin{figure}[htbp]
%	\centering
%	\includegraphics[scale=0.45]{PIEEE/fig/training/sSGD.pdf}
	%\vspace{-10pt}
%	\caption{The architecture of the deep learning system based on SSGD protocol}\label{fig:SSGD}
	%\vspace{-5pt}
%\end{figure}

For the optimization problem, the challenge is to optimize the gradient of a shared model by the distributed gradient updates on mobile devices. On this issue, federated learning adopts stochastic gradient descent (SGD). SGD updates the gradient over extremely small subsets (mini-batch) of the whole dataset, which is a simple but widely-used gradient descent method.  Shokri et al. \cite{shokri2015privacy} design a selective stochastic gradient descent (SSGD) protocol, allowing the clients to train independently on their own datasets and selectively share small subsets of their models’ key parameters to the centralized aggregator. Since SGD is easy to be parallelized as well as asynchronously executed, SSGD targets both privacy and training loss. Specifically, while preserving clients’ own privacy, the training loss can be reduced by sharing the models among clients, comparing to training solely on their own inputs. A flaw of \cite{shokri2015privacy} is that it does not consider unbalanced and non-IID (none Independent Identical Distribution) data. As an extension, McMahan et al. \cite{mcmahan2016communication} advocate a decentralized approach, termed as federated learning, and present FedAvg method for federated learning with the deep neural network based on iterative model averaging. Here iterative model averaging means that the clients update the model locally with one-step SGD and then the server averages the resulting models with weights. The optimization on \cite{mcmahan2016communication} emphasizes the properties of unbalanced and non-IID since the distributed data may come from various sources. 

For the communication problem, it is the unreliable and unpredictable network that poses the challenge of communication efficiency. In federated learning, each client sends a full model or a full model update back to the server in a typical round. For large models, this step is likely to be the bottleneck due to the unreliable network connections. To decrease the number of rounds for training, Mcmahan et al. \cite{mcmahan2016communication} propose to increase the computation of local updates on clients. However, it is impractical when the clients are under severe computation resources constraint. In response to this issue, Kone${\Check{\text{c}}}$n${\Acute{\text{y}}}$ et al. \cite{konevcny2016federated} propose to reduce communication cost with two new update schemes, namely structured update and sketched update. In a structured update, the model directly learns an update from a restricted space parametrized using a smaller number of variables, e.g. either low-rank or a random mask. If using a sketched update, the model first learns a full model update and then compressed the update using a combination of quantization, random rotations, and subsampling before sending it to the server.

%\begin{figure}[htbp]
%	\centering
%	\includegraphics[scale=0.45]{PIEEE/fig/training/SFL@CCS'17.pdf}
	%\vspace{-10pt}
%	\caption{Federated Learning with Secure Aggregation}\label{fig:SFL}
	%\vspace{-5pt}
%\end{figure}

%Based on Federated Learning, there are advanced variants aiming to deal with different challenges when applying in different scenarios. 

%To further enhance the privacy, Bonawitz et al. \cite{bonawitz2017practical} design a protocol for secure aggregation of high-dimensional data in federated learning setting. With a constant number of rounds, the protocol has low communication overhead as well as robustness to failures, which enable Federated Learning more secure. Zhao et al. \cite{zhao2018privacy} identify the difficulty of applying Federated Learning and design a systematic solution, Zoo, to deal with it. Zoo supports users to easily construct, compose and deploy different machine learning models on edge computing environment and Federated Learning is included. Lynskey et al. \cite{lynskey2018intelligent} focus on the synchronization of distributed updates, and show the latency contribution of dataset sizes, accuracy requirement, communication bandwidth and computation resources. Besides, Lynskey et al. further develop a technique to alleviate the maximum delay for training a model.

Though Federated Learning technique exploits a new decentralized deep learning architecture, it is built upon a central server for aggregating local updates. Considering the scenario of training a DNN model over a fully decentralized network, i.e., a network without a central server,  Lalitha et al. \cite{lalithafully} propose a Bayesian-based distributed algorithm, in which each device updates its belief by aggregating information from its one-hop neighbors to train a model that best fits the observations over the entire network. Furthermore, with the emerging blockchain technique,  Kim et al. \cite{kim2018device} propose Blockchain Federated Learning (BlockFL) with the devices’ model update exchanged and verified by leveraging blockchain. BlockFL also works for a fully decentralized network, where machine learning model can be trained without any central coordination, even when some devices lack their own training data samples.

\subsubsection{Aggregation Frequency Control}
This method focuses on the optimization of communication overhead during the DNN model training. On training deep learning model in edge computing environment, a commonly adopted idea (e.g., federated learning) is to train distributed models locally first, and then aggregate updates centrally. In this case, the control of updates’ aggregation frequency significantly influences the communication overhead. Thus, the aggregation process, including aggregation content as well as aggregation frequency, should be controlled carefully.

Based on the above insight, Hsieh et al. \cite{hsieh2017gaia} develop Gaia system and the Approximate Synchronous Parallel (ASP) model for geo-distributed DNN model training. The basic idea of Gaia is to decouple the communication within a datacenter from the communication between datacenters, enabling different communication and consistency models for each. To this end, the ASP model %shown in Fig. \ref{fig:Gaia_ASP} 
is developed to dynamically eliminate insignificant communication between datacenters, where the aggregation frequency is controlled by the preset significance threshold. However, Gaia focuses on geo-distributed datacenters that are capacity-unlimited, making it is not generally applicable to edge computing nodes whose capacity is highly constrained.

%which is not a general solution for mobile edge computing systems with computation resource constraint. 

%\begin{figure}[htbp]
%	\centering
%	\includegraphics[scale=0.5]{PIEEE/fig/training/Gaia_system.pdf}
	%\vspace{-10pt}
%	\caption{System overview of Gaia}\label{fig:Gaia_system}
	%\vspace{-5pt}
%\end{figure}

%\begin{figure}[htbp]
%	\centering
%	\includegraphics[scale=0.7]{PIEEE/fig/training/Gaia_ASP.pdf}
	%\vspace{-10pt}
%	\caption{The synchronization mechanisms of ASP model}\label{fig:Gaia_ASP}
	%\vspace{-5pt}
%\end{figure}

\begin{figure}[htbp]
	\centering
	\includegraphics[scale=0.5]{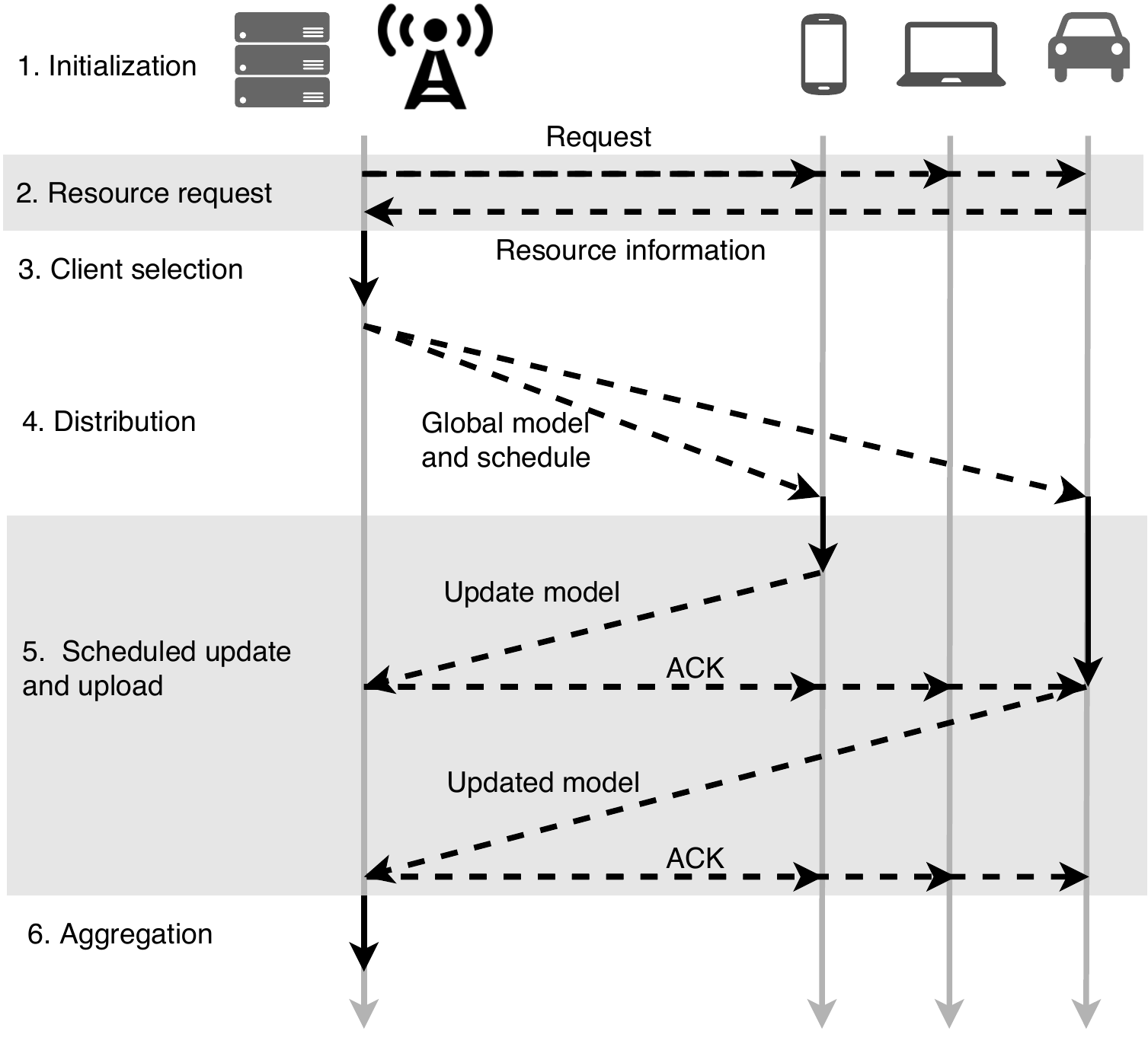}
	%\vspace{-10pt}
	\caption{Overview of FedCS protocol}\label{fig:CSFL}
	%\vspace{-5pt}
\end{figure}

To incorporate the capacity constraint of edge nodes, Wang et al. \cite{wang2018adaptive} propose a control algorithm that determines the best trade-off between local update and global parameter aggregation under a given resource budget. The algorithm is based on the convergence analysis of distributed gradient descent and can be applied to federated learning in edge computing with provable convergence. To implement federated learning in the capacity-limited edge computing environment, Nishio et al. \cite{nishio2018client} study the client selection problem with resource constraints. In particular, an update aggregation protocol named FedCS is developed to allow the centralized server to aggregate as many client updates as possible and to accelerate performance improvement in machine learning models. An illustration of FedCS is shown in Fig. \ref{fig:CSFL}.

\subsubsection{Gradient Compression}

To reduce the communication overhead incurred by decentralized training, gradient compression is another intuitive approach to compress the model update (i.e., gradient information). To this end, gradient quantization and gradient sparsification have been advocated. Specifically, gradient quantization performs lossy compression of the gradient vectors by quantizing each of their elements to a finite-bit low precision value. Gradient sparsification reduces the communication overhead by transmitting part of the gradient vectors. 

Lin et al. \cite{lin2017deep} observe that 99.9\% of the gradient exchange in distributed SGD are redundant, which demonstrates the power of gradient compression. Based on this observation, Lin et al. propose Deep Gradient Compression (DGC), which compresses the gradient by 270-600$\times$ for a wide range of CNNs and RNNs. To preserve accuracy during this compression, DGC employs four methods: momentum correction, local gradient clipping, momentum factor masking, and warm-up training. 

Inspired by the above work \cite{lin2017deep}, Tao et al. \cite{tao2018esgd} propose Edge Stochastic Gradient Descent (eSGD), a family of sparse schemes with both convergence and practical performance guarantees. To improve the first order gradient-based optimization of stochastic objective functions in edge computing, eSGD includes two mechanisms: (1) determine which gradient coordinates are important and only transmits these coordinates; (2) design momentum residual accumulation for tracking out-of-date residual gradient coordinates in order to avoid low convergence rate caused by sparse updates. A concise convergence analysis of sparsified SGD is given in \cite{stich2018sparsified}, where SGD are analyzed with k-sparsification or compression (e.g., top-k or random-k). The analysis shows that this scheme converges at the same rate as vanilla SGD when equipped with error compensation (keeping track of accumulated errors in memory). In other words, communication can be reduced by a factor of the dimension of the problem (sometimes even more) whilst still converging at the same rate.

Quantizing the gradients to low-precision values can also reduce the communication bandwidth. %In a `slow' network environment, there are two common techniques to reduce communication overhead: communication compression and decentralization. Combining both techniques, 
In this regard, Tang et al. \cite{tang2018communication} develop a framework of compressed, decentralized training and proposes two different algorithms, called extrapolation compression and difference compression respectively. The analysis on the two algorithms proves that both converge at the rate of $O(1/\sqrt{nT})$ where $n$ is the number of clients and $T$ is the number of iterations, matching the convergence rate for full precision, centralized training. Amiri et al. \cite{amiri2019machine} implement distributed stochastic gradient descent (DSGD) at the wireless edge with the help of a remote parameter server. Besides, Amiri et al. further develop DSGD in digital and analog scheme respectively. Digital DSGD (D-DSGD) assumes that the clients operate on the boundary of the Multiple Access Channel (MAC) capacity region at each iteration of DSGD algorithm, and employs gradient quantization and error accumulation to transmit their gradient estimates within the bit budget allowed by the employed power allocation. In Analog DSGD (A-DSGD), the clients first sparsify their gradient estimates with error accumulation and then project them to a lower dimensional space imposed by the available channel bandwidth. These projections are transmitted directly over the MAC without employing any digital code.

%Gradients compression is a technique that involves both software and hardware. In the literature most of the attention is paid on algorithmic advancement, but Li et al. \cite{li2018network} set out to reduce communication overhead by combining both two levels. Li et al. focus on Network Interface Cards (NICs) since they are the essential carrier of communication in edge computing. To maximize the benefits of in-network acceleration, Li et al. proposed INCEPTIONN (In-Network Computing to Exchange and Process Training Information Of Neural Networks), which includes a lightweight and hardware-friendly lossy-compression algorithm for floating-point gradients, and an aggregator-free training algorithm that exchanges gradients in both legs of communication in the group. INCEPTIONN reduces the communication time by 70.9\%-80.7\% and offers 2.2-3.1$\times$ speedup over the conventional model training while achieving the same level of accuracy.

%\subsubsection{Momentum Residual Accumulation}

\subsubsection{DNN Splitting}
\label{subsubsec:DNN_splitting}

The aim of DNN splitting is to protect privacy. DNN splitting protects user privacy by transmitting partially processed data rather than transmitting raw data. To enable a privacy-preserving edge-based training of DNN models, DNN splitting is conducted between the end devices and the edge server. This bases on the important observation that a DNN model can be split inside between two successive layers with two partitions deployed on different locations without losing accuracy.

An inevitable problem on DNN splitting is how to select the splitting point such that distributed DNN training is still under the latency requirement. On this problem, Mao et al. \cite{mao2018privacy} utilize the differentially private mechanism and partitions DNN after the first convolutional layer to minimize the cost of mobile devices. The proof in \cite{mao2018privacy} guarantees that applying differentially private mechanism on activations is feasible for outsourcing training tasks to untrusted edge servers. 
Wang et al. \cite{wang2018not} consider this problem across mobile devices and cloud datacenters. To benefit from the computation power of cloud datacenters without privacy risks, Wang et al. design Arden (privAte infeRence framework based on Deep nEural Networks), a framework which partitions the DNN model with a lightweight privacy-preserving mechanism. By arbitrary data nullification and random noise addition, Arden achieves privacy protection. Considering the negative impact of private perturbation to the original data, Wang et al. use a noisy training method to enhance the cloud-side network robustness to perturbed data.

Osia et al. \cite{osia2017hybrid} introduce a hybrid user-cloud framework on the privacy issue, which utilizes a private-feature extractor as its core component and breaking down large, complex deep models for cooperative, privacy-preserving analytics. In this framework, the feature extractor module is properly designed to output the private feature constrained to keeping the primary information while discarding all the other sensitive information. Three different techniques are employed to make sensitive measures unpredictable: dimensionality reduction, noise addition, and Siamese fine-tuning.

\begin{figure}[htbp]
	\centering
	\includegraphics[scale=0.5]{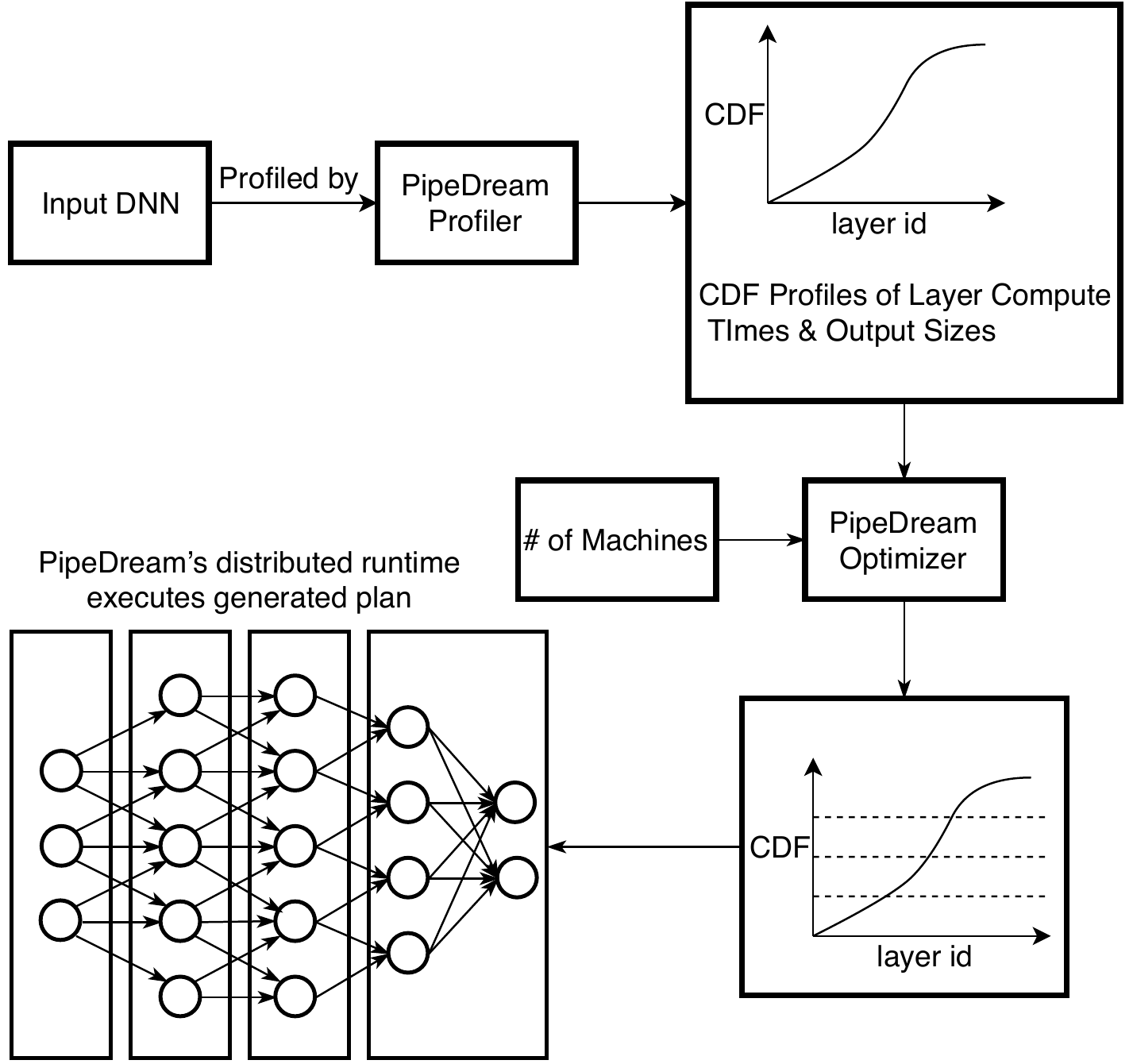}
	%\vspace{-10pt}
	\caption{PipeDream’s automated mechanism}\label{fig:PipeDream}
	%\vspace{-5pt}
\end{figure}

When applying DNN splitting for privacy-preserving, it is remarkable that this technique also works for dealing with the tremendous computation of DNN. Exploiting the fact that edge computing usually involves a large number of devices, parallelization approaches is usually employed to manage DNN computation. DNN training in parallel includes two kinds of parallelism, data parallelism and model parallelism. However, data parallelism may bring heavy overhead of communication while model parallelism usually leads to severe under-utilization of computation resources. To address these problems,  Harlap et al. \cite{harlappipedream} propose pipeline parallelism, an enhancement to model-parallelism, where multiple mini-batches are injected into the system at once to ensure efficient and concurrent use of computation resources. Based on pipeline parallelism, Harlap et al. design PipeDream, a system which supports pipelined training, and automatically determine how to systematically split a given model across the available computing nodes. PipeDream shows the advantage of reducing communication overhead and utilizing computing resource efficiently. The overview of PipeDream's automated mechanism is in Fig. \ref{fig:PipeDream}.

\subsubsection{Knowledge Transfer Learning}

Knowledge transfer learning, or transfer learning for simplicity, is closely connected with DNN splitting technique. In transfer learning, for the purpose of reducing DNN model training energy cost on edge devices, we first train a base network (teacher network) on a base dataset, and then we repurpose the learned features, i.e., transfer them to a second target network (student network) to be trained on a target dataset. This process will tend to work if the features are general (i.e., suitable to both base and target tasks) instead of specific to the base task. The transition involves a process from generality to specificity. 

%Yosinski et al. \cite{yosinski2014transferable} give an experimental quantification on the generality versus specificity of neurons in each layer of a CNN. By a series of experimental evaluations, Yosinski et al. draw that transferability is negatively affected by two distinct issues: (1) the specialization of higher layer neurons to their original task at the expense of performance on the target task, which was expected, and (2) optimization difficulties related to splitting networks between co-adapted neurons, which is not expected. Another interesting result is that initializing a network with transferred features from almost any number of layers can produce a boost to generalization that lingers even after fine-tuning to the target dataset. This shows the feasibility of Transfer Learning, which inspires the research on Transfer Learning on edge computing.

The approach of transfer learning seems to be promising for learning on edge devices since it has greatly reduced resource demand, but a thorough investigation on its effectiveness is lacking. To bridge this gap, Sharma et al. \cite{sharma2018existing} and Chen et al. \cite{chen2019data} provide extensive studies on the performance (in both accuracy and convergence speed) of Transfer Learning, considering different student network architectures and different techniques for transferring knowledge from teacher to student. The result varies with architectures and transfer techniques. %Different types of knowledge transfer techniques are shown in Fig. \ref{fig:KT} . 
A good performance improvement is obtained by transferring knowledge from both the intermediate layers and last layer of the teacher to a shallower student while other architectures and transfer techniques do not fare so well and some of them even lead to negative performance impact.

%\begin{figure}[htbp]
%	\centering
%	\includegraphics[scale=0.39]{PIEEE/fig/training/KT@HPDC'18.pdf}
	%\vspace{-10pt}
%	\caption{Different kinds of knowledge transferring techniques}\label{fig:KT}
	%\vspace{-5pt}
%\end{figure}

Transfer Learning technique regards the shallow layers of a pre-trained DNN on one dataset as a generic feature extractor that can be applied to other target tasks or datasets. With this feature, Transfer Learning is employed in many pieces of research and inspires the design of some frameworks. Osia et al. \cite{osia2017hybrid}, which we have mentioned on Sec. \ref{subsubsec:DNN_splitting} , use Transfer Learning to determine the degree of generality and particularity of a private feature. Arden, proposed in \cite{wang2018not}, partitions a deep neural network across the mobile device and the cloud data center, where the raw data is transformed by the shallow portions of the DNN on the mobile device side. As \cite{wang2018not} referred, the design of DNN splitting in Arden is inspired by Transfer Learning.

\subsubsection{Gossip Training}

\begin{figure}[htbp]
	\centering
	\includegraphics[scale=0.45]{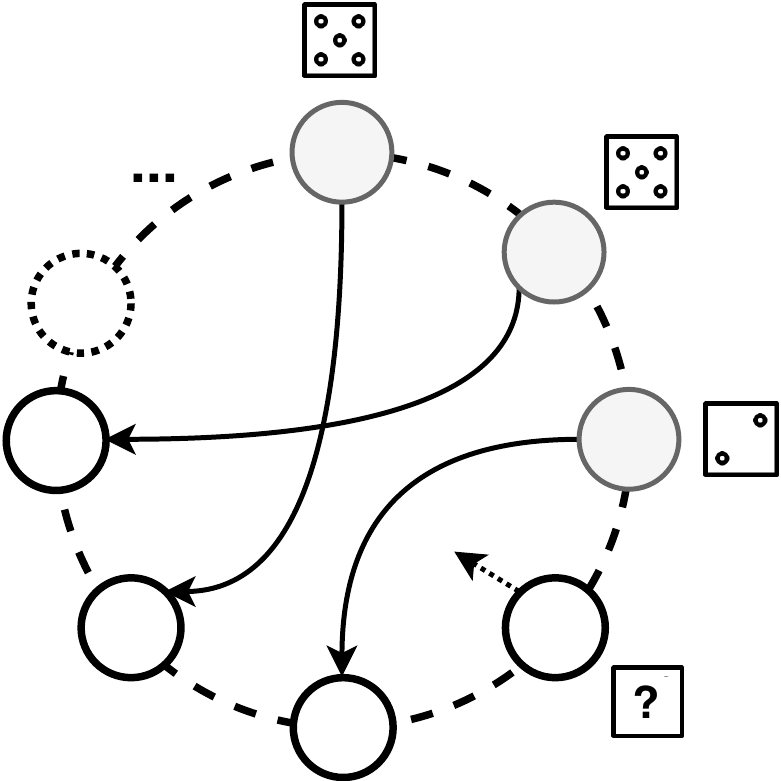}
	%\vspace{-10pt}
	\caption{Communication with randomly selected partner in gossip manner}\label{fig:gossip}
	%\vspace{-5pt}
\end{figure}

Aiming at shortening the training latency, gossip training is a new decentralized training method, which is built on randomized gossip algorithms. The early work on random gossip algorithms is gossip averaging \cite{boyd2006randomized}, which can fast converge towards a consensus among nodes by exchanging information peer-to-peer. The gossip distributed algorithms enjoy the advantage of full asynchronization and total decentralization as they have no requirement on centralized nodes or variables. Inspired by this, GoSGD (Gossip Stochastic Gradient Descent) \cite{blot2016gossip} is proposed to train DNN models in an asynchronous and decentralized way. GoSGD manages a group of independent nodes, where each of them hosts a DNN model and iteratively proceeds two steps: gradient update and mixing update. Specifically, each node updates its hosted DNN model locally in gradient update step and then shares its information with another randomly selected node in mixing update step, as shown in Fig. \ref{fig:gossip}. The steps repeat until all the DNN converge on a consensus. 
%The random communication partner is selected for sending model updates, as shown in Figure 2(b).

The aim of GoSGD is to address the issue of speeding up the training of convolutional networks. Instead, another gossip-based algorithm, gossiping SGD \cite{jin2016scale}, is designed to retain the positive features of both synchronous and asynchronous SGD methods. Gossiping SGD replaces the all-reduce collective operation of synchronous training with a gossip aggregation algorithm, achieving an asynchronous manner. 

Both \cite{blot2016gossip} and \cite{jin2016scale} apply gossip algorithms on the updates of SGD, but neither of them s performance convergence degradation at large scale. By deployment on large-scale systems, Daily et al. \cite{daily2018gossipgrad} show that the trivial gossip-based algorithms at scale lead to a communication imbalance, poor convergence and heavy communication overhead. To mitigate these issues, Daily et al. introduce GossipGraD, a gossip communication protocol based SGD algorithm which is practical for scaling deep learning algorithms on large scale systems. GossipGrad reduces the overall communication complexity from $\Theta(\log(p))$ to $O(1)$ and considers diffusion such that computing nodes exchange their updates (gradients) indirectly after every $\log(p)$ steps. It also considers the rotation of communication partners for facilitating direct diffusion of gradients and asynchronous distributed sample shuffling during the feedforward phase in SGD to prevent over-fitting.

\begin{table*}[]
	\tiny
	\caption{A Overview of Systems and Frameworks on EI Model Training} \label{tab:training_systems}
		\centering
		\begin{tabular}{c|c|c|c|c|c}
			\hline
			System or Framework & Architecture & EI Level & Objectives & Employed Technology & Effectiveness\\ \hline
			\hline

			FedAvg\cite{mcmahan2016communication}  & Hybrid & Level-4 &         
			\multicolumn{1}{l|}{\begin{minipage}{1.2in}
					\vskip 4pt
					\begin{itemize}
						\item Robustness to non-IID and unbalanced optimization
						\item Low communication cost
						\item Privacy preservation
					\end{itemize}
					\vskip 4pt
			\end{minipage}  }         &        
			\multicolumn{1}{l|}{\begin{minipage}{1.3in}
			\vskip 4pt
			\begin{itemize}
				\item Federated Learning
                \item Iterative model averaging
			\end{itemize}
			\vskip 4pt
		\end{minipage}  }          & 
		\multicolumn{1}{l}{\begin{minipage}{2in}
		\vskip 4pt
		\begin{itemize}
			\item Reduce communication rounds by 10-100 $\times$ as compared to synchronized stochastic gradient descent
		\end{itemize}
		\vskip 4pt
	\end{minipage}  }                    \\ \hline

			SSGD\cite{shokri2015privacy}  & Hybrid & Level-4 &         
			\multicolumn{1}{l|}{\begin{minipage}{1.2in}
					\vskip 4pt
					\begin{itemize}
						\item Jointly training an DNN model among clients
						\item Privacy preservation
					\end{itemize}
					\vskip 4pt
			\end{minipage}  }         &        
			\multicolumn{1}{l|}{\begin{minipage}{1.3in}
			\vskip 4pt
			\begin{itemize}
				\item Federated Learning
                \item Selective SGD
			\end{itemize}
			\vskip 4pt
		\end{minipage}  }          & 
		\multicolumn{1}{l}{\begin{minipage}{2in}
		\vskip 4pt
		\begin{itemize}
			\item Clients' privacy is preserved while the model accuracy beyond training solely
		\end{itemize}
		\vskip 4pt
	\end{minipage}  }                    \\ \hline

			Zoo\cite{zhao2018privacy} & Hybrid & Level-4 &         \multicolumn{1}{l|}{\begin{minipage}{1.2in}
					\vskip 4pt
					\begin{itemize}
						\item Reducing communication cost
						\item Privacy preservation
					\end{itemize}
					\vskip 4pt
			\end{minipage}  }        &       
			\multicolumn{1}{l|}{\begin{minipage}{1.3in}
			\vskip 4pt
			\begin{itemize}
				\item Federated Learning
				\item Composable services
			\end{itemize}
			\vskip 4pt
		\end{minipage}  }           &                
		\multicolumn{1}{l}{\begin{minipage}{2in}
		\vskip 4pt
		\begin{itemize}
			\item Processes each image within constant time despite the size difference of images
		\end{itemize}
		\vskip 4pt
	\end{minipage}  }                 \\ \hline
		
			BlockFL\cite{kim2018device} & Decentralized & Level-6 &         \multicolumn{1}{l|}{\begin{minipage}{1.2in}
					\vskip 4pt
					\begin{itemize}
						\item Federated Learning in decentralized manner
						\item Low latency
						\item Privacy preservation
					\end{itemize}
					\vskip 4pt
			\end{minipage}  }        &       
			\multicolumn{1}{l|}{\begin{minipage}{1.3in}
			\vskip 4pt
			\begin{itemize}
				\item Federated Learning
				\item Blockchain
			\end{itemize}
			\vskip 4pt
		\end{minipage}  }           &                
		\multicolumn{1}{l}{\begin{minipage}{2in}
		\vskip 4pt
		\begin{itemize}
			\item Latency increase up to 1.5\% to achieve the optimal block generation than the simulated minimum latency
		\end{itemize}
		\vskip 4pt
	\end{minipage}  }                 \\ \hline
	
			Gaia\cite{hsieh2017gaia} & Centralized & Cloud Intelligence & 
			\multicolumn{1}{l|}{\begin{minipage}{1.2in}
					\vskip 4pt
					\begin{itemize}
						\item Geo-distributed scalability 
						\item Intelligent communication mechanism over WANs
						\item Generic and flexible for most machine learning algorithms
					\end{itemize}
					\vskip 4pt
			\end{minipage}  }        &   
			\multicolumn{1}{l|}{\begin{minipage}{1.3in}
			\vskip 4pt
			\begin{itemize}
				\item Aggregation frequency control
				\item ASP model
			\end{itemize}
			\vskip 4pt
		\end{minipage}  }            &              
		\multicolumn{1}{l}{\begin{minipage}{2in}
		\vskip 4pt
		\begin{itemize}
			\item Speedup 1.8-53.5$\times$ over distributed machine learning systems
			\item Within 0.94-1.40$\times$ of the speed of running the same machine learning algorithm on  machines on a local area network (LAN).
		\end{itemize}
		\vskip 4pt
	\end{minipage}  }                   \\ \hline

			DGC\cite{lin2017deep} & N/A & N/A & 
			\multicolumn{1}{l|}{\begin{minipage}{1.2in}
					\vskip 4pt
					\begin{itemize}
						\item Reducing the communication bandwidth
						\item High compression rate without losing model accuracy
						\item Fast Convergence
					\end{itemize}
					\vskip 4pt
			\end{minipage}  }        &   
			\multicolumn{1}{l|}{\begin{minipage}{1.3in}
			\vskip 4pt
			\begin{itemize}
				\item Gradient Compression
				\item Momentum correction
				\item Local gradient clipping
				\item Momentum factor Masking
				\item Warm-up training
			\end{itemize}
			\vskip 4pt
		\end{minipage}  }            &              
		\multicolumn{1}{l}{\begin{minipage}{2in}
		\vskip 4pt
		\begin{itemize}
			\item Achieve a gradient compression ratio from 270$\times$ to 600$\times$ without losing accuracy
			\item Cut the gradient size of ResNet-50 from 97MB to 0.35MB and for DeepSpeech from 488MB to 0.74MB.
		\end{itemize}
		\vskip 4pt
	\end{minipage}  }                   \\ \hline

			eSGD\cite{tao2018esgd} & Hybrid & Level-4 & 
			\multicolumn{1}{l|}{\begin{minipage}{1.2in}
					\vskip 4pt
					\begin{itemize}
						\item Scaling up edge training of CNN
						\item Reducing communication cost
					\end{itemize}
					\vskip 4pt
			\end{minipage}  }        &   
			\multicolumn{1}{l|}{\begin{minipage}{1.3in}
			\vskip 4pt
			\begin{itemize}
				\item Selective transmit important gradient coordinates
				\item Momentum residual accumulation 
			\end{itemize}
			\vskip 4pt
		\end{minipage}  }            &              
		\multicolumn{1}{l}{\begin{minipage}{2in}
		\vskip 4pt
		\begin{itemize}
			\item Reach 91.2\%, 86.7\%, 81.5\% accuracy on MNIST data set with gradient drop ratio 50\%, 75\%, 87.5\% respectively
		\end{itemize}
		\vskip 4pt
	\end{minipage}  }                   \\ \hline

			INCEPTIONN\cite{li2018network} & Hybrid & Level-5 & 
			\multicolumn{1}{l|}{\begin{minipage}{1.2in}
					\vskip 4pt
					\begin{itemize}
						\item Maximizing the opportunities for Compression
						\item Avoiding the bottleneck at aggregators
					\end{itemize}
					\vskip 4pt
			\end{minipage}  }        &   
			\multicolumn{1}{l|}{\begin{minipage}{1.3in}
			\vskip 4pt
			\begin{itemize}
				\item Lossy gradient compression NIC-integrated compression accelerator
				\item Gradient-centric aggregator-free training
			\end{itemize}
			\vskip 4pt
		\end{minipage}  }            &              
		\multicolumn{1}{l}{\begin{minipage}{2in}
		\vskip 4pt
		\begin{itemize}
			\item Reduce the communication time by 70.9\%-80.7\% 
			\item Offer 2.2-3.1$\times$ speedup over the conventional training system while achieving the same level of accuracy.
		\end{itemize}
		\vskip 4pt
	\end{minipage}  }                   \\ \hline

			Arden\cite{wang2018not} & Centralized & Cloud Intelligence &
			\multicolumn{1}{l|}{\begin{minipage}{1.2in}
					\vskip 4pt
					\begin{itemize}
						\item Maximize utilization of computing resources
						\item Low latency
						\item Fast convergence
					\end{itemize}
					\vskip 4pt
			\end{minipage}  }           &        
			\multicolumn{1}{l|}{\begin{minipage}{1.3in}
			\vskip 4pt
			\begin{itemize}
				\item DNN splitting
				\item Arbitrary data nullification
				\item Random noise addition
			\end{itemize}
			\vskip 4pt
		\end{minipage}  }               &             
		\multicolumn{1}{l}{\begin{minipage}{2in}
		\vskip 4pt
		\begin{itemize}
			\item The average reductions compared with the other four DNNs in terms of time, memory, and energy are 60.10\%, 92.07\%, and 77.05\%, respectively
		\end{itemize}
		\vskip 4pt
	\end{minipage}  }                    \\ \hline
	
			PipeDream\cite{harlappipedream} & Hybrid & Level-5 &        
			\multicolumn{1}{l|}{\begin{minipage}{1.2in}
					\vskip 4pt
					\begin{itemize}
						\item Maximizing utilization of computing resources
						\item Low latency 
					    \item Fast convergence
					\end{itemize}
					\vskip 4pt
			\end{minipage}  }          &        
			\multicolumn{1}{l|}{\begin{minipage}{1.3in}
			\vskip 4pt
			\begin{itemize}
				\item DNN splitting
				\item Pipeline parallelism
			\end{itemize}
			\vskip 4pt
		\end{minipage}  }          & 
		\multicolumn{1}{l}{\begin{minipage}{2in}
		\vskip 4pt
		\begin{itemize}
			\item Using 4 machines to train the $>$ 100 million parameter VGG16 on the ImageNet1K dataset, PipeDream converges 2.5$\times$ faster than using a single machine and 3$\times$ faster than data parallel training
		\end{itemize}
		\vskip 4pt
	\end{minipage}  }                   \\ \hline

			GoSGD\cite{blot2016gossip} & Decentralied & Level-6 &        \multicolumn{1}{l|}{\begin{minipage}{1.2in}
					\vskip 4pt
					\begin{itemize}
						\item Speeding up DNN training
						\item Fast convergence
					\end{itemize}
					\vskip 4pt
			\end{minipage}  }          &        
			\multicolumn{1}{l|}{\begin{minipage}{1.3in}
			\vskip 4pt
			\begin{itemize}
			    \item Gossip Training
			\end{itemize}
			\vskip 4pt
		\end{minipage}  }          &
		\multicolumn{1}{l}{\begin{minipage}{2in}
		\vskip 4pt
		\begin{itemize}
			\item Do a better use of the exchanges comparing to EASGD
			\item Converge a lot faster comparing to EASGD
		\end{itemize}
		\vskip 4pt
	\end{minipage}  }                \\ \hline

			Gossiping SGD\cite{jin2016scale} & Decentralied & Level-6 &        \multicolumn{1}{l|}{\begin{minipage}{1.2in}
					\vskip 4pt
					\begin{itemize}
						\item Speeding up DNN training
						\item Scaling up DNN training
						\item Asynchronous training
					\end{itemize}
					\vskip 4pt
			\end{minipage}  }          &        
			\multicolumn{1}{l|}{\begin{minipage}{1.3in}
			\vskip 4pt
			\begin{itemize}
			    \item Gossip Training
				\item Model partition
			\end{itemize}
			\vskip 4pt
		\end{minipage}  }          &
		\multicolumn{1}{l}{\begin{minipage}{2in}
		\vskip 4pt
		\begin{itemize}
			\item One iteration of gossiping SGD is faster than one iteration of all-reduce SGD 
			\item Work quickly at the initial step size.
		\end{itemize}
		\vskip 4pt
	\end{minipage}  }                \\ \hline
	
			GossipGraD\cite{daily2018gossipgrad} & Decentralied & Level-6 &        \multicolumn{1}{l|}{\begin{minipage}{1.2in}
					\vskip 4pt
					\begin{itemize}
						\item Reducing communication complexity
						\item Fast convergence
					    \item Privacy preservation
					\end{itemize}
					\vskip 4pt
			\end{minipage}  }          &        
			\multicolumn{1}{l|}{\begin{minipage}{1.3in}
			\vskip 4pt
			\begin{itemize}
			    \item Gossip Training
				\item Model partition
			\end{itemize}
			\vskip 4pt
		\end{minipage}  }          &
		\multicolumn{1}{l}{\begin{minipage}{2in}
		\vskip 4pt
		\begin{itemize}
			\item Achieve about 100\% compute efficiency for ResNet50 using 128 NVIDIA Pascal P100 GPUs while matching the top-1 classification accuracy published in literature.
		\end{itemize}
		\vskip 4pt
	\end{minipage}  }                \\ \hline

		\end{tabular}
	
\end{table*}

\subsection{Summary of Existing Systems and Frameworks}

In this subsection, we summary the systems and framework for distributed EI model training on the edge. An overview of the above-mentioned existing systems and frameworks is given in Table. \ref{tab:training_systems}, including the architecture, EI level, objectives, employed technologies and effectiveness.

%High performance, low resources consumption, and robustness are the common pursuit of those systems and frameworks as listed in Table \ref{tab:training_systems}. Specifically, high performance represents the low loss function value that the DNN model achieves by distributed training. To evaluate the performance more accurately, the influence factors, including the whole latency, the testing DNN model and the available computation resource, are fixed in the experimental setup. Low resources consumption is a board statement, where the resources range from network bandwidth, energy power to computation capability. We expect the system or the framework to consume as little as resources, since it significantly influences the total economic viability of distributed DNN training, especially for commercial companies. Robustness implies the ability that the system or the framework cope with errors. A robust system is able to identify invalid input and properly deal with it. In distributed DNN training, robustness requires the system to train data with special properties such as non-IID, perturbance, and noise. Robustness is practically significant for commercial products since users’ inputs are often versatile and noisy. 

In general, a key challenge for distributed EI model training is the data privacy issue. It is because the distributed data sources may originate from individual persons and different organizations. For users, they may be sensitive to their own private data, not allowing any private information to be shared. For companies, they have to consider the privacy policy to avoid legal subpoenas and extra-judicial surveillance. Therefore, the design of distributed training systems needs to carefully consider privacy preservation. Systems considering privacy issue in Table. \ref{tab:training_systems} include FedAvg, BlockFL, GossipGraD and so on. The decentralized architecture is naturally friendly to users’ privacy, for which the systems that are based on the decentralized architecture such as BlockFL and GossipGraD typically preserve privacy better. As a contrast, the centralized architecture involves a centralized data collection operation, and the hybrid architecture requires a data transmission operation. For this reason, the systems based on these two architectures would implement more extra efforts in data privacy protection.

Compared with the DNN training under cloud-based framework, the DNN training under edge-based framework pays more attention to protecting users' privacy and training an available deep learning model faster. Under cloud-based training, a large amount of raw data generated at the client side is directly transmitted to the cloud data center through the long WAN, which not only causes hidden dangers of user privacy leakage but also consumes huge bandwidth resources. Moreover, in some scenarios such as military and disaster applications when the access to the cloud center is impossible, the edge-based training will be highly desirable. On the other hand, the cloud data center can collect a larger amount of data and train an AI model with more powerful resources, and hence the advantage of cloud intelligence is that it can train a much larger-scale and more accurate model.
%There is another feature, convergence, that need to be considered. This feature is specialized for systems using decentralized methods and parallel methods. The decentralized method, such as gossip-based methods, train DNN models by iteratively updating the gradients on devices without aggregating them. The parallel methods parallelize training data or DNN model on multiple threads. These methods make sense if the updates or the threads converge on a consensus. For example, there are two systems, PipeDream and GossipGraD, set out to achieve fast convergence in Table. \ref{tab:training_systems}. PipeDream is the system based on the hybrid architecture which employs pipeline parallelism and GossipGraD scales up distributed DNN training based on decentralized gossip training.
\section{Edge Intelligence Model Inference}
\label{edgeinference}
After the distributed training of deep learning model, then the efficient implementation of model inference at the edge will be critical for enabling high-quality edge intelligence service deployment. In this section, we discuss the DNN model inference at the edge, including the architectures, key performance indicators, enabling techniques and existing systems \& frameworks.

\subsection{Architectures}
\begin{figure*}[htbp]
	\centering
	
	\subfigure[Edge-based Mode]{
		\includegraphics[scale=0.4]{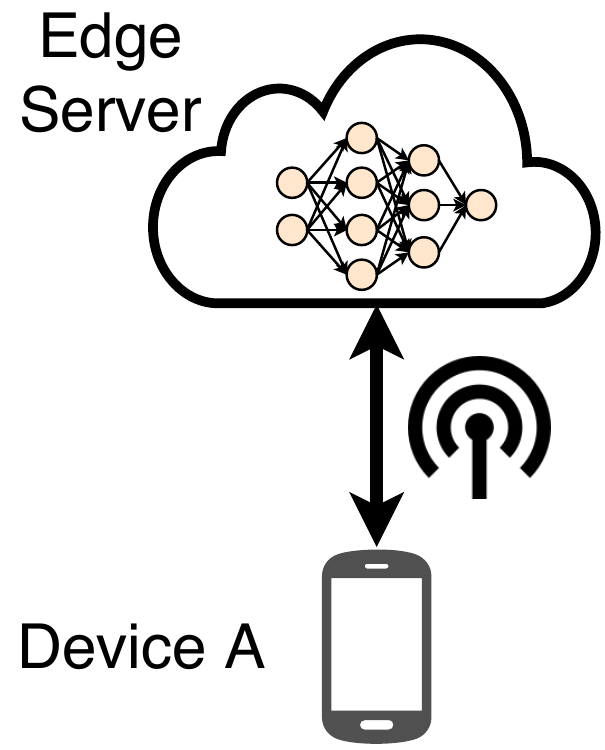}\label{mode a}
	}
	\subfigure[Device-based Mode]{
		\includegraphics[scale=0.4]{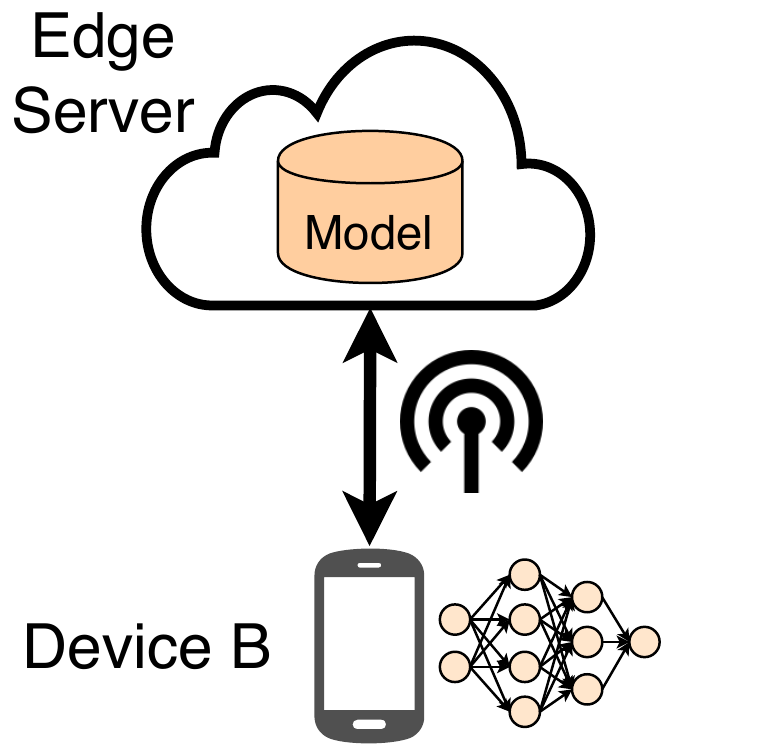}\label{mode b}
	}
	\subfigure[Edge-Device Mode]{
		\includegraphics[scale=0.4]{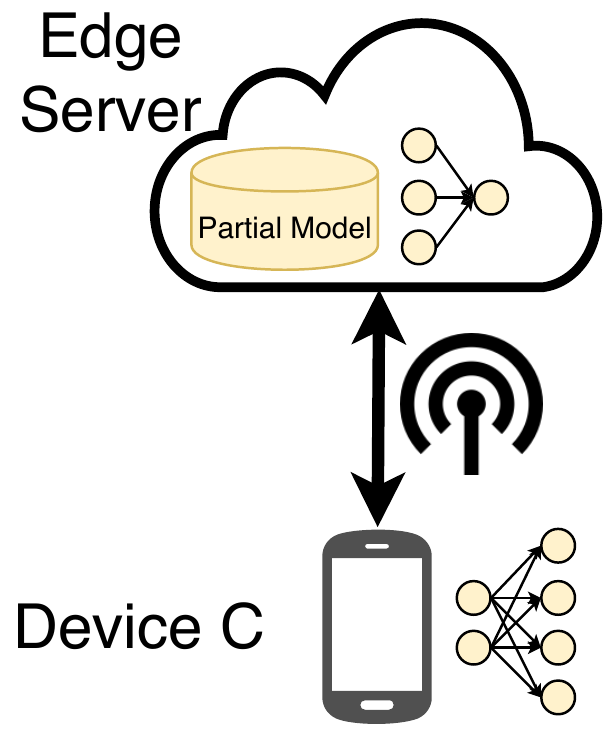}\label{mode c}
	}
	\subfigure[Edge-Cloud Mode]{
		\includegraphics[scale=0.4]{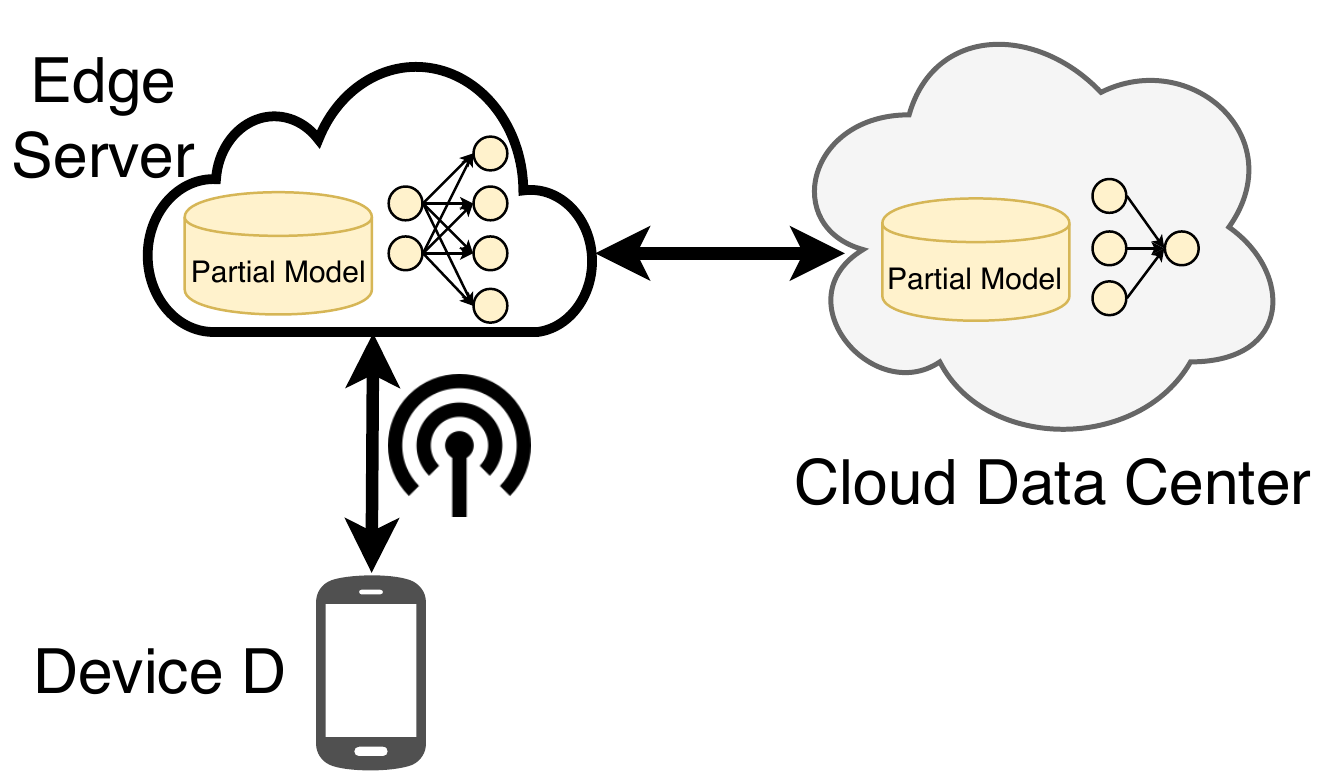}\label{mode d}
	}
	%	\vspace{-10pt}
	\caption{Major edge-centric inference modes: edge-based, device-based, edge-device and edge-cloud}\label{edge intelligence inference framework}
	%	\vspace{-10pt}
\end{figure*}

Besides the common cloud-based and device-cloud inference architectures, we further define several major edge-centric inference architectures and classify them into four modes, namely edge-based, device-based, edge-device and edge-cloud, which are illustrated in Fig. \ref{edge intelligence inference framework}. 

In Fig. \ref{edge intelligence inference framework}, we represent different four DNN model inference modes respectively. We describe the main workflow of each mode as follows.

%\begin{figure*}[!ht]
%	\centering
%	\includegraphics[scale=0.5]{fig/dnn_inference/edge_intelligence_framework.pdf}\\
	%\vspace{-10pt}
%	\caption{Edge intelligence modes: edge-based, device-based, edge-device and edge-synergy}\label{edge intelligence inference framework}
	%\vspace{-5pt}
%\end{figure*}

%Two main components of these modes shown in Fig. \ref{edge intelligence inference framework} are edge device and edge server. The device receives data inputs and preprocesses them locally. Then the device can send data either to the DNN model at the edge server or to run the DNN model locally on device. After the DNN model inference, the prediction results will be sent back to the  device and presented to the user. 

%In Fig. \ref{edge intelligence inference framework}, we represent different four DNN model inference modes respectively. We describe the main workflow of each mode as follows.
%The edge server obtains DNN model from the remote cloud data center and sends it to the mobile device. The edge server can well execute DNN model and provide APIs for mobile devices to help inference processing.

\subsubsection{Edge-based} In Fig. \ref{mode a}, Device A is in the edge-based mode, which means that the device receives the input data then send them to the edge server. When the DNN model inference is done at the edge server, the prediction results will be returned to the device. Under this inference mode, since the DNN model is on the edge server, it is easy to implement the application on different mobile platforms. But the main disadvantage is that the inference performance depends on network bandwidth between the device and the edge server.

\subsubsection{Device-based} In Fig. \ref{mode b}, Device B is in the device-based mode. The mobile device obtains the DNN model from the edge server and performs the model inference locally. During the inference process, the mobile device does not communicate with the edge server. So the inference is reliable, but it requires a large amount of resources such as CPU, GPU, and RAM on the mobile device. The performance depends on the local device itself.

\subsubsection{Edge-device} In Fig. \ref{mode c}, Device C is in the edge-device mode. Under the edge-device mode, the device first partitions the DNN model into multiple parts according to the current system environmental factors such as network bandwidth, device resource and edge server workload. Then, the device will execute the DNN model up to a specific layer and send the intermediate data to the edge server. The edge server will execute the remain layers and sends the prediction results to the device. Compared to the edge-based mode and the device-based mode, the edge-device mode is more reliable and flexible. It may also require huge resource on the mobile device because the convolution layers at the front position of a DNN model is computational-intensive generally.

\subsubsection{Edge-cloud} In Fig. \ref{mode d}, Device D is in the edge-cloud mode. It is similar to the edge-device mode and is suitable for the case that the device is highly resource constrained. In this mode, the device is responsible for input data collection and the DNN model is executed through edge-cloud synergy. The performance of this model heavily depends on the network connection quality.

We should emphasize that the four edge-centric inference modes above can be adopted in a system simultaneously to carry out complex AI model inference tasks (e.g., Cloud-Edge-Device hierarchy), by efficiently pooing heterogeneous resources across a multitude of end devices, edge nodes and clouds.

\subsection{Key Performance Indicators}
To describe the service quality of edge intelligence model inference, we introduce the following five metrics.
\subsubsection{Latency}
Latency refers to the time spent in the whole inference process, including pre-processing, model inference, data transmission, and post-processing. For some real-time intelligent mobile applications (e.g., AR/VR mobile gaming and intelligent robots), they usually have stringent deadline requirement such as 100ms latency. Latency indicator is affected by many factors, including the resources on edge devices, the way of data transmission and the way to execute the DNN model.

\subsubsection{Accuracy}
Accuracy refers to the ratio of the number of the input samples that get the correct predictions from inference to the total number of input samples, reflecting the performance of the DNN models. For some mobile applications requiring a high level of reliability, such as self-driving car and face authentication, they demand the ultrahigh accuracy on the DNN model inference. Besides the DNN model's own inference capability, the inference accuracy depends on the speed of feeding the input data to the DNN model. For a video analytics application, under a fast feeding rate, some input sample may be skipped due to the edge device's constraint resources, causing a drop in accuracy.

\subsubsection{Energy}
To execute a DNN model, compared with the edge server and the cloud data center, the end devices are usually battery-limited. The computation and communication overheads of DNN model inference bring a large amount of energy consumption. For an edge intelligence application, energy efficiency is of great importance and is affected by the size of DNN model and the resources on edge devices.

\subsubsection{Privacy}
The IoT and mobile devices generate a huge amount of data, which could be privacy sensitive. Thus, it is also important to protect privacy and data security near the data source for an edge intelligence application during the model inference stage. Privacy protection depends on the way of processing the original data.

\subsubsection{Communication overhead}
Except for the device-based mode, the communication overhead affects the inference performance of the other modes greatly. It is necessary to minimize the overhead during the DNN model inference in an edge intelligence application, particularly the expensive wide-area network bandwidth usage for the cloud. Communication overhead here mainly depends on the mode of DNN inference and the available bandwidth.

\subsubsection{Memory Footprint}
Optimizing the memory footprint of performing deep neural network model inference on mobile devices is very necessary. On the one hand, typically, a high-precision deep neural network model is accompanied by millions of parameters, which is very hungry for the hardware resources of mobile devices. On the other hand, unlike high-performance discrete GPUs on the cloud data center, there is no dedicated high-bandwidth memory for mobile GPUs on mobile devices \cite{8675201}. Moreover, mobile CPUs and GPUs typically compete for shared and scarce memory bandwidth. For the optimization of the DNN inference at the edge side, memory footprint is a non-negligible indicator. Memory footprint is mainly affected by the size of the original DNN model and the way of loading the tremendous DNN parameters.

\subsection{Enabling Technologies}
\label{tech introduce}

In this subsection, we review the enabling technologies for improving one or more of the aforementioned key performance indicators for edge intelligence model inference. Table \ref{tech highlights} summarizes the highlights of each enabling technology.
\begin{table*}[!h]
	\caption{Technologies for distributed DNN inference at the  edge}\label{tech highlights}
	\centering
	\begin{tabular}{c|c|c}
		\hline
		\textbf{Technology}            & \textbf{Highlights} & \textbf{Related Work} \\ \hline
		\hline
		Model Compression              &           \multicolumn{1}{l|}{\begin{minipage}{3.5in}
			\vskip 4pt
			\begin{itemize}
				\item Weight pruning and quantization to reduce storage and computation
			\end{itemize}
		\vskip 4pt
		\end{minipage}  }       &         \cite{han2015learning, han2015deep, chen2016eyeriss, yang2017designing, reagen2016minerva, liu2018demand, pact2018Going,  lane2016deepx}           \\ \hline
		
		Model Partition                &           \multicolumn{1}{l|}{ \begin{minipage}{3.5in}
			\vskip 4pt
			\begin{itemize}
				\item Computation offloading to the edge server or mobile devices
				\item Latency- and energy-oriented optimization
			\end{itemize}
			\vskip 4pt
		\end{minipage} }          &           \cite{zeng2019boomerang, kang2017neurosurgeon, Li2018edgeintelligence, dan2019infocom,  li2018jalad, jeong2018ionn, ko2018edge, mao2017mednn, mao2017modnn, zhao2018deepthings}         \\ \hline
		
		Model Early-Exit               &           \multicolumn{1}{l|}{ \begin{minipage}{3.5in}
				\vskip 4pt
				\begin{itemize}
					\item Partial DNNs model inference
					\item Accuracy-aware
				\end{itemize}
				\vskip 4pt
		\end{minipage} }           &           \cite{Li2018edgeintelligence, zeng2019boomerang, teerapittayanon2016branchynet, teerapittayanon2017distributed, bolukbasi2017adaptive, li2018deep, lo2017dynamic, leroux2017cascading}         \\ \hline
		
		Edge Caching                 &           \multicolumn{1}{l|}{ \begin{minipage}{3.5in}
				\vskip 4pt
				\begin{itemize}
					\item Fast response towards reusing the previous results of the same task
				\end{itemize}
				\vskip 4pt
		\end{minipage} }           &          \cite{chen2015glimpse, drolia2017cachier, drolia2017precog, guo2018foggycache, venugopal2018shadow}          \\ \hline
		
		Input Filtering                &           \multicolumn{1}{l|}{ \begin{minipage}{3.5in}
				\vskip 4pt
				\begin{itemize}
					\item Detecting difference between inputs, avoiding abundant computation
				\end{itemize}
				\vskip 4pt
		\end{minipage} }           &          \cite{kang2017noscope, wang2018bandwidth, jain2018rexcam, zhang2018ffs, canelpicking}          \\ \hline
		
		Model Selection                &           \multicolumn{1}{l|}{ \begin{minipage}{3.5in}
				\vskip 4pt
				\begin{itemize}
					\item Inputs-oriented optimization
					\item Accuracy-aware
				\end{itemize}
				\vskip 4pt
		\end{minipage} }           &         \cite{park2015big, taylor2018adaptive, jiang2018chameleon, shu2018if, stamoulis2018designing}           \\ \hline
		
		Support for Multi-Tenancy      &          \multicolumn{1}{l|}{ \begin{minipage}{3.5in}
				\vskip 4pt
				\begin{itemize}
					\item Scheduling multiple DNN-based task
					\item Resource-efficient
				\end{itemize}
				\vskip 4pt
		\end{minipage} }            &          \cite{fang2018nestdnn, mathur2017deepeye, fang2018mitigating, hung2018videoedge, jiang2018mainstream, jiang2018chameleon, narayananaccelerating}          \\ \hline
		
		Application-specific Optimization &          \multicolumn{1}{l|}{ \begin{minipage}{3.5in}
				\vskip 4pt
				\begin{itemize}
					\item Optimizations for the specific DNN-based application
					\item Resource-efficient
				\end{itemize}
				\vskip 4pt
		\end{minipage} }            &           \cite{ ran2018deepdecision, jiang2018chameleon}         \\ \hline
	\end{tabular}
\end{table*}

\subsubsection{Model Compression}
To alleviate the tension between resource hungry DNNs and resource-poor end devices, DNN compression has been commonly adopted to reduce the model complexity and resource requirement, enabling local, on-device inference which in turn reduces the response latency and has fewer privacy concerns. That is, model compression method optimizes the above four indicators, latency, energy, privacy and memory footprint. Various DNN compression techniques have been proposed, including weight pruning, data quantization, and compact architecture design.

Weight pruning represents the most widely adopted technique of model compression. This technique removes redundant weights (i.e., connections between neurons) from a trained DNN. Specifically, it first ranks the neurons in the DNN according to how much the neuron contributes, and then removes the low-ranking neurons to reduce the model size. Since removing neurons damages the accuracy of the DNN, then how to reduce the network size meanwhile preserving the accuracy is the key challenge. For modern large-scale DNNs, a pilot research \cite{han2015learning} in 2015 tackled this challenge by applying a magnitude-based weight pruning method. The basic idea of this method is as follows: first remove small weights whose magnitudes are below a threshold (e.g., 0.001) and then fine-tune the model to restore the accuracy. For AlexNet and VGG-16, this method can reduce the number of weights by 9x and 13x with no loss of accuracy on ImageNet. The follow-up work “Deep Compression” \cite{han2015deep} which blends the advantages of pruning, weight sharing and Huffman coding to compress DNNs, further pushes the compression ratio to 35-49x.

However, for energy-constrained end devices, the above magnitude-based weight pruning method may not be directly applicable, since empirical measurements show that the reduction of the number of weights does not necessarily translate into significant energy saving \cite{chen2016eyeriss}. This is because for DNNs as exemplified by AlexNet, the energy of the convolutional layers dominates the total energy cost, while the number in the fully-connected layers contributes most of the total number of weights in the DNN. This suggests that the number of weights may not be a good indicator for energy, and the weight pruning should be directly energy-aware for end devices. As the first step towards this end, an online DNN energy estimation tool (https://energyestimation.mit.edu/) has been developed by MIT to enable fast and easy DNN energy estimation. This fine-grained tool profiles the energy for the data movement from different levels of the memory hierarchy, the number of MACs, and the data sparsity at the granularity of DNN layer. Based on this energy estimation tool, an energy-aware pruning method called EAP \cite{yang2017designing} is proposed. 

Another mainstream technique for model compression is data quantization. Instead of adopting the 32-bit floating point format, this technique uses a more compact format to represent layer inputs, weights, or both. Since representing a number with fewer bits reduces memory footprint and accelerates computation, data quantization improves overall computation and energy efficiency. %Table~\ref{tab:num-repr} surveys the most commonly adopted number representations used by DNNs, here they are categorized into 4 types: floating-point, fixed-point, exponent and binary. The above four types can be represented by the the canonical number format based on IEEE 754 Standard, as illustrated in Fig. \ref{fig:ieee-754}. The canonical format consists of four fields: sign (S), exponent (E), fraction (F) and bias (B), as shown in Fig.~\ref{fig:ieee-754}. 
Most prior proposals for quantization tune the bit-width only for a fixed number type in an ad hoc manner, which may lead to a suboptimal result. To address this issue, the recent work \cite{pact2018Going} investigated the problem of optimal number representations at the layer granularity, in terms of finding the optimal bit-width for the canonical format based on IEEE 754 Standard. This problem is challenging due to the combinatorial explosion of feasible number formats. In response, the authors developed a portable API called Number abstract data type (ADT). it enables users to declare the data to be quantized in a layer (e.g., inputs, weights, or both) as Number type. By doing so, ADT encapsulates the internal representation of a number, thus separating the concern for developing an effective DNN from the concern of optimizing the number representation at a bit level.

While most existing efforts use a single compression technique, they may not suffice to meet the diverse requirements and constraints on accuracy, latency, storage, and energy imposed by some IoT devices. Emerging studies have shown how different compression techniques can be coordinated to maximally compress DNN models. For example, both Deep Compression \cite{han2015deep} and Minerva \cite{reagen2016minerva} combine weight pruning and data quantization to enable fast, low-power and highly-accurate DNN inference. More recently, researchers argue that for a given DNN, the combination of compression techniques should be selected on demand, i.e., adapting to the application driven system performance (e.g. accuracy, latency and energy) and the varying resource availability across platforms (e.g. storage and processing capability). To this end, the proposed automatic optimization framework AdaDeep \cite{liu2018demand} systematically formulates the goals and constraints on accuracy, latency, storage and energy into a unified optimization problem, and leverages deep reinforcement learning (DRL) to effectively find a good combination of compression techniques. %Extensive evaluations on five public datasets and across twelve mobile devices demonstrate that AdaDeep achieves up to 9.8x latency reduction, 4.3x energy efficiency improvement, and 38x storage reduction in DNNs while incurring negligible accuracy loss.

\subsubsection{Model Partition}
\begin{figure}[!ht]
	\centering
	\includegraphics[scale=0.32]{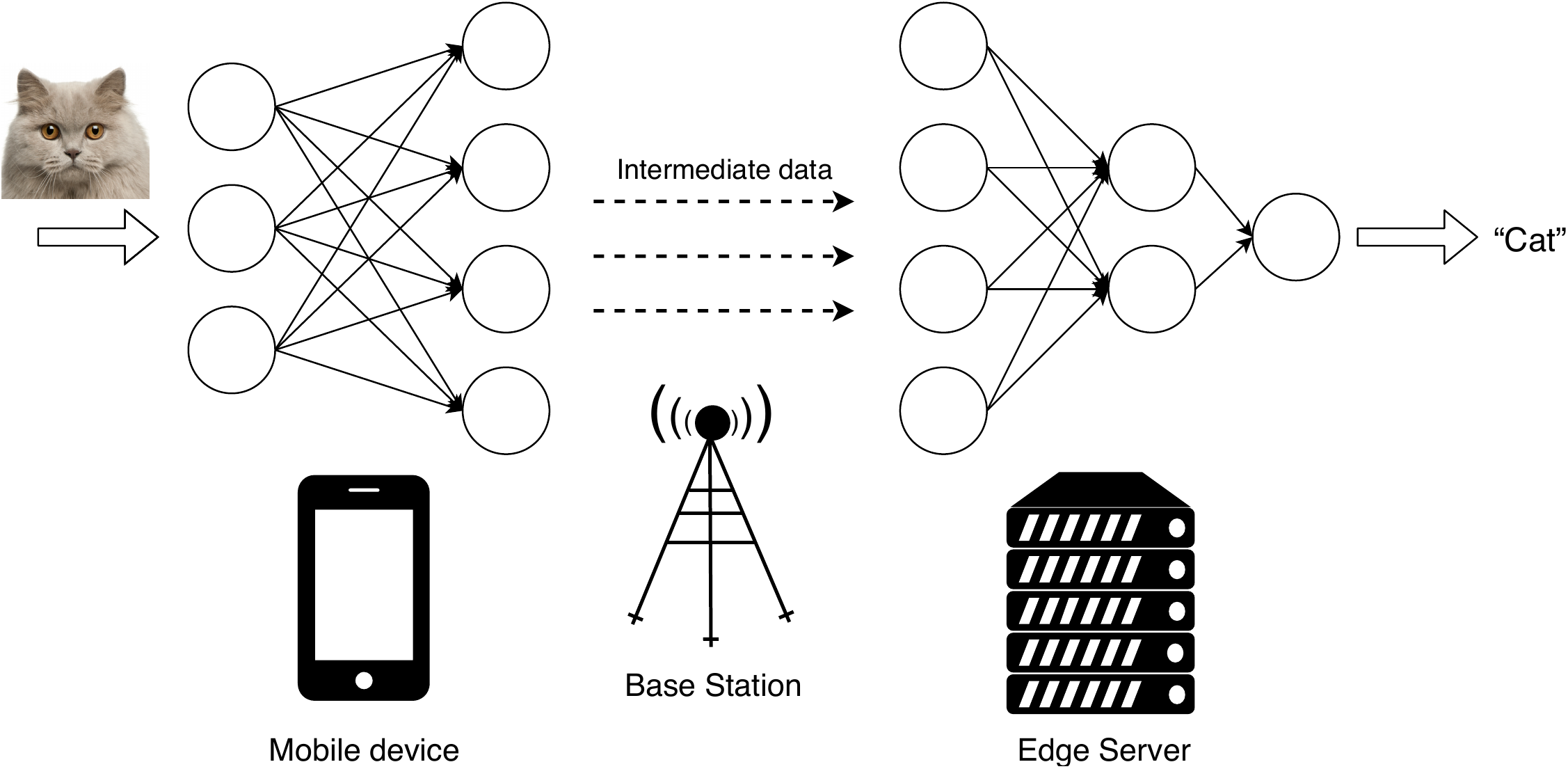}\\
	%\vspace{-10pt}
	\caption{An illustration for model partition between devices and edge server}\label{tech model partition}
	%\vspace{-5pt}
\end{figure}

To alleviate the pressure of the edge intelligence application execution on end devices, as shown in Fig. \ref{tech model partition}, one intuitive idea is the model partition, offloading the computational-intensive part to the edge server or the nearby mobile devices, obtaining a better model inference performance. Model partition mainly cares about the issues of latency, energy and privacy.

The model partition can be divided into two types, partition between server and device and partition between devices. For the model partition between server and device, Neurosurgeon \cite{kang2017neurosurgeon} represents an iconic effort. In Neurosurgeon, DNN model is partitioned between the device and the server, the key challenge is to figure out one suitable partition point to get the optimal model inference performance. Considering from latency aspect and energy efficiency aspect respectively, the authors propose a regression-based method to estimate the latency of each layer in the DNN model and return an optimal partition point which makes the model inference meet latency requirement or energy requirement. %When optimizing for best latency performance, Neurosurgeon achieves a latency speedup of 3.1x on average and up to 40.7x over the cloud-based approach. When optimizing for best energy consumption, Neurosurgeon achieves on average a 59.5\% reduction in mobile energy and up to 94.7\% reduction over the cloud-based approach.

Hereafter, Ko et al. propose an edge-host partitioning method \cite{ko2018edge}, which combines model partition with lossy feature encoding. That is, the intermediate data after model partition will be compressed using lossy feature encoding before transmission. %The experiment shows that the proposed approach improves the system energy efficiency and throughput by 15.3x and 16.5x compared to the edge-based method, and 2.3x and 2.5x compared to the device-based method. 
Also jointly leverages model partition and lossy feature encoding, the JALAD \cite{li2018jalad} framework formulates the model partition as an integer linear programming (ILP) problem to minimize the model inference latency under a guaranteed accuracy constraint. %The simulation demonstrates that the method can speed up model inference while guaranteeing the accuracy loss within a user-defined requirement. 
For DNNs those are characterized by a directed acyclic graph (DAG) rather than a chain, optimizing the model partition to minimize the latency is proven to be NP-hard in general. In response, Hu \emph{et al.}  \cite{dan2019infocom} propose an approximation algorithm that provides worst-case performance guarantee, based on the graph min-cut method. 
The above frameworks all have an assumption that the server has the DNN model of the edge intelligence application. IONN \cite{jeong2018ionn} propose an incremental offloading technique for edge intelligence application. IONN partitions the DNN layers and incrementally uploads them to allow collaborative DNN model inference by mobile devices and the edge server. Compared to the approach that uploads the entire model, IONN significantly improves query performance and energy consumption during DNN model uploading.

Another type of model partition is the partition between devices. As the pioneering effort of model partition between devices, MoDNN \cite{mao2017modnn} introduces WiFi Direct technique to build a micro-scale computing cluster in WLAN with multiple authorized WiFi-enabled mobile devices for partitioned DNN model inference. The mobile device that carries the DNN task will be the group owner and the others act as the worker nodes. Two partition schemes are proposed in MoDNN to accelerate DNN layer execution. The experiment shows that with  2 to 4  worker nodes, MoDNN accelerates DNN model inference by 2.17-4.28x. In the follow-up work MeDNN \cite{mao2017mednn}, greedy two-dimensional partition is proposed to adaptively partition DNN model onto multiple mobile devices and utilize a structured sparsity pruning technique to compress DNN model. MeDNN improves DNN model inference by 1.86-2.44x with 2-4 worker nodes and saves 26.5\% of additional computing time and 14.2\% of extra communication time. Note that DNN layers are partitioned horizontally in MoDNN and MeDNN, in contrast, DeepThings \cite{zhao2018deepthings} employs a fused tile partitioning method that partitions the DNN layers vertically to reduce the memory footprint. %The Fused Tile partitioning method can reduce memory usage by more than 68\% without accuracy loss and DeepThings speedups DNN model inference of 1.7x-3.5x on 2-6 edge devices with no more than 23MB memory each. 

%A special case for model partition between devices is DeepX . 
DeepX \cite{lane2016deepx} also tries to partition DNN models but it only partitions the DNN model into several sub-models and distributes them on local processors. DeepX proposes two schemes: Runtime Layer Compression (RLC) and Deep Architecture Decomposition (DAD). The layer after compression will be executed by specific local processors (CPU, GPU, and DSP). An additional note is that when we have multiple tasks of model partition, we need to make optimization for the scheduler. LEO \cite{georgiev2016leo} is a novel sensing algorithm scheduler that maximizes the performance for multiple continuous mobile sensor applications by partitioning the sensing algorithm execution and distributing tasks on CPU, co-processor, GPU and the cloud. %And OpenVDAP \cite{zhang2018openvdap} proposes a heterogeneous vehicle computing platform, including a tasks scheduling framework. The framework schedules the tasks to specific acceleration hardware based on task computing characteristics and hardware utilization.

\subsubsection{Model Early-Exit}
A DNN model with a high accuracy usually has a deep structure. It consumes a large number of resources to execute such a DNN model on the end device. To accelerate model inference, model early-exit method leverages output data of early layer to get the classification result, which means that the inference process is finished by using partial DNN model. Latency is the optimization target of model early-exit.

\begin{figure}[!ht]
	\centering
	\includegraphics[scale=0.22]{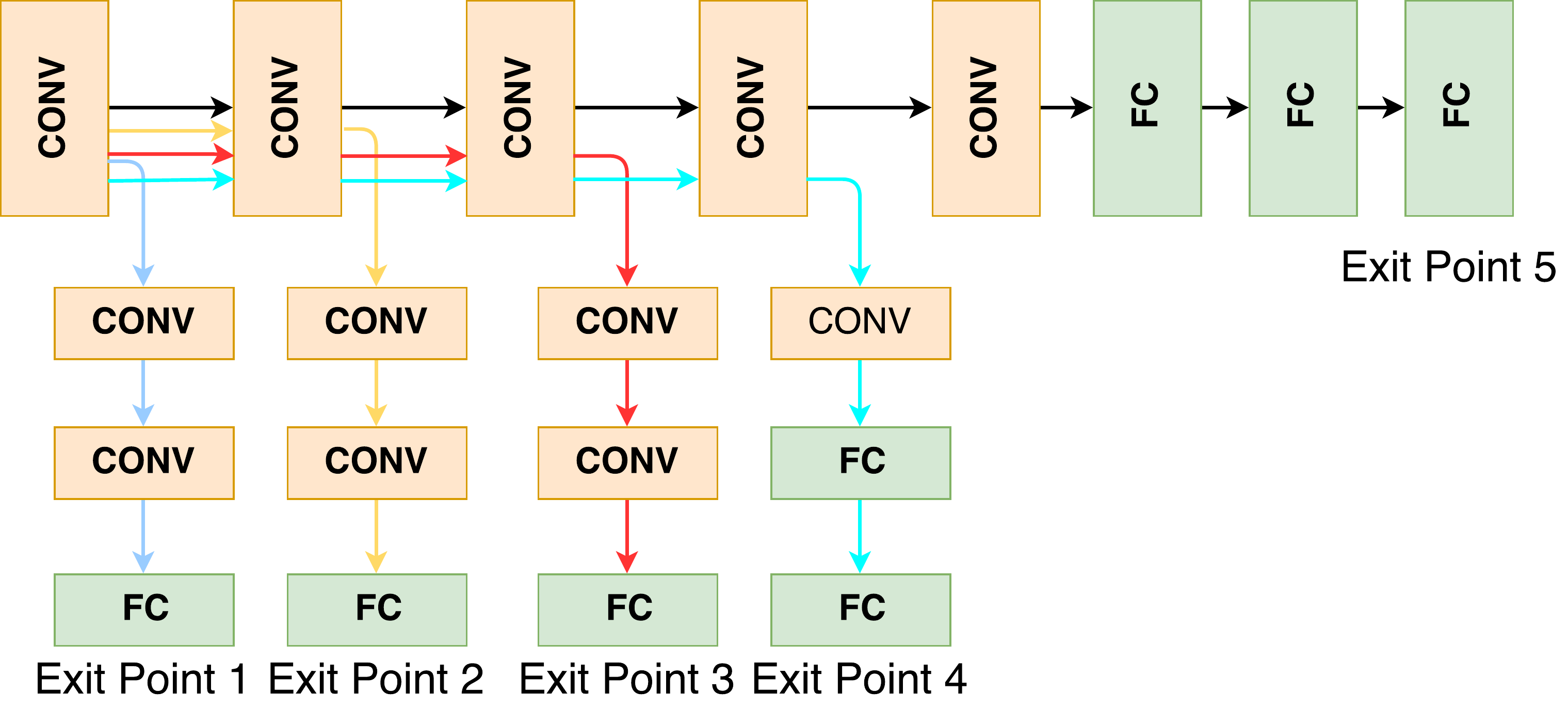}\\
	%\vspace{-10pt}
	\caption{A CNN model with five exit points}\label{tech model earlyexit}
	%\vspace{-5pt}
\end{figure}

BranchyNet \cite{teerapittayanon2016branchynet} is a programming framework that implements the model early-exit mechanism. With BranchyNet, the standard DNN model structure is modified by adding exit branches at certain layer locations. Each exit branch is an exit point and shares part of DNN layers with the standard DNN model. Fig. \ref{tech model earlyexit} shows a CNN model with three five points. The input data can be classified at these diverse early exit point. %The BranchyNet model can be viewed as a DNN model with multi-tasks, therefore the training of BranchyNet is a joint optimization problem. 
%For the trained BranchyNet model, the mobile devices can execute the branches at early exit points, if the accuracy cannot meet a requirement, the intermediate data of the early exit point will be sent to the edge server and be executed by the later exit point. The experiments show that BranchyNet speedups 2x-6x on both CPU and GPU. 

Based on BranchyNet, a framework named DDNNs \cite{teerapittayanon2017distributed} for distributed deep neural networks across the cloud, edge and devices is proposed. DDNNs has a three-layer structure framework, including device layer, edge server layer and cloud layer. Each layer represents an exit point of BranchyNet. Three aggregation methods including max pooling (MP), average pooling (AP) and concatenation (CC) are proposed. The aggregation methods work when multiple mobile devices send intermediate to an edge server or when multiple edge servers send intermediate data to the cloud data center. MP aggregates the data vectors by taking the max of each component. AP aggregates the data vectors by taking the average of each component. CC just simply concatenates the data vectors as one vector. Also built on top of BranchyNet, Edgent \cite{Li2018edgeintelligence} is proposed to navigate the accuracy-latency tradeoff when jointly applying model early-exit and model partition. The basic idea of Edgent is to maximize the accuracy under a given latency requirement, via the regression-based layer latency prediction model.

In addition to BranchyNet, there are different methods to implement model early-exit. For example, Cascading network \cite{leroux2017cascading} simply adds max pooling layer and fully-connected layer to the standard DNN model and achieves a speedup of 20\%. DeepIns \cite{li2018deep} proposes a manufacture inspection system for the smart industry using DNN model early-exit. In DeepIns, edge devices are responsible for data collection, the edge server acts as the first exit point and the cloud data center acts as the second exit point. Then Lo et al. \cite{lo2017dynamic} proposes adding an authentic operation (AO) unit to the basic BranchyNet model. The AO unit determines whether an input has to be transferred to the edge server or cloud data center for further execution by setting different threshold criteria of confidence level for different DNN model output classes. And Bolukbasi et al. \cite{bolukbasi2017adaptive} trains a policy that determines whether the current samples should proceed to the next layer by adding regularization to the evaluation latency of the DNN model.

\subsubsection{Edge Caching}
\begin{figure}[!ht]
	\centering
	\includegraphics[scale=0.6]{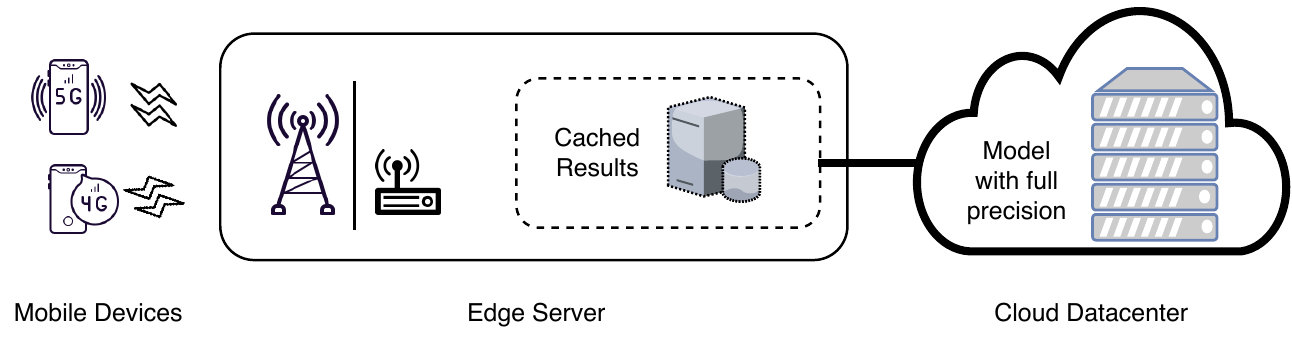}\\
	%\vspace{-10pt}
	\caption{The process of semantic cache technique}\label{tech semantic cache}
	%\vspace{-5pt}
\end{figure}

Edge caching is a new kind of method used to accelerate DNN model inference, i.e., optimizing the latency issue, by caching the DNN inference results. The core idea of edge caching is to cache and reuse the task results such as the prediction of image classification at the network edge, reducing the querying latency of edge intelligence application. Fig. \ref{tech semantic cache} shows the basic process of semantic cache technique, if the request from mobile devices hit the cached results stored in the edge server, the edge server will return the result, otherwise, the request will be transferred to the cloud data center for inference with the model of full precision.

Glimpse \cite{chen2015glimpse} is a pioneering effort to introduce cache technique to DNN inference task. For an object detection application, Glimpse proposes to reusing the stale detection result to detect the object on current frames. The results of the detected object of the stale frames are cached on the mobile devices, then Glimpse extracts a subset of these cached results and computes the optical flow of features between processed frames and the current frame. And the computing results of optical flow will guide us to move the bounding box to the right location in the current frame. Glimpse achieves an acceleration of 1.6-5.5x.

But caching results locally does not scale beyond tens of images, then Cachier \cite{drolia2017cachier} is proposed to achieve recognition of thousands of objects. In Cachier, results of edge intelligence application are cached in the edge server, storing the features of input (e.g., image) and the corresponding task results. Then Cachier uses the least frequently used (LFS) as the cache replacement strategy. If the input can not hit the cache, the edge server will transfer the input to the cloud data center. Cachier can increase responsiveness by 3x or more. Precog \cite{drolia2017precog} is the extension of Cachier. In Precog, the cached data is not only stored in the edge server but also in the mobile device. Precog uses predictions of Markov chains to prefetch data onto the mobile device and reach a speedup of 5x. In addition, Precog also proposes to dynamically adjust the cached feature extraction model on the mobile device according to the environment information. Shadow Puppets is another improved version of Cachier. Cachier extracts features from input using the standard feature extraction like locality sensitive hashing (LSH), but these features may not reflect the similarity as precise as the human dose. Then in Shadow Puppets \cite{venugopal2018shadow}, it uses a small-footprint DNN to generate hash codes to represent the input data and get a remarkable latency improvement of 5-10x.

Considering the application scenario that the same application runs on multiple devices in close proximity and the DNN model often processes similar input data, then FoggyCache \cite{guo2018foggycache} are proposed to minimize these redundant computations. There are two challenges in FoggyCache: one is that the input data distribution is unknown so the problem is how to index the input data with a constant lookup quality, and the other is how to represent the similarity of the input data. To address these two challenges, FoggyCache proposes adaptive locality sensitive hashing (A-LSH) and homogenized kNN (H-kNN) schemes, respectively. FoggyCache reduces computation latency and energy consumption by a factor of 3x to 10x.

\subsubsection{Input Filtering}
\begin{figure}[!ht]
	\centering
	\includegraphics[scale=0.7]{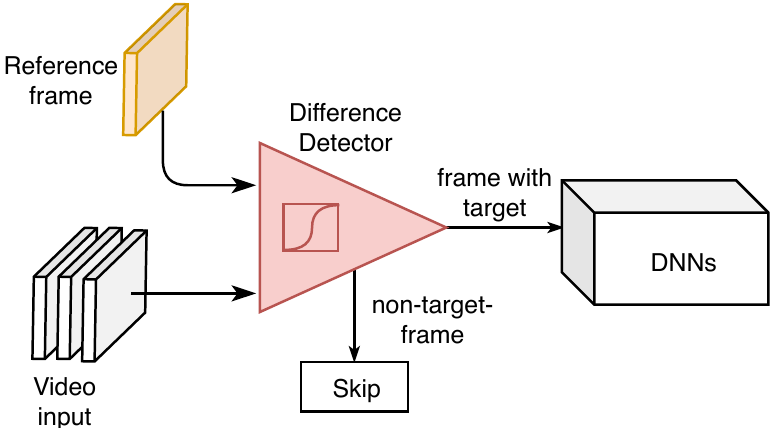}\\
	%\vspace{-10pt}
	\caption{The workflow of input filtering}\label{tech input filtering}
	%\vspace{-5pt}
\end{figure}

Input filtering is an efficient method to accelerate DNN model inference, especially for the video analytics. As shown in Fig. \ref{tech input filtering}, the key idea of input filtering is to remove the non-target-object frames of input data, avoiding redundant computation of DNN model inference, so that improving inference accuracy, shortening inference latency and reducing energy cost.

NoScope \cite{kang2017noscope} is proposed to accelerate video analysis by skipping the frames that have little change. To this end, NoScope implements a difference detector that highlights temporal differences across frames, for example, the detector monitors the frames to check whether cars appear in the frames and the frame with cars will be processed in DNN model inference. The difference is detected by using lightweight binary classifiers. Under a scenario of continuous video transmission from a swarm of drones, Wang et al. \cite{wang2018bandwidth} optimize for the first hop wireless bandwidth of DNN inference. In particular, four strategies are proposed to reduce total transmission: EarlyDiscard, Just-in-Time-Learning (JITL), Reachback and Context-Aware.

%Wang at el. \cite{wang2018bandwidth} optimize for the first hop wireless bandwidth of DNN inference. Under a scenario of continuous video transmission from a swarm of drones, Wang at el. propose four different strategies to reduce total transmission.%: EarlyDiscard, Just-in-Time-Learning (JITL), Reachback and Context-Aware. EarlyDiscard is a content-based filtering method. It uses a trained lightweight DNN like MobileNet (classification task) to filter-out uninteresting frames. But the EarlyDiscard has a low accuracy due to the lightweight classifier, then JITL adds a cascade filter to reduce false-positive frames transmitted by EarlyDiscard. The cascade filter is trained periodically and pushed to the drone. The Reachback is that the edge server can fetch any filtered frames by previous two strategies from the drone on-demand and the Context-Aware performs task-specific optimizations such as detecting pixels with certain colors.

FFS-VA \cite{zhang2018ffs} is a pipelined system for multi-stage video analytic. There are three stages to build the filtering system of FFS-VA. The first is a stream-specialized different detector (SDD) which is used to remove the frames only containing a background. The second is a stream-specialized network model (SNM) to identify target-object frames. And the third is a Tiny-YOLO-Voc (T-YOLO) model to remove the frames whose target objects are fewer than a threshold. Canel et al. \cite{canelpicking} proposes a two-stage filtering system for video analytics. It first extracts the semantic content of the frames by outputting the intermediate data of DNN, then these output features are accumulated in a frame buffer. The buffer is viewed as a directed acyclic graph and the filtering system uses Euclidean distance as the similarity metric to figure out top-k interesting frames.

The above frameworks focus on filtering uninteresting frames of a video stream for a single camera. ReXCam \cite{jain2018rexcam} accelerates DNN model inference on cross-camera analytics. The ReXCam leverages a learned spatiotemporal model to filter video frames. ReXCam reduces computation workload by 4.6x and improves DNN model inference accuracy by 27\%.

\subsubsection{Model Selection}
Model selection method is proposed to optimize the DNN inference issue of latency, accuracy and energy. The main idea of model selection is that we can first train a set of DNN models for the same task with various model size offline, and then adaptively select the model for inference online. Model selection is similar to the model early-exit, and the exit point of model early-exit mechanism can be viewed as a DNN model. But the key difference is that the exit point shares part of DNN layers with the main branch model and the models in the model selection mechanism are independent.

%\begin{figure}[!ht]
%	\centering
%	\includegraphics[scale=0.5]{fig/dnn_inference/modelselection.pdf}\\
	%\vspace{-10pt}
%	\caption{The process of DNN model selection}\label{tech model selection}
	%\vspace{-5pt}
%\end{figure}

Park et al. \cite{park2015big} proposes a big/little DNN model selection framework. That is, a little and fast model is used to try to classify the input data and the big model is only used when the confidence of the little model is less than a predefined threshold. Taylor et al. \cite{taylor2018adaptive} points out that different DNN models (e.g., MobileNet, ResNet, Inception) reach lowest inference latency or highest accuracy on different evaluation metrics (top-1 or top-5) for different images. Then they propose a framework for selecting the best DNN in terms of latency and accuracy. In this framework, a model selector is trained to select the best DNN for different input images. Similarly, IF-CNN \cite{shu2018if} also trains a model selector called recognition predictor (RP) to change the model used in the task. RP is a DNN model of multi-task, meaning that RP has multiple outputs. The output of RP represents the probability of top-1 label of each candidate DNN model. The input of RP is the image and if the output of RP is over the predefined threshold, the corresponding DNN model will be selected.

%Also considering selecting DNN according to the input data, Chameleon \cite{jiang2018chameleon} proposes a periodic update method to change the DNN model used for video analytics. In Chameleon, considering the cross-video correlation, the related video sources (e.g., cameras) are grouped, then the leader of the group will search for top-k best configuration and share the top-k configuration to its followers in the same group. A configuration includes frame resolution, DNN model, etc. %Chameleon achieves 20-50\% higher accuracy with the same amount of resources or achieves the same accuracy with only 30-50\% of the resources. 

Besides the optimization for DNN model inference latency, aiming at energy saving, Stamoulis et al. \cite{stamoulis2018designing} cast the adaptive DNN model selection issue as a hyper-parameter optimization problem by taking into account the accuracy and communication constraints imposed by the devices. Then a Bayesian Optimization (BO) is adopted to solve this problem, achieving an improvement by up to 6x in terms of minimum energy per image under accuracy constraints.

\subsubsection{Support for Multi-Tenancy}
In practice, an end or edge device typically runs more than one
DNN applications concurrently. For example, the advanced driver assistance system (ADAS) for internet vehicles simultaneously runs DNN programs for vehicle detection, pedestrian detection, traffic sign recognition, and lane line detection. In this case, multiple DNN applications would compete for the limited resource. Without careful support for multi-tenancy, i.e., resource allocation and task scheduling for those concurrent applications, the global efficiency would be greatly deteriorated. The support for multi-tenancy focuses on the optimization of energy and memory footprint.

%The support for multi-tenancy is to enable efficient scheduling for multi-user or multi-task edge intelligence application scenarios. 
%As shown in Fig. \ref{tech multi tenancy}, 
%Mobile device usually runs multiple DNN application concurrently and the available resource of the device is dynamic at runtime.

Taking the dynamics of runtime resources into consideration, NestDNN \cite{fang2018nestdnn} is proposed to offering flexible resource-accuracy trade-offs for each DNN model. 
%NestDNN addresses two challenges: one is how to enable a DNN model to provide flexible resources-accuracy trade-offs, and the other is how to make a resource-accuracy trade-off for each concurrent DNN model. 
NestDNN implements a new model pruning and recovery scheme, transforming the DNN model into a single compact multi-capacity model which consists of a set of descendent models. Each descendent model offers a unique resource-accuracy trade-off. For each concurrent descendent model, NestDNN encodes its accuracy and latency into a cost function, then NestDNN builds a resource-accuracy runtime scheduler to make the optimal trade-off for each concurrent descendent model. %NestDNN achieves an improvement of 4.2\% in inference accuracy, 2x in video frame processing rate and a reduction of 1.7x in energy consumption. 
Also addressing the challenge of enabling flexible trade-offs, Mainstream \cite{jiang2018mainstream} uses the popular transfer-learning DNN training method to train multiple DNN models with different degrees of accuracy and implements a greedy approach to find the optimal scheduler that fits the cost budget. For multiple DNN model executions on one single device, HiveMind \cite{narayananaccelerating} is proposed to improve the GPU utilization for these concurrent workloads. HiveMind consists of two key components: a compiler and a runtime module. The compiler optimizes the data transmission, data preprocessing and computation across the workloads, and then the runtime module transforms the optimized models into an execution DAG, which will be executed on the GPU while trying to extract as much concurrency as possible.

At a finer granularity, DeepEye \cite{mathur2017deepeye} is proposed to optimizing the inference of multi-task on the mobile device by scheduling the executions of heterogeneous DNN layers. DeepEye first segregates DNN layers of all task into two pools: convolution layers and fully-connected layers. For the convolution layers, a FIFO queue based execution strategy is employed. For the fully-connected layers, DeepEye adopts a greedy approach for caching the parameters of the fully-connected layers to maximize memory utilization.

\subsubsection{Application-Specific Optimization}
While the above optimizing techniques are generally applicable to EI applications, application-specific optimization can be exploited to further optimize the performance of EI applications, i.e., accuracy, latency, energy and memory footprint. For example, for video-based applications, two knobs, i.e., frame rate and resolution can be flexibly adjusted to reduce resource demand. However, since such resource-sensitive knobs also deteriorate the inference accuracy, they naturally incur a cost-accuracy tradeoff. This requires us to strike a nice balance between the resource cost and inference accuracy when tuning the video frame rate and resolution.

Towards the above goal, Chameleon \cite{jiang2018chameleon} adjusts the knobs for each video analytic task by sharing the best top-k configuration between each task. In Chameleon, the video tasks are grouped according to the spatial correlation, then leader of the group will search for the best top-k configurations and share them with the followers. DeepDecision \cite{ran2018deepdecision} formulates the knob-tunning problem as a multiple-choice multiple-constraint knapsack program and solves it with an improved brute-force search method. %VideoStorm \cite{zhang2017live}

It is also worth noting that, in the computer architecture community, hardware acceleration for efficient DNN inference has been a very hot topic and gathered extensive research efforts. Interested readers are encouraged to refer to the recent monograph \cite{sze2017efficient} for more comprehensive discussions about recent advancements on hardware acceleration for DNN processing.

\subsection{Summary of Existing Systems and Frameworks}
To showcase the application of the above enabling techniques for edge intelligence model inference, the relevant systems and frameworks are summarized in Table \ref{framework summary}, including the perspectives of target applications, architecture and EI level, optimization objectives and adopted techniques, as well as effectiveness.

%For a better understanding of the above systems and frameworks, we now briefly introduce some representative systems and frameworks as follows.

%\textbf{Neurosurgeon} \cite{kang2017neurosurgeon} explores a model partition idea that automatically partitioning DNN computation between mobile devices and cloud data center at the granularity of DNN layers. The best partition point for a DNN model depends on its topology and the network condition between mobile devices and the cloud data center. Fig. \ref{system neurosurgeon} shows the workflow of Neurosurgeon, which includes two stages: deployment and runtime. At the deployment stage, Neurosurgeon profiles the mobile device and the server to generate performance (e.g., latency or energy) prediction regression models for different types of DNN layers. The profiling is done only once \emph{offline}, given specific mobile devices and the server. Then these regression models are stored on mobile devices and the server, respectively, and will be used in runtime stage for the prediction of DNN layer latency or energy consumption. At the runtime stage, Neurosurgeon extracts configuration (e.g., layer type) of DNN model, then uses stored regression models to estimate the latency and energy consumption for executing each layer on the mobile device and the server, and calculates the data transmission latency based on the current bandwidth condition. By doing so, Neurosurgeon selects the best partition point through iteration.

Clearly, existing systems and frameworks have adopted different subsets of enabling techniques tailored to specific edge intelligence applications and requirements. To maximize the overall performance of a generic edge intelligence system, comprehensive enabling techniques and various optimization methods should work in a cooperative manner to provide rich design flexibility. Nevertheless, we would face a high dimensional configuration problem that is required to determine a large number of performance-critical configuration parameters in real time. Taking video analytics, for example, the high dimensional configuration parameters can include video frame rate, resolution, model selection and model early-exit, etc. Due to the combinatorial nature, high dimensional configuration problem involves a huge search space of parameters and is very challenging to tackle.

\begin{table*}[]
	\tiny
	\caption{An Overview of Systems and Frameworks on EI Model Infernce} \label{framework summary}

		\centering
		\begin{tabular}{c|c|c|c|c|c|c|c}
			\hline
			\textbf{System or Framework} & \textbf{Application} & \textbf{Architecture} & \textbf{EI Level} & \textbf{Objectives} & \textbf{Optimization Technology} & \textbf{Online/Offline} & \textbf{Effectiveness}\\ \hline
			\hline
			VideoEdge \cite{hung2018videoedge}  &        Video Analytics         &        Cloud-Edge-Device          &       Level-1         &        \multicolumn{1}{l|}{\begin{minipage}{0.7in}
					\vskip 4pt
					\begin{itemize}
						\item Accuracy
						\item Resource cost
					\end{itemize}
					\vskip 4pt
			\end{minipage}  }         &        \multicolumn{1}{l|}{\begin{minipage}{0.9in}
			\vskip 4pt
			\begin{itemize}
				\item Frame rate adaptation
				\item Resolution adaptation
				\item Multi-tenancy
				\item Service placement
			\end{itemize}
			\vskip 4pt
		\end{minipage}  }          &            Online               &             \multicolumn{1}{l}{\begin{minipage}{1in}
		\vskip 4pt
		\begin{itemize}
			\item Accuracy improvement: 5.4-25.4$\times$
		\end{itemize}
		\vskip 4pt
	\end{minipage}  }                    \\ \hline
		
			Chameleon \cite{jiang2018chameleon}        &        Video Analytics         &         Device-cloud         &        Level-1        &         \multicolumn{1}{l|}{\begin{minipage}{0.7in}
					\vskip 4pt
					\begin{itemize}
						\item Accuracy
						\item Resource cost
					\end{itemize}
					\vskip 4pt
			\end{minipage}  }        &       \multicolumn{1}{l|}{\begin{minipage}{0.9in}
			\vskip 4pt
			\begin{itemize}
				\item Frame rate adaptation
				\item Resolution adaptation
				\item Model selection
			\end{itemize}
			\vskip 4pt
		\end{minipage}  }           &              Online             &                \multicolumn{1}{l}{\begin{minipage}{1in}
		\vskip 4pt
		\begin{itemize}
			\item Resource reduction: 2-3$\times$
		\end{itemize}
		\vskip 4pt
	\end{minipage}  }                 \\ \hline
	
			DeepX  \cite{lane2016deepx}      &        Mobile Sensing Apps         &        On Device          &        Level-2        &        \multicolumn{1}{l|}{\begin{minipage}{0.7in}
					\vskip 4pt
					\begin{itemize}
						\item Accuracy
						\item Energy
					\end{itemize}
					\vskip 4pt
			\end{minipage}  }        &         \multicolumn{1}{l|}{\begin{minipage}{0.9in}
			\vskip 4pt
			\begin{itemize}
				\item Model compression
				\item Model partition
			\end{itemize}
			\vskip 4pt
		\end{minipage}  }         &             Online              &              \multicolumn{1}{l}{\begin{minipage}{1in}
		\vskip 4pt
		\begin{itemize}
			\item Energy reduction: 7.12-26.7$\times$
		\end{itemize}
		\vskip 4pt
	\end{minipage}  }                   \\ \hline
	
	Edgent  \cite{Li2018edgeintelligence}      &        N/A         &        Device-Edge          &        Level-2        &        \multicolumn{1}{l|}{\begin{minipage}{0.7in}
					\vskip 4pt
					\begin{itemize}
						\item Accuracy
						\item Latency
					\end{itemize}
					\vskip 4pt
			\end{minipage}  }        &         \multicolumn{1}{l|}{\begin{minipage}{0.9in}
			\vskip 4pt
			\begin{itemize}
				\item Early-exit
				\item Model partition
			\end{itemize}
			\vskip 4pt
		\end{minipage}  }         &             Offline              &              \multicolumn{1}{l}{\begin{minipage}{1in}
		\vskip 4pt
		\begin{itemize}
			\item Accuracy improvement
		\end{itemize}
		\vskip 4pt
	\end{minipage}  }                   \\ \hline
	
			AdaDeep \cite{liu2018demand}       &        N/A         &         On Device         &        Level-3        &       \multicolumn{1}{l|}{\begin{minipage}{0.7in}
					\vskip 4pt
					\begin{itemize}
						\item Accuracy
						\item Energy
						\item Storage
					\end{itemize}
					\vskip 4pt
			\end{minipage}  }           &        \multicolumn{1}{l|}{\begin{minipage}{0.9in}
			\vskip 4pt
			\begin{itemize}
				\item Model compression
			\end{itemize}
			\vskip 4pt
		\end{minipage}  }          &            Online               &             \multicolumn{1}{l}{\begin{minipage}{1in}
		\vskip 4pt
		\begin{itemize}
			\item Latency reduction: 9.8$\times$
			\item Energy reduction: 4.3$\times$
			\item Storage reduction: 38$\times$
		\end{itemize}
		\vskip 4pt
	\end{minipage}  }                    \\ \hline
	
			DeepIns \cite{li2018deep}      &         IIoT        &        Edge-Cloud          &         Level-1       &       \multicolumn{1}{l|}{\begin{minipage}{0.7in}
					\vskip 4pt
					\begin{itemize}
						\item Accuracy
						\item Latency
					\end{itemize}
					\vskip 4pt
			\end{minipage}  }          &        \multicolumn{1}{l|}{\begin{minipage}{0.9in}
			\vskip 4pt
			\begin{itemize}
				\item Early-exit
			\end{itemize}
			\vskip 4pt
		\end{minipage}  }          &             Offline              &               \multicolumn{1}{l}{\begin{minipage}{1in}
		\vskip 4pt
		\begin{itemize}
			\item Latency reduction: 0.98-1.21x
		\end{itemize}
		\vskip 4pt
	\end{minipage}  }                   \\ \hline
	
			Neurosurgeon \cite{kang2017neurosurgeon} &        N/A         &          Device-Cloud        &       Level-1         &        \multicolumn{1}{l|}{\begin{minipage}{0.7in}
					\vskip 4pt
					\begin{itemize}
						\item Latency
						\item Energy
					\end{itemize}
					\vskip 4pt
			\end{minipage}  }          &        \multicolumn{1}{l|}{\begin{minipage}{0.9in}
			\vskip 4pt
			\begin{itemize}
				\item Model partition
			\end{itemize}
			\vskip 4pt
		\end{minipage}  }          &             N/A              &                  \multicolumn{1}{l}{\begin{minipage}{1in}
		\vskip 4pt
		\begin{itemize}
			\item Latency reduction: 3.1-40.7$\times$
			\item Energy reduction: 59.5\%-94.7\%
		\end{itemize}
		\vskip 4pt
	\end{minipage}  }                \\ \hline
	
	Minerva  \cite{reagen2016minerva}   &        N/A         &        On Device         &       Level-3         &        \multicolumn{1}{l|}{\begin{minipage}{0.7in}
					\vskip 4pt
					\begin{itemize}
						\item Energy
					\end{itemize}
					\vskip 4pt
			\end{minipage}  }          &        \multicolumn{1}{l|}{\begin{minipage}{0.9in}
			\vskip 4pt
			\begin{itemize}
				\item Hardware Acceleration
				\item Model Compression
			\end{itemize}
			\vskip 4pt
		\end{minipage}  }          &             Offline              &               \multicolumn{1}{l}{\begin{minipage}{1in}
		\vskip 4pt
		\begin{itemize}
			\item Energy saving: 8$\times$
		\end{itemize}
		\vskip 4pt
	\end{minipage}  }                   \\ \hline
	
%			OpenVDAP \cite{zhang2018openvdap} &        Connected  Vehicles        &        Cloud-Edge-Device          &       Level-1         &        \multicolumn{1}{l|}{\begin{minipage}{0.7in}
%					\vskip 4pt
%					\begin{itemize}
%						\item Privacy
%						\item Latency
%						\item Resource cost
%					\end{itemize}
%					\vskip 4pt
%			\end{minipage}  }         &       \multicolumn{1}{l|}{\begin{minipage}{0.9in}
%			\vskip 4pt
%			\begin{itemize}
%				\item Model compression
%				\item Multi-tenancy
%			\end{itemize}
%			\vskip 4pt
%		\end{minipage}  }           &              N/A             &        %      \multicolumn{1}{l}{\begin{minipage}{1in}
%		\vskip 4pt
%		\begin{itemize}
%			\item N/A
%		\end{itemize}
%		\vskip 4pt
%	\end{minipage}  }                    \\ \hline
	
			FoggyCache \cite{guo2018foggycache}    &        IIoT         &        Device-Edge          &       Level-2         &        \multicolumn{1}{l|}{\begin{minipage}{0.7in}
					\vskip 4pt
					\begin{itemize}
						\item Accuracy
						\item Latency
					\end{itemize}
					\vskip 4pt
			\end{minipage}  }          &          \multicolumn{1}{l|}{\begin{minipage}{0.9in}
			\vskip 4pt
			\begin{itemize}
				\item Edge Caching
			\end{itemize}
			\vskip 4pt
		\end{minipage}  }        &             Online              &               \multicolumn{1}{l}{\begin{minipage}{1in}
		\vskip 4pt
		\begin{itemize}
			\item Latency reduction: 3-10$\times$x
			\item Energy reduction: 3-10$\times$
		\end{itemize}
		\vskip 4pt
	\end{minipage}  }                   \\ \hline
	
			NoScope  \cite{kang2017noscope}   &         Video Analytics        &        Cloud          &        Cloud Intelligence        &        \multicolumn{1}{l|}{\begin{minipage}{0.7in}
					\vskip 4pt
					\begin{itemize}
						\item Latency
					\end{itemize}
					\vskip 4pt
			\end{minipage}  }          &         \multicolumn{1}{l|}{\begin{minipage}{0.9in}
			\vskip 4pt
			\begin{itemize}
				\item Input filtering
			\end{itemize}
			\vskip 4pt
		\end{minipage}  }         &            N/A               &                 \multicolumn{1}{l}{\begin{minipage}{1in}
		\vskip 4pt
		\begin{itemize}
			\item Latency reduction: 265-15500$\times$
		\end{itemize}
		\vskip 4pt
	\end{minipage}  }                 \\ \hline
	
			JALAD  \cite{li2018jalad}   &        N/A         &         Device-Cloud         &       Level-1         &        \multicolumn{1}{l|}{\begin{minipage}{0.7in}
					\vskip 4pt
					\begin{itemize}
						\item Latency
					\end{itemize}
					\vskip 4pt
			\end{minipage}  }          &        \multicolumn{1}{l|}{\begin{minipage}{0.9in}
			\vskip 4pt
			\begin{itemize}
				\item Model compression
				\item Model partition
			\end{itemize}
			\vskip 4pt
		\end{minipage}  }          &             Offline              &               \multicolumn{1}{l}{\begin{minipage}{1in}
		\vskip 4pt
		\begin{itemize}
			\item Latency reduction: 1-25.1$\times$
		\end{itemize}
		\vskip 4pt
	\end{minipage}  }                   \\ \hline
	
	DDNNs  \cite{teerapittayanon2017distributed}   &        N/A         &         Cloud-Edge-Device         &       Level-1         &        \multicolumn{1}{l|}{\begin{minipage}{0.7in}
					\vskip 4pt
					\begin{itemize}
						\item Latency
						\item Accuracy
					\end{itemize}
					\vskip 4pt
			\end{minipage}  }          &        \multicolumn{1}{l|}{\begin{minipage}{0.9in}
			\vskip 4pt
			\begin{itemize}
				\item Model selection
			\end{itemize}
			\vskip 4pt
		\end{minipage}  }          &             N/A              &               \multicolumn{1}{l}{\begin{minipage}{1in}
		\vskip 4pt
		\begin{itemize}
			\item Latency reduction: over 20$\times$
		\end{itemize}
		\vskip 4pt
	\end{minipage}  }                   \\ \hline
	
	FFS-VA   \cite{zhang2018ffs}  &        Video Analytics         &         On Device         &       Level-3         &        \multicolumn{1}{l|}{\begin{minipage}{0.7in}
					\vskip 4pt
					\begin{itemize}
						\item Latency
					\end{itemize}
					\vskip 4pt
			\end{minipage}  }          &        \multicolumn{1}{l|}{\begin{minipage}{0.7in}
			\vskip 4pt
			\begin{itemize}
				\item Input filtering
				\item Multi-tenancy
			\end{itemize}
			\vskip 4pt
		\end{minipage}  }          &             N/A              &               \multicolumn{1}{l}{\begin{minipage}{1in}
		\vskip 4pt
		\begin{itemize}
			\item Latency reduction: 3$\times$
			\item Throughput improvement: more than 7$\times$
		\end{itemize}
		\vskip 4pt
	\end{minipage}  }                   \\ \hline
	
	Cachier  \cite{drolia2017cachier}   &        N/A         &         Cloud-Edge         &       Level-1         &        \multicolumn{1}{l|}{\begin{minipage}{0.7in}
					\vskip 4pt
					\begin{itemize}
						\item Throughput
					\end{itemize}
					\vskip 4pt
			\end{minipage}  }          &        \multicolumn{1}{l|}{\begin{minipage}{0.9in}
			\vskip 4pt
			\begin{itemize}
				\item Edge Caching
			\end{itemize}
			\vskip 4pt
		\end{minipage}  }          &             N/A              &               \multicolumn{1}{l}{\begin{minipage}{1in}
		\vskip 4pt
		\begin{itemize}
			\item Throughput improvement: more than 3$\times$
		\end{itemize}
		\vskip 4pt
	\end{minipage}  }                   \\ \hline
	
	Taylor et al.  \cite{taylor2018adaptive}   &        N/A         &         On Device         &       Level-3         &        \multicolumn{1}{l|}{\begin{minipage}{0.7in}
					\vskip 4pt
					\begin{itemize}
						\item Accuracy
						\item Latency
					\end{itemize}
					\vskip 4pt
			\end{minipage}  }          &        \multicolumn{1}{l|}{\begin{minipage}{0.9in}
			\vskip 4pt
			\begin{itemize}
				\item Input filtering
				\item Model selection
			\end{itemize}
			\vskip 4pt
		\end{minipage}  }          &             N/A              &               \multicolumn{1}{l}{\begin{minipage}{1in}
		\vskip 4pt
		\begin{itemize}
			\item Accuracy improvement: 7.52\%
			\item Latency reduction: 1.8$\times$
		\end{itemize}
		\vskip 4pt
	\end{minipage}  }                   \\ \hline
	
	DeepDecision  \cite{ran2018deepdecision}   &        Video Analytics         &         Cloud-Edge        &       Level-1         &        \multicolumn{1}{l|}{\begin{minipage}{0.7in}
					\vskip 4pt
					\begin{itemize}
						\item Accuracy
						\item Latency
						\item Energy
					\end{itemize}
					\vskip 4pt
			\end{minipage}  }          &        \multicolumn{1}{l|}{\begin{minipage}{0.9in}
			\vskip 4pt
			\begin{itemize}
				\item Application-level optimization
				\item Model selection
			\end{itemize}
			\vskip 4pt
		\end{minipage}  }          &             N/A              &               \multicolumn{1}{l}{\begin{minipage}{1in}
		\vskip 4pt
		\begin{itemize}
			\item Latency reduction: 2-10$\times$
		\end{itemize}
		\vskip 4pt
	\end{minipage}  }                   \\ \hline
	
		\end{tabular}
	
\end{table*}
\section{Future Research Directions}
Based on the comprehensive discussions above on existing efforts, we now articulate the key open challenges and future research directions for edge intelligence (EI).

\subsection{Programming and Software Platforms}
Currently many companies around the world focus on the AI cloud computing service provisioning. Some leading companies are also starting to provide programming/software platforms to deliver edge computing services, such as Amazon's Greengrass, Microsoft's Azure IoT Edge and Google's Cloud IoT Edge. Nevertheless, currently, most of these platforms mainly serve as relays for connecting to the powerful cloud data centers. 

As more and more AI-driven computation-intensive mobile and IoT applications are emerging, edge intelligence as a service (EIaaS) can become a pervasive paradigm and EI platforms with powerful edge AI functionalities will be developed and deployed. This is substantially different from machine learning as a service (MLaaS) provided by public clouds. Essentially, MLaaS belongs to cloud intelligence and it focuses on selecting the proper server configuration and machine learning framework to train model in the cloud in a cost-efficient manner. While in a sharp contrast, EIaaS concerns more about how to perform model training and inference in resource-constrained and privacy-sensitive edge computing environments. To fully realize the potential of EI services, there are several key challenges to overcome. First of all, the EI platforms should be heterogeneity-compatible. In the future, there are many dispersive EI service providers/vendors, and the common open standard should be set such that users can enjoy seamless and smooth services across heterogeneous EI platforms anywhere and anytime. Second, there are many AI programming frameworks available (e.g., Tensorflow, Torch and Caffe). In the future, the portability of the edge AI models trained by different programming frameworks across heterogeneously distributed edge nodes should be supported. Third, there are many programming frameworks designed specifically for edge devices (e.g., TensorFlow Lite, Caffe2, CoreML and MXNet), however, empirical measurements \cite{zhang2018pcamp} show that there is no single winner that can outperform other frameworks in all metrics. A framework that performs efficiently on more metrics can be expected in the future. Last but not least, lightweight virtualization and computing techniques such as container and function computing should be further explored to enable efficient EI service placement and migration over resource-constrained edge environments.

\subsection{Resource-friendly Edge AI Model Design}
Many existing AI models such as CNN and LSTM were originally designed for applications such as computer vision and natural language processing. Most of deep learning based AI models are highly resource-intensive, which means that powerful computing capability supported by abundant hardware resources (e.g., GPU, FPGA, TPU) is an important boost the performance of these AI models. Therefore, as mentioned above there are many studies to exploit model compression techniques (e.g., weight pruning) to resize the AI models, making them more resource-friendly for edge deployment. 

Along with a different line, we can promote a resource-aware edge AI model design. Instead of utilizing the existing resource-intensive AI models, we can leverage the AutoML idea \cite{he2018amc} and the Neural Architecture Search (NAS) techniques \cite{zoph2016neural} to devise resource-efficient edge AI models tailored to the hardware resource constraints of the underlying edge devices and servers. For example, methods such as reinforcement learning, genetic algorithm, and Bayesian optimization can be adopted to efficiently search over the AI model design parameter space (i.e., AI model components and their connections) by taking into account the impact of hardware resource (e.g., CPU, memory) constraints on the performance metrics such as execution latency and energy overhead. 

\subsection{Computation-aware Networking Techniques}  
For EI, computation-intensive AI-based applications are typically run on the distributed edge computing environment. As a result, advanced networking solutions with computation awareness is highly desirable, such that the computation results and data can be efficiently shared across different edge nodes. 

For the future 5G networks, the Ultra-Reliable Low-Latency Communication (URLLC) has been defined for mission-critical application scenarios that demand low delay and high reliability. Therefore, it will be promising to integrate the 5G URLLC capability with edge computing to provide Ultra-Reliable Low-Latency EI (URLL-EI) services. Also, advanced techniques such as software-defined network and network function virtualization will be adopted in 5G. These techniques will enable flexible control over the network resources for supporting on-demand interconnections across different edge nodes for computation-intensive AI applications. 

On the other hand, autonomous networking mechanism design is important to achieve efficient EI service provisioning under dynamic heterogeneous network coexistence (e.g., LTE/5G/WiFi/LoRa), allowing newly added edge nodes and devices to self-configure in the plug and play manner.  Also, the computation-aware communication techniques are starting to draw attention, such as gradient coding \cite{tandon2017gradient} to mitigate straggler effect in distributed learning and over-the-air computation for distributed stochastic gradient descent \cite{zhu2018low}, which can be useful for edge AI model training acceleration.

\subsection{Trade-off Design with Various DNN Performance Metrics}
For an edge intelligence application with a specific mission, there is usually a series of DNN model candidates that are capable of finishing the task. However, it is difficult for software developers to choose an appropriate DNN model for the EI application because the standard performance indicators such as top-k accuracy or mean average precision fail to reflect the runtime performance of DNN model inference on edge devices. For instance, during the EI application deployment phase, beside accuracy, inference speed and resource usage are also critical metrics. We need to explore the trade-offs between these metrics and identify the factors that affect them. 

In the effort \cite{Huang_2017_CVPR}, for the object recognition application, the authors investigate the influence of the main factors, e.g., number of proposals, input image size and the selection of feature extractor, on inference speed and accuracy. Based on their experiment results, a new combination of these factors is found to outperform the state-of-the-art method. Therefore, it is necessary to explore the trade-offs between different metrics, helping the improve the efficiency of deploying EI application.

\subsection{Smart Service and Resource Management}
By the distributed nature of edge computing, edge devices and nodes that offer EI functionality are dispersive across diverse geo-locations and regions. Different edge devices and nodes may run different AI models and deploy different specific AI tasks. Therefore, it is important to design efficient service discovery protocols such that users can identify and locate the relevant EI service providers to fulfill their need in a timely manner. Also, to fully exploit the dispersive resource across edge nodes and devices, the partition of complex edge AI models into small subtasks and efficiently offloading these tasks among the edge nodes and devices for collaborative executions are essential. 

%On one hand, we need to take into account both communication and computation resource constraints on model partition and offloading. On the other hand, we should also consider the different characteristics between model training and inference tasks. For instance, model training generally involves forward and backward message passing for gradient updates and is more delay-tolerant. How to jointly optimize both EI model training and inference task offloadings will be an important and challenging topic. 

Since for many EI application scenarios (e.g., smart cities), the service environments are of high dynamics and it is hard to accurately predict future events. As a result, it would require the outstanding capability of online edge resource orchestration and provisioning to continuously accommodate massive EI tasks. Real-time joint optimization of heterogeneous computation, communication, and cache resource allocations and the high dimensional system parameter configuration (e.g., choosing the proper model training and inference techniques) tailored to diverse task demands is critical. To tackle the algorithm design complexity, an emerging research direction is to leverage the AI techniques such as deep reinforcement learning to adapt efficient resource allocation policy in a data-driven self-learning way.

\subsection{Security and Privacy Issues} 
The open nature of edge computing imposes that the decentralized trust is required such that the EI services provided by different entities are trustworthy \cite{DBLP:conf/edge/LiZLX18}. Thus, lightweight and distributed security mechanism designs are critical to ensure user authentication and access control, model and data integrity, and mutual platform verification for EI. Also, it is important to study novel secure routing schemes and trust network topologies for EI service delivery when considering the coexistence of trusted edge nodes with malicious ones.

On the other hand, the end users and devices would generate a massive volume of data at the network edge, and these data can be privacy sensitive since they may contain users’ location data, health or activities records, or manufacturing information, among many others. Subject to the privacy protection requirement, e.g., EU's General Data Protection Regulation (GDPR), directly sharing the original datasets among multiple edge nodes can have a high risk of privacy leakage. Thus, federated learning can be a feasible paradigm for privacy-friendly distributed data training such that the original datasets are kept in their generated devices/nodes and the edge AI model parameters are shared. To further enhance the data privacy, more and more research efforts are devoted to utilizing the tools of differential privacy, homomorphic encryption and secure multi-party computation for designing privacy-preserving AI model parameter sharing schemes \cite{du2018big}.

\subsection{Incentive and Business Models}
EI ecosystem will be a grand open consortium that consists of EI service providers and users, which can include but not limited to: platform providers (e.g., Amazon), AI software providers (e.g., SenseTime), edge device providers (e.g., Hikvision), network operators (e.g., AT\&T), data generators (e.g., IoT and mobile device owners), and service consumers (i.e., EI users). The high-efficiency operation of EI services may require close collaboration and integration across different service providers, e.g., for implementing expanded resource sharing and smooth service handover. Thus, proper incentive mechanism and business model are essential for stimulate effective and efficient cooperation among all members of EI ecosystem. Also, for EI service, a user can be a service consumer and meanwhile a data generator as well. In this case, a novel smart pricing scheme is needed to factorize user’s service consumption and the value of its data contribution. 

As a means for decentralized collaboration, blockchain with a smart contract may be integrated into EI service by running on decentralized edge servers. It is worthwhile to do research on how to smartly charge the price and properly distribute the revenue among the members in the EI ecosystem according to their proof of work. Also, designing resource-friendly lightweight blockchain consensus protocol for edge intelligence is highly desirable.

\section{Concluding Remarks}
Driving by the flourishing of both AI and IoT, there is a stringent need to pushing the AI frontier from the cloud to the network edge. To fulfill this trend, edge computing has been widely recognized as a promising solution to support computation-intensive AI applications in resource-constrained environments. The nexus between edge computing and AI gives birth to the novel paradigm of edge intelligence. 

In this paper, we conduct a comprehensive survey of the recent research efforts %as well as industrial progresses 
on edge intelligence. Specifically, we first review the background and motivation for artificial intelligence running at the network edge. We then provide an overview of the overarching architectures, frameworks and emerging key technologies for deep learning model towards training and inference at the network edge. Finally, we discuss the open challenges and future research directions on edge intelligence. We hope this survey is able to elicit escalating attentions, stimulate fruitful discussions and inspire further research ideas on edge intelligence.

%\Theta

\bibliographystyle{IEEEtran}
\bibliography{ref}
\vspace{-10pt}
\begin{IEEEbiography}[{\includegraphics[width=1in,height=1.25in,clip,keepaspectratio]{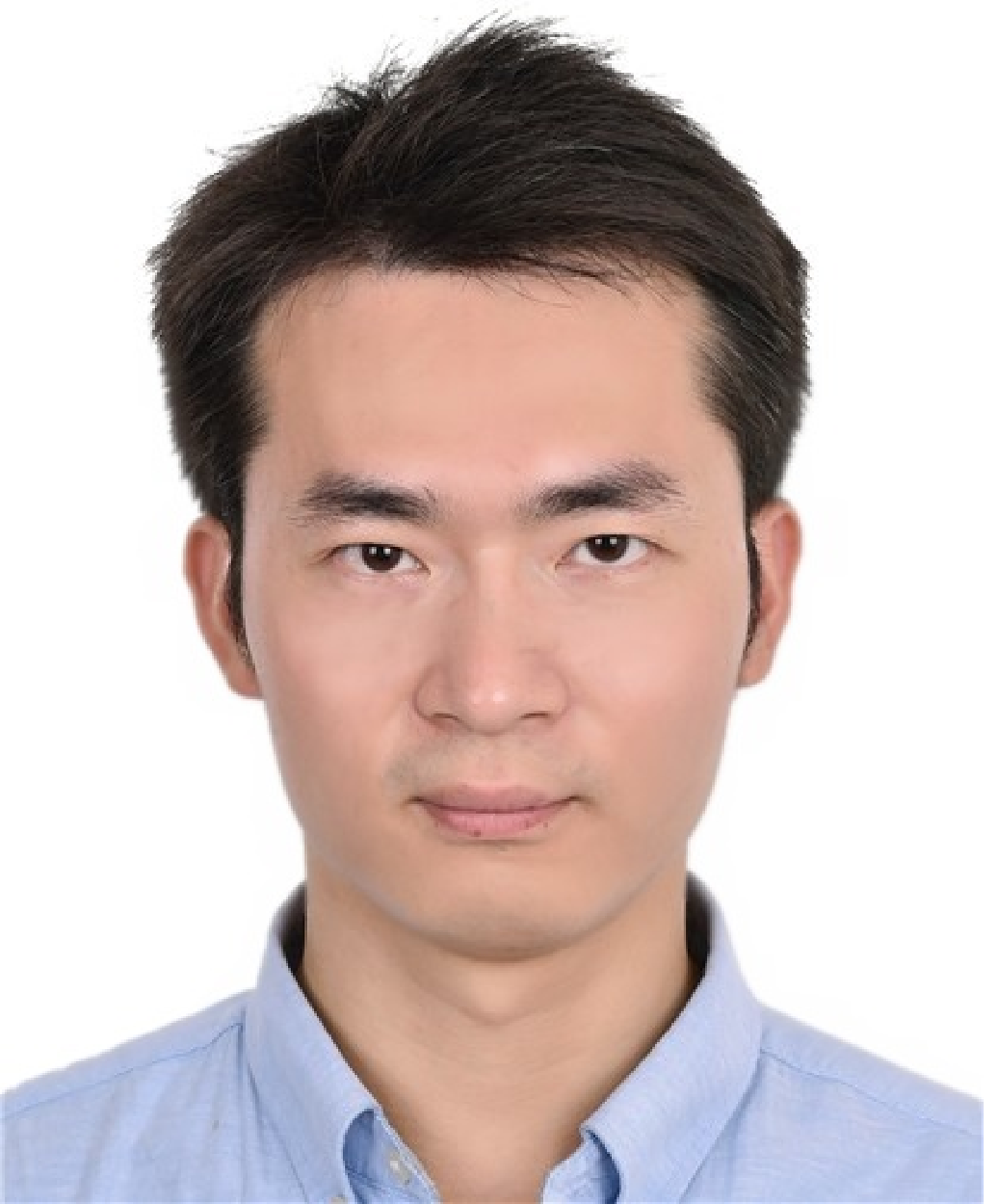}}]{Zhi Zhou}
received the B.S., M.E. and Ph.D. degrees in 2012, 2014 and 2017, respectively, all from the School of Computer Science and Technology, Huazhong University of Science and Technology (HUST), Wuhan, China. He is currently a research fellow in School of Data and Computer Science, Sun Yat-sen University, Guangzhou, China. In 2016, he has been a Visiting Scholar at University of Gottingen. He was the sole recipient of 2018 ACM Wuhan \& Hubei Computer Society Doctoral Dissertation Award, a recipient of the Best Paper Award of IEEE UIC 2018, and a general co-chair of 2018 International Workshop on Intelligent Cloud Computing and Networking (ICCN). His research interests include edge computing, cloud computing and distributed systems.
\end{IEEEbiography}

\vspace{-5pt}

\begin{IEEEbiography}[{\includegraphics[width=1in,height=1.25in,clip,keepaspectratio]{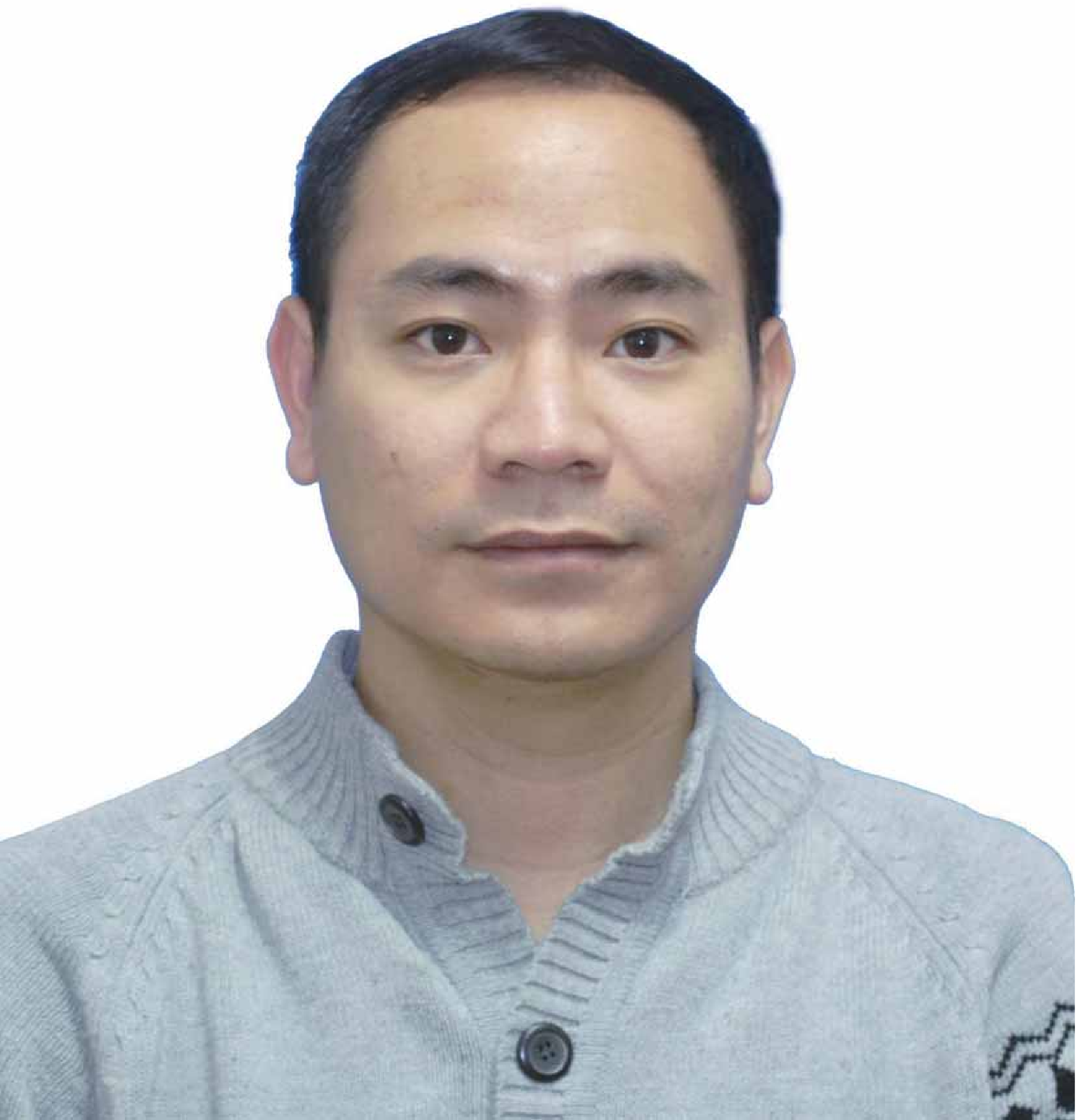}}]{Xu Chen}
is a Full Professor with Sun Yat-sen University, Guangzhou, China, and the vice director of National and Local Joint Engineering Laboratory of Digital Home Interactive Applications. He received the Ph.D. degree in information engineering from the Chinese University of Hong Kong in 2012, and worked as a Postdoctoral Research Associate at Arizona State University, Tempe, USA from 2012 to 2014, and a Humboldt Scholar Fellow at Institute of Computer Science of University of Goettingen, Germany from 2014 to 2016. He received the prestigious Humboldt research fellowship awarded by Alexander von Humboldt Foundation of Germany, 2014 Hong Kong Young Scientist Runner-up Award, 2016 Thousand Talents Plan Award for Young Professionals of China, 2017 IEEE Communication Society Asia-Pacific Outstanding Young Researcher Award, 2017 IEEE ComSoc Young Professional Best Paper Award, Honorable Mention Award of 2010 IEEE international conference on Intelligence and Security Informatics (ISI), Best Paper Runner-up Award of 2014 IEEE International Conference on Computer Communications (INFOCOM), and Best Paper Award of 2017 IEEE International Conference on Communications (ICC). He is currently an Associate Editor of IEEE Internet of Things Journal and IEEE Journal on Selected Areas in Communications (JSAC) Series on Network Softwarization and Enablers.
 \end{IEEEbiography}

\vspace{-5pt}

\begin{IEEEbiography}[{\includegraphics[width=1in,height=1.25in,clip,keepaspectratio]{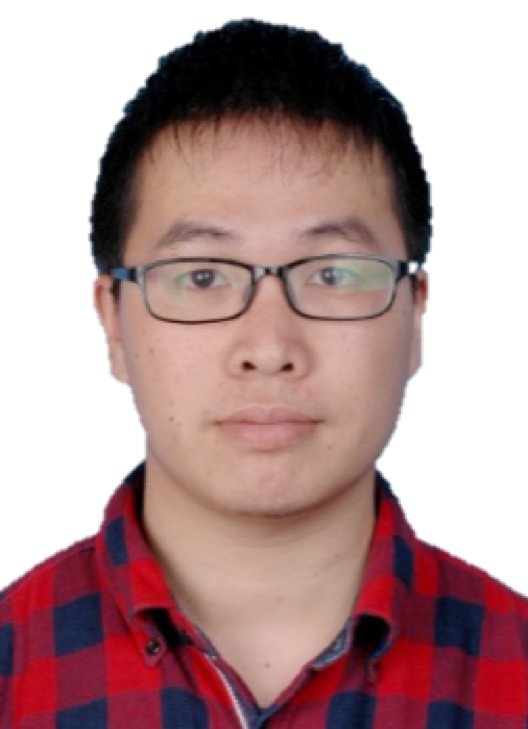}}]{En Li} received the B.S. degree in communication engineering from the School of Physics \& Telecommunication Engineering, South China Normal University (SCNU), Guangzhou, China in 2017. He is currently pursuing the master’s degree with the School of Data and Computer Science, Sun Yat-sen University, Guangzhou, China. His research interests include mobile deep computing, edge intelligence, deep learning.
\end{IEEEbiography}

\vspace{-5pt}

\begin{IEEEbiography}[{\includegraphics[width=1in,height=1.25in,clip,keepaspectratio]{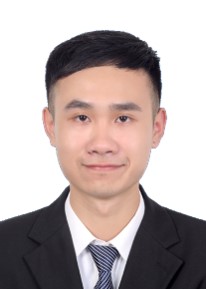}}]{Liekang Zeng} received the B.S. degree in computer science from the School of Data and Computer Science, Sun Yat-sen University (SYSU), Guangzhou, China in 2018. He is currently pursuing the master’s degree with the School of Data and Computer Science, Sun Yat-sen University, Guangzhou, China. His research interests include mobile edge computing, deep learning, distributed computing.
\end{IEEEbiography}

\vspace{-5pt}

\begin{IEEEbiography}[{\includegraphics[width=1in,height=1.25in,clip,keepaspectratio]{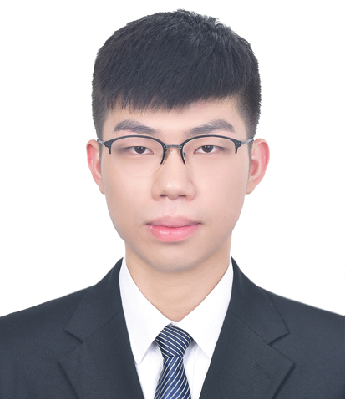}}]{Ke Luo}
received the B.S. degree in computer science from the School of Data and Computer Science, Sun Yat-sen University (SYSU), Guangzhou, China in 2018. He is currently working towards the Ph.D. degree in the School of Data and Computer Science, SYSU. His primary research interests include cloud computing, mobile edge computing, and distributed systems.
\end{IEEEbiography}
 
\vspace{-5pt}

\begin{IEEEbiography}[{\includegraphics[width=1in,height=1.25in,clip,keepaspectratio]{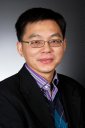}}]{Junshan Zhang}
received the Ph.D. degree from the School of ECE at Purdue University in 2000. He joined the School of ECEE, Arizona State University in August 2000, where he has been the Fulton Chair Professor since 2015. His research interests are in the general field of information networks and data science, including communication networks, Internet of Things (IoT), Fog Computing, social networks, and smart grid. He is a Fellow of the IEEE, and a recipient of the ONR Young Investigator Award in 2005 and
the NSF CAREER award in 2003. He received the IEEE Wireless
Communication Technical Committee Recognition Award in 2016. His papers have won several awards, including the Kenneth C. Sevcik Outstanding Student Paper Award of ACM SIGMETRICS/IFIP Performance 2016, the Best Paper Runner-up Award of IEEE INFOCOM 2009 and IEEE INFOCOM 2014, and the Best Paper Award at IEEE ICC 2008 and ICC 2017. He was TPC co-chair for a number of major conferences in communication networks, including IEEE INFOCOM 2012 and ACM MOBIHOC 2015. He was the general chair for ACM/
IEEE SEC 2017, WiOPT 2016, and IEEE Communication Theory Workshop 2007. He was a Distinguished Lecturer of the IEEE Communications Society. He was an associate editor for IEEE Transactions on Wireless Communications, an editor for the Computer Networks journal, and an editor for IEEE Wireless Communication Magazine. He is currently serving as the editor-in-chief for IEEE Transactions on Wireless Communications, an editor-at-large for IEEE/ACM Transactions on Networking, and an editor for IEEE Network.
\end{IEEEbiography}
 
\end{document}